# מיתרים וממברנות לא קונבנציונליים והאינטראקציות שלהם

מחקר לשם מילוי חלקי של הדרישות לקבלת תואר ״דוקטור לפילוסופיה״

מאת

טטיאנה    וולפס

הוגש לסינאט אוניברסיטת בן גוריון בנגב

18אוקטובר 2020    30 תשרי 5781

באר שבע

מיתרים וממברנות לא קונבנציונליים והאינטראקציות שלהם

מחקר לשם מילוי חלקי של הדרישות לקבלת תואר "דוקטור לפילוסופיה"

מאת

טטיאנה     וולפס

הוגש לסינאט אוניברסיטת בן גוריון בנגב

אישור המנחה
אישור דיקן בית הספר ללימודי מחקר מתקדמים ע"ש קרייטמן

18אוקטוב 2020     30 תשרי 5781

באר שבע

העבודה נעשתה בהדרכת פרופ אדוארדו גנדלמן, פרופ דוד אייכלר

במחלקה לפיזיקה

<u>הצהרת תלמיד המחקר עם הגשת עבודת הדוקטור לשיפוט</u>

אני החתום מטה מצהירה בזאת:

חיברתי את חיבורי בעצמי, להוציא עזרת ההדרכה שקיבלתי מאת המנחים.

תאריך 18אוקטוב 2020        שם התלמידה טטיאנה וולפס        חתימה

# Non-conventional Strings and Branes and Their Interactions.

Thesis submitted in partial fulfillment
of the requirements for the degree of
"DOCTOR OF PHILOSOPHY"

by

Tatiana Vulfs

Submitted to the Senate of Ben-Gurion University
of the Negev

18 October 2020

Beer-Sheva

# Non-conventional Strings and Branes and Their Interactions.

Thesis submitted in partial fulfillment
of the requirements for the degree of
"DOCTOR OF PHILOSOPHY"

by

Tatiana    Vulfs

Submitted to the Senate of Ben-Gurion University
of the Negev

Approved by the advisor

Approved by the Dean of the Kreitman School of Advanced Graduate Studies

18 October 2020

Beer-Sheva

This work was carried out under the supervision of Prof. Eduardo Guendelman
and Prof. David Eichler

In the Physics Department

Research-Student's Affidavit when Submitting the Doctoral Thesis for Judgment

I, Tatiana Vulfs, whose signature appears below, hereby declare that

I have written this Thesis by myself, except for the help and guidance offered by my Thesis Advisors.

Date: 18 October 2020     Student's name: Tatiana Vulfs     Signature:

# Acknowledgements and thanks

I wish to thank

Prof. Eduardo Guendelman,

Prof. David Eichler,

Prof. Horst Stoecker,

Prof. David Owen

and

the Ministry of Aliyah and Integration (IL),

the Israel Science Foundation.

None of this would have been possible without you.

# The Table of Content

# The List of Illustrations

Figure 1. The string meson. Circles are the charged endpoints. $T$ is the tension of the string. $e_1$ and $e_2$ are charges with the condition $e_1 = -e_2$.

Figure 2. The string baryon. Dotted and curved lines denote two strings, $X$ and $Y$, respectively. A cross is the intersection point. $T_1$, $T_2$, $T_3$ are the tensions. The charge's number corresponds to the number of its location, for example, the charge $e_3$ is located at the point $\sigma_3$.

# Hebrew Abstract

תורת המיתרים היא תיאוריה מובילה של כל החומר והאינטראקציות. למרבה התיאוריה רחוקה מלהיות שלמה. תיאוריות המיתרים המתוקנות של המידה, חלקן ידועות וחלקן חדשות לחלוטין, הן הנושא העיקרי במחקר שלנו.

כל מערכת פיזית מאופיינת בפעולה שלה, $S$. כאשר אנו דורשים שפעולת המחרוזת תהיה לורנץ והפרמטריזציה קבועה ושתהיה מוגבלת לא רלוונטית נכונה, אנו מקבלים את פעולת הנמבו-גוטו.

$$S_{Nambu-Goto} = -T \int d\tau d\sigma \sqrt{-\gamma}, \tag{1}$$

כאשר $\tau$ ו- $\sigma$ הם הפרמטרים של גליון העולם, $\gamma$, הוא הקובע של המדד המושרה ו- $T$ הוא מתח המיתרים.

למרות שהתוכן הפיזי של המערכת חייב להישאר קבוע, הפעולה עשויה ללבוש צורות שונות. כאשר אנו דורשים בנוסף על פעולת המחרוזת שלא להכיל שורש ריבועי של המדד, אנו מקבלים את פעולת ה- מודל סיגמא

$$S_{sigma-model} = -\frac{T}{2} \int d\tau d\sigma \sqrt{-h} h^{ab} \partial_a X^\mu \partial_b X^\nu g_{\mu\nu}, \tag{2}$$

כאשר $h^{ab}$ הוא המדד המהותי בגליון העולמי, $h$, הוא הקובע שלו, $g_{\mu\nu}$, הוא מדד המרחב הזמן ו- $X^\mu$ הם שדות גליון העולם הבוסוני.

המדדים הלטיניים הם $0,1$, המדדים היווניים הם $0,1,\ldots,D$, כאשר D הוא הממד של זמן החלל. הסיכום על מדדים חוזרים מובן.

מדד האינטגרציה הסטנדרטי הוא $\sqrt{-h}$. הוא נבחר על סמך דרישה יחידה שזו חייבת להיות צפיפות תחת טרנספורמציות דיאומורפיות. לכן $\sqrt{-h}$, לא יכול להיות בחירה ייחודית.

המטרה העיקרית שלנו היא להציע פעולה שונה ולבדוק אותה למספר משימות.

ראשית, אנו מציגים מדד חדש, $\Phi(\chi)$, הבנוי משדה סקלרי אחד, $\chi$.

$$\Phi(\chi) = \partial_a(\sqrt{-h} h^{ab} \partial_b \chi). \tag{3}$$

השדה הסקלרי $\chi$ נקרא שדה סקלרי של גליאון מכיוון ש- $\Phi(\chi)$ הוא משתנה תחת סימטריה של משמרת גליאון:

$$\partial_a \chi \to \partial_a \chi + b_a, \qquad \chi \to \chi + b_a \sigma^a, \tag{4}$$

כאשר $b_a$ הוא וקטור קבוע ו- $\sigma^a = (\tau, \sigma)$.

פעולת המידה של גליאון, $S_{GM}$, היא

$$S_{GM} = -\int d\tau d\sigma (h^{ab} \partial_a X^\mu \partial_b X^\nu g_{\mu\nu} - \frac{\epsilon^{cd}}{\sqrt{-h}} F_{cd}) \partial_e (h^{ef} \sqrt{-h} \partial_f \chi), \tag{5}$$

כאשר $\epsilon^{cd}$ הוא סמל לוי-סיוויטה, $F_{cd} = \partial_c A_d + \partial_d A_c$ הוא חוזק השדה של $A_c$.

הקריטריון לקבלת הפעולה כתקפה הוא כי משוואות התנועה המתקבלות זהות לאלו של פעולת מודל הסיגמה. נקודת המפתח היא שכמות משוואות התנועה תלויה בכמות המשתנים הדינמיים עשויה להיות שונה. ומתאפשר לקבל את המשמעות לקבועים שהונחו ביד לפני כן.

תוך כדי ניסוח תיאוריה מנסים לחסל את כל הפרמטרים או לפחות כמה שיותר. פחות פרמטרים יכולים להפוך את התיאוריה לעקבית וייחודית יותר. פעולת הנמבו-גוטו ופעולת מודל הסיגמה מכילות את מתח המיתרים, $T$, במפורש. זה מביא קנה מידה לתיאוריה שהיא לחילופין אינווריאנט בקנה מידה.

לסיכום, למחרוזת המדידה של גלילאון יש שני מאפיינים קריטיים:

1. המתח במיתרים אינו לשים ביד אלא נראה כמידת חופש דינמית נוספת;

2. מוצגת הסימטריה החדשה וכל הדרישות של תורת הנגזרת הגבוהה יותר של גלילאון ברמת הפעולה מתקיימות, בעוד שמשוואות התנועה עדיין מהסדר השני.

שנית, אנו מרחיבים את תורת מיתרי הגלילאון לתורת מיתרי העל במדידת הגלילאון.

אין עדויות ניסיוניות לכך שקיימת סופר-סימטריה. עם זאת, אנו מבינים שכל תיאוריה אינה שלמה בלי להכניס את הבוזונים. יש להכניס פרמיונים. כשכוללים את הפרמיונים, נדרשת הסופר-סימטריה. ישנן שתי גישות סטנדרטיות, הגישה של רמונד - נבו - שווארץ וגישת הירוק - שווארץ. ההבדל העיקרי ביניהם נובע מהמקום בו אנו מניחים שמתקיים סופר-סימטריה: על הזמן חלל או על גליון העולם.

הפעולה העל-סימטרית של גליון העולם, $S_{RNS\_SUSY}$, היא

$$S_{RNS\_SUSY} = -\frac{1}{2\pi} \int d\tau d\sigma (\partial_a X_\mu \partial^a X^\mu + \bar{\psi}^\mu \rho^a \partial_a \psi_\mu - B_\mu B^\mu), \qquad (6)$$

כאשר $\psi^\mu$ הם שדות הגליונות העולמיים הפרמיוניים, $\rho^a$ הם מטריצות ה- דיראק הדו-ממדיות ו- $B_\mu$ הוא שדה העזר.

הפעולה העל-סימטרית בחלל, $S_{Green-Schwarz SUSY}$, היא

$$S_{Green-Schwarz SUSY} = -T \int d\tau d\sigma (\mathcal{L}_{simpleSUSY} + \mathcal{L}_{additionalSUSY}), \qquad (7)$$

כאשר

$$\mathcal{L}_{simpleSUSY} = \frac{1}{2}\sqrt{-h}h^{ab}\Pi_a^\mu \Pi_{b\mu}, \qquad \Pi_a^\mu = \partial_a X^\mu - i(\Theta^A \Gamma^\mu \partial_a \Theta^A), \qquad (8)$$

$$\mathcal{L}_{additionalSUSY} = i\epsilon^{ab}\partial_a X^\mu (\bar{\Theta}^1 \Gamma_\mu \partial_b \Theta^1 + \bar{\Theta}^2 \Gamma_\mu \partial_b \Theta^2) - \epsilon^{ab}\bar{\Theta}^1 \Gamma^\mu \partial_a \Theta^1 \bar{\Theta}^2 \Gamma_\mu \partial_b \Theta^2, \qquad (9)$$

כאשר $\Theta^A$ הם הקוארדינטות הפרמיוניות, $\Gamma^\mu \equiv \Gamma^\mu_{\alpha\beta}$ הם המטריצות של דיראק.

הפעולה (27) אינה משתנה תחת טרנספורמציות סופר-פואנקרה ותחת דיפרומורפיזמים. כדי לקחת בחשבון את הסימטריה הפרמיונית המקומית הנוספת (או סימטריית הקאפה), מניחים וס-זומינו, $\mathcal{L}_{additionalSUSY}$. הבעיה היא ש $\mathcal{L}_{additionalSUSY}$ הוא משתנה תחת טרנספורמציות גלובליות רק עד סך הנגזרות.

זה נפתר על ידי וו. סיגל ב- [hep-th / 9403144]. ניסוח מחדש של הפעולה מבוסס על רעיון הזרמים העל-סימטריים. הפעולה העל-סימטרית של סיגל, $S_{Siegel\_SUSY}$, היא

$$S_{Siegel\_SUSY} = -T \int d\tau d\sigma \sqrt{-h}(\mathcal{L}_{simpleSiegel} + \mathcal{L}_{additionalSiegel}), \qquad (10)$$

כאשר

$$\mathcal{L}_{simpleSiegel} = \frac{1}{2} h^{ab} \Pi_a^\mu \Pi_{b\mu}, \qquad \mathcal{L}_{additionalSiegel} = i \frac{\epsilon^{cd}}{\sqrt{-h}} J_c^\alpha J_{\alpha d}, \qquad (11)$$

$$J_a^\alpha = \partial_a \Theta^\alpha, \qquad J_{\alpha a} = \partial_a \phi_\alpha - 2i(\partial_a X^\mu)\Gamma_{\mu\alpha\beta}\Theta^\beta - \frac{2}{3}(\partial_a \Theta^\beta)\Gamma_{\beta\delta}^\mu \Theta^\delta \Gamma_{\mu\alpha\epsilon}\Theta^\epsilon. \qquad (12)$$

הסימטריה העולמית מדויקת כעת בגלל $\phi_\alpha$ המאפשר לבטא את המונח וס-זומינו בצורה העל-סימטרית בעליל.

עם זאת, בעיה חדשה היא ששדות $\phi_\alpha$ אינם דינמיים. זו הסיבה לערב את תיאוריות המידה המתוקנות אשר יעילות ביותר במתן משמעות לקבועים הפתאומיים. מיתר העל של שני המידות כבר נחקר, מיתר הגליליאון חדש לחלוטין.

פעולת מיתר העל של גליליאון, $S_{GM\_SUSY}$, היא

$$S_{GM\_SUSY} = -\int d\tau d\sigma \partial_a(h^{ab}\sqrt{-h}\partial_b\chi)(\mathcal{L}_{simpleSiegel} + \mathcal{L}_{additionalSiegel}). \qquad (13)$$

מה שחשוב, יש לנו את משוואת התנועה הנוספת שקובעת $\phi_\alpha$.

לסיכום, למיתר העל של גליליאון יש שני מאפיינים קריטיים:

1. המונח וס-זומינו, $\mathcal{L}_{additionalSiegel}$, הוא על-סימטרי בעליל,

2. הפעולה מוצגת עם כל המונחים הנגזרים ממשוואות התנועה.

שלישית, אנו זזים הצידה ומציגים את הרקע של תיאוריית שני המידות הכרוכות בפרספקטיבה ידועה ונקודת מבט חדשה על אי-שינוי הקנה המידה.

הממד הדו-ממדי, $\Phi(\varphi)$, בשני ממדים הוא

$$\Phi(\varphi) = \epsilon^{ab}\epsilon_{ij}\partial_a\varphi^i\partial_b\varphi^j, \qquad (14)$$

כאשר $\varphi^i, \varphi^j$ הם שדות סקלריים שאין להם שום קשר ל- $\phi$ הבא.

מערכת השדה הסקלרית השנייה היא

$$S = \frac{1}{2}\int \Phi(\varphi)\mathcal{L}d^2x, \qquad \mathcal{L} = \frac{1}{2}(\partial_\mu\phi_1\partial_\nu\phi_1 g^{\mu\nu} + \partial_\mu\phi_2\partial_\nu\phi_2 g^{\mu\nu}), \tag{15}$$

כאשר $\phi_1$ הוא השדה הסקלרי לשעבר $\phi$ ו- $\phi_2$ הוא השלב המוסף.

התוצאות הן

1. מעצורים טבעיים בשדה הסקלרי שלנו, שאז מתייחסים אליהם כאל סקלר נולד-אינפלד.

$$S_{eff} = \int \sqrt{const - \partial_\mu\phi_1\partial^\mu\phi_1}d^2x; \tag{16}$$

2. הפרה דינמית באופן טבעי הנובעת מסטיות המידה;

3. הדגמה שלמרות היותה סימטריה, אי-שינוי הקנה המידה אינו מוביל לשימור מטען האבנית.

שתי שדות סקלריים סקלריים עם פוטנציאל מוחלים על כוח המשיכה וכבר ידועים.

רביעית, אנו מציגים את שני מודלי המיתרים החדשים של הדרונים במסגרת תיאוריית הדו-מדד. במודל מחרוזת מזון יש מחרוזת אחת ושתי מטענים מנוגדים בנקודות הקצה. במודל בריון המיתרים ישנם שני מיתרים $X,$ ו- $Y$, שני זוגות מטענים מנוגדים בנקודות הקצה ומטען נוסף בנקודת הצומת, הנקודה שבה מחברים שני מיתרים אלה.

הפעולה המסדירה את תצורת מסון המחרוזת היא

$$S_{meson} = -\int d\sigma d\tau \Phi(\varphi)[\frac{1}{2}h^{ab}\partial_a X^\mu \partial_b X^\nu g_{\mu\nu} - \frac{\epsilon^{cd}}{2\sqrt{-h}}F_{cd}] + \int d\sigma d\tau A_i j^i, \tag{17}$$

כאשר $j^i$ הוא הנוכחי של חיובים דמויי נקודה.

הפעולה המסדירה את תצורת בריון המיתרים היא

$$S_{baryon} = -\int d\tau d\sigma \Phi(\varphi)_X[\frac{1}{2}h_X^{ab}\partial_a X^\mu \partial_b X^\nu g_{\mu\nu} - \frac{\epsilon^{cd}}{2\sqrt{-h_X}}F_{cd}] + \sum_{i=1,2}\int d\tau d\sigma A_i j_A^i +$$

$$-\int d\tau d\sigma \Phi(\varphi)_Y[\frac{1}{2}h_Y^{ab}\partial_a Y^\mu \partial_b Y^\nu g_{\mu\nu} - \frac{\epsilon^{cd}}{2\sqrt{-h_Y}}F_{cd}] + \sum_{j=3,4,5}\int d\tau d\sigma B_j j_B^j +$$

$$+ \int d\tau d\sigma (\lambda_1\sqrt{-h_X}h_X^{ab} + \lambda_2\sqrt{-h_Y}h_Y^{ab})\partial_a(\frac{\Phi(\varphi)_X}{\sqrt{-h_X}})\partial_b(\frac{\Phi(\varphi)_Y}{\sqrt{-h_Y}})V(X,Y), \tag{18}$$

כאשר $\lambda_1, \lambda_2$ הם מקדמים חיוביים ו- $V(X,Y)$ הוא פוטנציאל שבצורתו הפשוטה ביותר הוא $V = (X-Y)^2$.

היתרונות של הדגמים שלנו הם

1. תנאי הגבול של נויטן, $\partial_\sigma X^\mu(\tau,\sigma)$, מתקבלים באופן דינמי בכל נקודה בה נמצא המטען;

2. תנאי הגבול של הדיריכלט, $X^\mu|_\sigma = Y^\mu|_\sigma$ נוצרים באופן טבעי בנקודת הצומת.

3. בעיית האי-לוקליות נפתרה. המדד המתוקן מוביל למתח הדינמי; המתח הדינמי מוביל לתנאי הגבול של נויטן בנקודת הצומת; תנאי הגבול של נויטן בנקודת הצומת מובילים לפתרון בעיית האי-לוקליות שעליה הצביע על ידי ג׳ יט הופט ב- [hep-th / 0408148] לגישה הסטנדרטית יותר.

לשם השלמות מוצגים הפתרונות של משוואות התנועה. בהנחה שכל נקודת קצה היא החלקיק הדינמי חסר המסה, מתקבל מסלול רג׳ה עם פרמטר השיפוע התלוי בשלושה מתחים שונים.

חמישית, אנו מרחיבים את המחקר לאובייקטים מורחבים בעלי ממדים גבוהים יותר. האובייקטים המורחבים הדו-ממדים, $\Phi(\varphi)$, ידועים כבר. המדד החדש של גליאון, $\Phi(\chi)$, אובייקטים מורחבים התמודדו עם הבעיה. רק מיתרים, שהם אובייקטים דו-ממדים, הם בעלי סימטריה של גליאון והם קבועים בפורמט בו זמנית. לכן אנו בונים אובייקט מורחב קבוע כלשהו עם אחד האמצעים שהשתמשנו

$$\Phi(\chi) = \partial_u(h^{ux}\sqrt{-h}\partial_x\chi(-2h^{yz}\partial_y\chi\partial_z\chi)^{\frac{D-2}{2}}), \qquad \Phi(\chi) = \partial_u(h^{ux}\sqrt{-h}\partial_x\chi(F_{yz}F^{yz})^{\frac{D-2}{4}}), \qquad (19)$$

כאשר $u,x,y,z = 0,1,\ldots,p$ ו- $p$ הוא ממד הבריין

או בלתי-קונפורמי שאינו קונפורמי אך עדיין עם הסימטריה של הגליאון. עם זאת, משוואות התנועה המקוריות אינן מתקבלות.

# A List of Keywords in Hebrew

מידה שונה; מדד גלילאון; שתי מידות; מיתר העל של סיגל; מיתר על מיתר שונה; אי-שינוי בקנה מידה; סקלר נולד-אינפלד; אנרגיית ואקום שאינה מכבידה; הדרונים מיתרים; דגם מזון מחרוזת; דגם בריון מחרוזת; דגם הברון של הופט; אי-מקומיות; עצמים מורחבים ממדיים גבוהים יותר

# English abstract

String theory is a leading theory of all matter and interactions. It is far from complete. In String theory there are still some classical issues that require to be understood better. For example, the origin of string tension and how to make a string tension a dynamical variable, different ways to understand the string splitting, building mesons and baryons out of strings by taking advantage of the dynamical tension theory which is available to us through the modified measure approach.

The modified measure string theories, some known and some completely new, are the main topic of our research. The whole thesis deals only with the classical actions.

Every physical system is characterized by its action, $S$. When we require the string action to be Lorentz and reparameterization invariant and have a correct nonrelativistic limit, we obtain the Nambu-Goto action.

$$S_{Nambu-Goto} = -T \int d\tau d\sigma \sqrt{-\gamma}, \tag{20}$$

where $\tau$ and $\sigma$ are the worldsheet parameters, $\gamma$ is the determinant of the induced metric and $T$ is the string tension.

Even though the physical content of the system must remain constant, the action may take different forms. When we additionally require the string action not to contain a square root of the metric, we obtain the sigma-model action

$$S_{sigma-model} = -\frac{T}{2} \int d\tau d\sigma \sqrt{-h} h^{ab} \partial_a X^\mu \partial_b X^\nu g_{\mu\nu}, \tag{21}$$

where $h^{ab}$ is the intrinsic metric on the worldsheet, $h$ is its determinant, $g_{\mu\nu}$ is the spacetime metric and $X^\mu$ are the bosonic worldsheet fields.

The Latin indices are $0, 1$, the Greek indices are $0, 1, \ldots, D$, where D is the dimension of spacetime. The summation over repeated indices is understood.

The standard measure of integration is $\sqrt{-h}$. It is chosen on the basis of a single requirement that it must be a density under diffeomorphic transformations. Therefore, $\sqrt{-h}$ may not be a unique choice.

Our main purpose is to suggest a modified action and to test it for several tasks.

First, we introduce a new measure, $\Phi(\chi)$, which is constructed out of one scalar field, $\chi$.

$$\Phi(\chi) = \partial_a(\sqrt{-h} h^{ab} \partial_b \chi). \tag{22}$$

The scalar field $\chi$ is called a Galileon scalar field since $\Phi(\chi)$ is invariant under a Galileon shift symmetry:

$$\partial_a \chi \to \partial_a \chi + b_a, \qquad \chi \to \chi + b_a \sigma^a, \tag{23}$$

where $b_a$ is a constant vector and $\sigma^a = (\tau, \sigma)$.

The Galileon measure action, $S_{GM}$, is

$$S_{GM} = -\int d\tau d\sigma (h^{ab}\partial_a X^\mu \partial_b X^\nu g_{\mu\nu} - \frac{\epsilon^{cd}}{\sqrt{-h}}F_{cd})\partial_e(h^{ef}\sqrt{-h}\partial_f \chi), \tag{24}$$

where $\epsilon^{cd}$ is the Levi-Civita symbol, $F_{cd} = \partial_c A_d - \partial_d A_c$ is the field-strength of $A_c$.

The criterion for accepting the action as valid is that the resulting equations of motion are the same as those of the sigma-model action. The key point is that the number of lagrangian variables can differ. Therefore, the number of the equations of motion which is equal to the number of lagrangian variables also differ in different formulations. It becomes possible to prescribe the meaning to the constants that were put by hand before.

While formulating a theory, one tries to eliminate all the parameters or at least as many as possible. The fewer number of parameters can make the theory more consistent and unique. The Nambu-Goto action and the sigma-model action contain the string tension, $T$, explicitly. It brings a scale in the otherwise scale invariant theory.

In sum, the Galileon measure string has two critical properties:

1. the string tension isn't put by hand but appears as an additional dynamical degree of freedom;

The string tension is generated even without having charges. Even for the closed string. The equation for the internal gauge field $A_a$ gives us the result

$$\frac{\Phi}{\sqrt{-h}} = const. \tag{25}$$

This constant is identified as the string tension. No charges have been introduced in this case. The charges only introduce discontinuities of the string tension. This constant is a scale that is generated dynamically.

2. the new symmetry is introduced, and all requirements of the Galileon higher derivative theory at the action level are satisfied, while the equations of motion are still of the second order.

Second, we extend the Galileon string theory to the superstring theory with the Galileon measure.

There is no experimental evidence that supersymmetry exists. However, we realize that every theory is only half full if only the bosons are presented. Fermions must be introduced. When including the fermions, the supersymmetry is required. There are two standard approaches, the Ramond - Neveu - Schwarz approach and the Green - Schwarz approach. The main difference between them is where we assume the supersymmetry to be, namely, on the spacetime or on the worldsheet.

The worldsheet supersymmetric action, $S_{RNS\_SUSY}$, is

$$S_{RNS\_SUSY} = -\frac{1}{2\pi}\int d\tau d\sigma(\partial_a X_\mu \partial^a X^\mu + \bar\psi^\mu \rho^a \partial_a \psi_\mu - B_\mu B^\mu), \tag{26}$$

where $\psi^\mu$ are the fermionic worldsheet fields, $\rho^a$ are the two-dimensional Dirac matrices and $B_\mu$ is the auxiliary field.

The spacetime supersymmetric action, $S_{Green-SchwarzSUSY}$, is

$$S_{Green-SchwarzSUSY} = -T\int d\tau d\sigma(\mathcal{L}_{simpleSUSY} + \mathcal{L}_{additionalSUSY}), \tag{27}$$

where

$$\mathcal{L}_{simpleSUSY} = \frac{1}{2}\sqrt{-h}h^{ab}\Pi_a^\mu \Pi_{b\mu}, \qquad \Pi_a^\mu = \partial_a X^\mu - i(\bar\Theta^A \Gamma^\mu \partial_a \Theta^A), \tag{28}$$

$$\mathcal{L}_{additionalSUSY} = i\epsilon^{ab}\partial_a X^\mu(\bar\Theta^1 \Gamma_\mu \partial_b \Theta^1 + \bar\Theta^2 \Gamma_\mu \partial_b \Theta^2) - \epsilon^{ab}\bar\Theta^1\Gamma^\mu\partial_a\Theta^1\bar\Theta^2\Gamma_\mu\partial_b\Theta^2, \tag{29}$$

where $\Theta^A$ are the fermionic coordinates, $\Gamma^\mu \equiv \Gamma^\mu_{\alpha\beta}$ are the Dirac matrices.

The action (27) is invariant under super-Poincare transformations and under diffeomorphisms. To take the additional local fermionic symmetry (or the kappa symmetry) into account, the Wess-Zumino term, $\mathcal{L}_{additionalSUSY}$, is introduced. The problem is that $\mathcal{L}_{additionalSUSY}$ is invariant under global transformations only up to total derivatives.

It is solved by W. Siegel in [hep-th/9403144]. The reformulation of the action is based on the idea of the supersymmetric currents. The Siegel supersymmetric action, $S_{Siegel\_SUSY}$, is

$$S_{Siegel\_SUSY} = -T\int d\tau d\sigma \sqrt{-h}(\mathcal{L}_{simpleSiegel} + \mathcal{L}_{additionalSiegel}), \tag{30}$$

where

$$\mathcal{L}_{simpleSiegel} = \frac{1}{2}h^{ab}\Pi_a^\mu \Pi_{b\mu}, \qquad \mathcal{L}_{additionalSiegel} = i\frac{\epsilon^{cd}}{\sqrt{-h}}J_c^\alpha J_{\alpha d}, \tag{31}$$

$$J_a^\alpha = \partial_a \Theta^\alpha, \qquad J_{\alpha a} = \partial_a \phi_\alpha - 2i(\partial_a X^\mu)\Gamma_{\mu\alpha\beta}\Theta^\beta - \frac{2}{3}(\partial_a \Theta^\beta)\Gamma^\mu_{\beta\delta}\Theta^\delta \Gamma_{\mu\alpha\epsilon}\Theta^\epsilon. \tag{32}$$

The global symmetry is exact now because of $\phi_\alpha$ that allows the Wess-Zumino term to be expressed in the manifestly supersymmetric way.

However, a new problem is that the $\phi_\alpha$ fields are not dynamical. That is the reason to involve the modified measure theories which are extremely useful in giving meaning to the yet unknown constants. The two-measure superstring was already studied, the Galileon superstring is completely new.

The Galileon measure superstring action, $S_{GM\_SUSY}$, is

$$S_{GM\_SUSY} = -\int d\tau d\sigma \partial_a(h^{ab}\sqrt{-h}\partial_b\chi)(\mathcal{L}_{simpleSiegel} + \mathcal{L}_{additionalSiegel}). \tag{33}$$

What matters, we have the additional equation of motion that determines $\phi_\alpha$.

In sum, the Galileon superstring has two critical properties:

1. the Wess-Zumino term, $\mathcal{L}_{additionalSiegel}$, is manifestly supersymmetric,

2. the action is presented with all terms being derived from the equations of motion.

Third, we step aside and present some known and some new aspects of the two-measure theory with the focus on the scale invariance.

The two-measure, $\Phi(\varphi)$, in two dimensions is

$$\Phi(\varphi) = \epsilon^{ab}\epsilon_{ij}\partial_a\varphi^i\partial_b\varphi^j, \tag{34}$$

where $\varphi^i, \varphi^j$ are scalar fields that have nothing to do with the following $\phi$.

The two scalar field system is

$$S = \frac{1}{2}\int \Phi(\varphi)\mathcal{L}d^2x, \qquad \mathcal{L} = \frac{1}{2}(\partial_\mu\phi_1\partial_\nu\phi_1 g^{\mu\nu} + \partial_\mu\phi_2\partial_\nu\phi_2 g^{\mu\nu}), \tag{35}$$

where $\phi_1$ is the former scalar field $\phi$ and $\phi_2$ is the supplemented one.

The outcomes are

1. a naturally arising restraints on our scalar field, that is then treated as a Born-Infeld scalar

$$S_{eff} = \int \sqrt{const - \partial_\mu\phi_1\partial^\mu\phi_1}d^2x; \tag{36}$$

2. a naturally arising dynamical violation of the scale invariance;

3. a demonstration that despite being a symmetry, the scale invariance does not lead to the conservation of the scale charge.

The two scalar field system with potentials is applied to gravity and is already known.

Fourth, we present the two new string models of hadrons in the framework of the two-measure theory. In the string meson model there are one string and two opposite charges at the endpoints. In the string baryon model there are two strings, $X$ and $Y$, two pairs of opposite charges at the endpoints and one additional charge at the intersection point, the point where these two strings are connected.

The action governing the string meson configuration is

$$S_{meson} = -\int d\sigma d\tau \Phi(\varphi)[\frac{1}{2}h^{ab}\partial_a X^\mu \partial_b X^\nu g_{\mu\nu} - \frac{\epsilon^{cd}}{2\sqrt{-h}}F_{cd}] + \int d\sigma d\tau A_i j^i, \qquad (37)$$

where $j^i$ is the current of point-like charges.

The action governing the string baryon configuration is

$$S_{baryon} = -\int d\tau d\sigma \Phi(\varphi)_X[\frac{1}{2}h_X^{ab}\partial_a X^\mu \partial_b X^\nu g_{\mu\nu} - \frac{\epsilon^{cd}}{2\sqrt{-h_X}}F_{cd}] + \sum_{i=1,2}\int d\tau d\sigma A_i j_A^i +$$

$$-\int d\tau d\sigma \Phi(\varphi)_Y[\frac{1}{2}h_Y^{ab}\partial_a Y^\mu \partial_b Y^\nu g_{\mu\nu} - \frac{\epsilon^{cd}}{2\sqrt{-h_Y}}F_{cd}] + \sum_{j=3,4,5}\int d\tau d\sigma B_j j_B^j +$$

$$+\int d\tau d\sigma (\lambda_1\sqrt{-h_X}h_X^{ab} + \lambda_2\sqrt{-h_Y}h_Y^{ab})\partial_a(\frac{\Phi(\varphi)_X}{\sqrt{-h_X}})\partial_b(\frac{\Phi(\varphi)_Y}{\sqrt{-h_Y}})V(X,Y), \qquad (38)$$

where $\lambda_1, \lambda_2$ are positive coefficients and $V(X,Y)$ is a potential which in its most simple form is $V = (X-Y)^2$.

The benefits of our models are

1. the Neumann boundary conditions, $\partial_\sigma X^\mu(\tau,\sigma) = 0$, are obtained dynamically at every point where the charge is located;

2. the Dirichlet boundary conditions, $X^\mu|_\sigma = Y^\mu|_\sigma$, arise naturally at the intersection point.

3. The problem of nonlocality is resolved. The modified measure leads to the dynamical tension; the dynamical tension leads to the Neumann boundary conditions at the intersection point; the Neumann boundary conditions at the intersection point leads to the resolution of the nonlocality problem that was pointed out by G. 't Hooft in [hep-th/0408148] for the more standard approach.

For the sake of completeness the solutions of the equations of motion are presented. Assuming that each endpoint is the dynamical massless particle, the Regge trajectory with the slope parameter that depends on three different tensions is obtained.

Fifth, we extend the research to higher dimensional extended objects. The two-measure, $\Phi(\varphi)$, extended objects are already known. The new Galileon measure, $\Phi(\chi)$, extended objects face the problem. Only strings, being two-dimensional objects, have the Galileon symmetry and are conformal invariant at the same time. Therefore, we construct either conformal invariant extended object with one of the modified measures

$$\Phi(\chi) = \partial_u(h^{ux}\sqrt{-h}\partial_x\chi(-2h^{yz}\partial_y\chi\partial_z\chi)^{\frac{D-2}{2}}), \qquad \Phi(\chi) = \partial_u(h^{ux}\sqrt{-h}\partial_x\chi(F_{yz}F^{yz})^{\frac{D-2}{4}}), \qquad (39)$$

where $u, x, y, z = 0, 1, \ldots, p$ and $p$ is the dimension of the brane or non-conformal invariant extended object but still with the Galileon symmetry. However, the original equations of motion are not obtained.

# A List of Keywords in English

modified measure; Galileon measure; two-measure; Siegel superstring; modified measure superstring; scale invariance; Born-Infeld scalar; non-gravitating vacuum energy; stringy hadrons; string meson model; string baryon model; 't Hooft baryon model; nonlocality; higher dimensional extended objects

# Scientific publications

1. T.O. Vulfs, E.I. Guendelman, "Galileon string measure and other modified measure extended objects", Mod.Phys.Lett. A32, no.38, 1750211, (2017).

2. T.O. Vulfs, E.I. Guendelman, "Superstrings with the Galileon Measure", Annals Phys. 398, 138-145, (2018).

3. T.O. Vulfs, E.I. Guendelman, "String Model with Mesons and Baryons in Modified Measure Theory", Int.J.Mod.Phys. A34, no.31, 1950204, (2019).

4. T.O. Vulfs, E.I. Guendelman, "A Born-Infeld Scalar and a Dynamical Violation of the Scale Invariance from the Modified Measure Action", Mod.Phys.Lett.A 35, 24, 2050198, (2020).

# 1   Introduction.

String theory [1, 2, 3, 4] is a leading theory of all matter and interactions. It is far from complete.

Its main areas of research include, for example, gauge theory/string theory dualities [5, 6, 7, 8, 9, 10, 11, 12, 13, 14, 15, 16, 17, 18, 19, 20, 21, 22] that is the equivalence between string theories and theories based on local fields for certain background geometries. Another direction of research concerns the additional dimensions which is the distinctive feature of String theory. Even the most simple case of the bosonic string requires 26 dimensions. The flux compactification being a particular way to deal with extra dimensions is extensively studied in [23, 24, 25, 26, 27]. Black holes and the information loss puzzle, singularities and their thermodynamic description are studied in [28, 29, 30, 31, 32, 33].

In String theory there are still some classical issues that require to be understood better. For example, the origin of string tension and how to make a string tension a dynamical variable, different ways to understand the string splitting, building mesons and baryons out of strings by taking advantage of the dynamical tension theory which is available to us through the modified measure approach.

String Theory begins with the discovery of a 4-particle scattering amplitude for open strings in [34]. The recognition that these amplitudes (including multiparticle and close-string amplitudes) actually describe one-dimensional extended objects was made in [35, 36, 37]. The derivation of the string action as the area of the worldsheet was introduced in [38, 39, 40].

The main topic of our research is the modified measure string theories, some known and some completely new. Our main purpose is to suggest a modified action and to test it for several tasks. The whole thesis deals only with the classical actions.

When considering the action formulation of a theory, the standard measure of the integration, $\sqrt{-g}$ (where $g$ is the determinant of the metric), is usually used. It must be a density under diffeomorphic transformations and therefore, $\sqrt{-g}$ is not a unique choice.

The modified measure theory was mostly applied to gravity problems. One of the problems is the cosmological constant problem. It was considered in [41]. The Einstein field equations for the general theory of relativity are

$$R_{\mu\nu} - \frac{1}{2}Rg_{\mu\nu} - \Lambda g_{\mu\nu} = \kappa T_{\mu\nu}, \tag{40}$$

where $R_{\mu\nu}$ is the Ricci curvature tensor, $R$ is the scalar curvature, $T_{\mu\nu}$ is the energy-momentum tensor, $\kappa = 16\pi G$ and $\Lambda$ is the cosmological constant.

Quantum field theory (QFT) predicts the existence of a vacuum energy due to the zero point fluctuations. It gives an infinite contribution to $T_{\mu\nu}$ which is indistinguishable from the $\Lambda$-term. The essence of the cosmological constant problem is that QFT prediction for the value of zero-point energy is more than $10^{120}$ times larger than the observed value of vacuum energy density.

The cosmological constant in the Einstein frame is a dynamical constant that can be made very small, for example, using a see-saw mechanism [42]. A see-saw mechanism can provide a long lived almost constant vacuum energy for a long period of time, which can be small if $\frac{f_1^2}{4f_2}$ is small, where $f_1$ is a coefficient of a potential that appears in the modified measure and $f_2$ is a coefficient of a potential that appears in the normal measure. If we take $f_1$ determined by the electroweak scale and $f_2$ determined by the Planck scale, then we get the right magnitude of the observed vacuum energy, $\frac{(M_{EW})^8}{(M_{Planck})^4}$.

Another gravity problem is the fifth force problem which was considered in [43].

A long range force can be originated from the coupling of the matter to a scalar field if the mass of the scalar particles is very small. Such fifth force could affect the results of tests of General relativity.

The modified measure model not only explains why all attempts to discover a scalar force correction to Newtonian gravity were unsuccessful so far but also predicts that in the near future there is no chance to detect such corrections in the astronomical measurements as well as in the specially designed fifth force experiments on intermediate, short (like millimeter) and even ultrashort (a few nanometer) ranges.

A stable emerging universe scenario was introduced in [44]. The inflationary phase of the early universe is a substantial concept in modern cosmology. However, in the context of the inflationary scenario one encounters the initial singularity problem which remains unsolved, showing that the universe necessarily had a beginning for generic inflationary cosmological model.

A modified measure scale invariant model which includes a $R^2$ term was proposed. It has made it possible to avoid instabilities and to achieve an inflationary phase.

Dark energy and dark matter are among the fundamental problems of the theoretical physics. Standard model does not contain them but it is mostly clear from the observations that they do exist. In the framework of the modified measure theory the dark energy/dark matter scenarios were considered in [45, 46]. Generalized ideas of unified dark matter and dark energy in the context of dynamical space time theories with a diffusive transfer of energy were proposed.

In our research we consider the modified measure string theories. Our research is divided into five parts.

The goal of the first part is to present a new Galileon string action and discuss its properties.

We start with a couple of textbook actions of a point particle and look through the basic principles and methods. Then we move on to a string. A worldline turns to the worldsheet. A proper time turns to the proper area. The analogy with the point particle leads to the Nambu-Goto action and further to the sigma-model action.

Then we look at the modified measure strings. The two-measure string was studied in [47, 48, 49]. The main benefit of this model is the absence of an ad hoc parameter, i.e. nothing is put by hand but appears as an additional degree of freedom. The string tension is derived from the equations of motion.

We take the modified theory further, and introduce a totally new measure of integration, the Galileon measure. It keeps the two-measure string property and contains a new symmetry.

The Galileon symmetry have been discussed in gravitational theories. We show it works much simpler in string theory. In particular, the Galileon symmetry is well defined in the conformal frame as opposed as in gravity theories.

The idea to modify a measure in the Galileon way is originated from the Galileon modification of gravity [50, 51, 52]. The Galileon gravity is a scalar-tensor theory with a nonminimal coupling of the special scalar Galileon to curvature, with the second order equations of motion and nonetheless, without ghosts. This theory is invariant under a Galileon shift symmetry. This very symmetry is a base for our new measure.

The goal of the second part is to present a new Galileon superstring and discuss its properties.

We start with the textbook string symmetries. Then we move on to the supersymmetry. First, we look at the supersymmetry of a standard measure string. In the Green-Schwarz formulation [53] the Wess-Zumino term is invariant only up to a total divergence. In Siegel reformulation [54] this term becomes manifestly supersymmetric. But the price to pay is that now it consists of some vector fields that in principle are not determined by the equations of motion.

Then we look at the two-measure string which solves this problem. Again, not a single parameter is put by hand.

We introduce the new Galileon supersymmetric string action which also solves this problem and in addition has a Galileon symmetry.

The goal of the third part is to consider the two scalar field system in two measure theory. We step aside from strings and consider the modified measure theory by itself. The two scalar field system with potentials in two measure theory was studied in [55, 56].

We look for a sector which can be presented in the form of the Born-Infeld scalar. The Born-Infeld theory was originated as a specific theory of nonlinear electrodynamics in [57], it put limitations on the self-energy of a point charge. Later it reappeared in string theory to describe the electromagnetic fields on the world-volumes of D-branes as it guarantees that the energy of the string is finite in [58, 59]. Recently, to bring limits on scalar fields in cosmology, the Born-Infeld scalar was considered in [60, 61]. This integration was developed later in [62, 63, 64, 65, 66, 67, 68, 69].

Then we consider our initial action for the scale invariance which, however, gets spontaneously broken. Moreover, in addition to having spontaneous symmetry breaking, our physical system serves as an example of a system with the symmetry that does not lead to the conserved charge.

The anomalous infrared behavior of the conserved chiral current in the presence of instantons was discussed in [70]. The conclusion was made that in this case there was no conserved $U(1)$ charge and Goldstone's theorem therefore failed to solve the $U(1)$ problem in QCD. The case of global scale invariance in the presence of a modified measure was considered in [71, 72, 73, 74, 75, 76, 77, 78, 79] and the dilatation currents were calculated in a special model in [42], where the current was shown to be singular in the infrared. Here also, the resulting scale current produces a nonzero flux of the dilatation current to infinity, so once again, although there is a conserved current, there is no conserved scalar charge.

A consideration of a scale invariance in cosmology started in [80, 81] and was continued in [82, 83]. When the scale symmetry is spontaneously broken, there is a conserved current. And since no singular behavior of the conserved current is obtained, there is a conserved scale charge and the Goldstone theorem holds.

The goal of the fourth part is to present a new string baryon model and a new string meson model.

String models of hadrons are circulating in the literature [84, 85, 86, 87, 88, 89, 90, 91, 92, 93, 94, 95] since 1974. While there is only one possible string configuration to represent a meson, that is, a single string with the opposite charges at both endpoints, the baryons have more freedom. Three strings with the charges at each endpoint can be arranged, for example, in $\Delta$-model, in the $Y$-shaped model which requires a vertex. A one-string quark-diquark model [96] is also possible as the limit of the $Y$-configuration.

Our string meson model consists of an open string with the opposite charged endpoints. These charges signify the discontinuity of the string tension and therefore, in this case the termination of the string. Contrary to the standard string theory we put the charges at first, then we see that they must be opposite and then Neumann boundary conditions are obtained. We do not put these conditions at the endpoints but we derive them.

Our string baryon model is constructed out of two strings with the opposite charged endpoints each. However, one of them has an additional charge. This charge brings the alteration to the tension. But instead of termination, the string changes its tension value and continues. The difference from the previous models, besides the number of strings used, is that the Dirichlet boundary conditions arise naturally at the intersection point. In [91] boundary conditions are enforced by the Lagrange multipliers and differ from ours. In our model both Dirichlet and Neumann boundary conditions come from the measure initially modified.

The goal of the fifth part is to consider the p-brane. In this chapter we look through a textbook p-brane, look at the two-measure p-brane and discuss the generalization of the Galileon measure string. The p-branes in the two-measure theory were studied in [97, 98, 99, 100].

# 2 The Standard and Modified Measure String Theory.

The Latin indices are $0, 1$, except $u, x, y, z$, which are $0, 1, \ldots, p$, where $p$ is the dimension of the brane, the Greek indices are $0, 1, \ldots, D$, where $D$ is the dimension of spacetime. The summation over repeated indices is understood.

Every physical system is characterized by its action, $S$. The most general form of the action is

$$S_{general} = \int_{t_1}^{t_2} L(\vec{q}(t), \dot{\vec{q}}(t), t) dt, \qquad (41)$$

where the arguments $\vec{q} = (q_1, q_2, \ldots, q_N)$ are $N$ generalized coordinates and the function $L$ is the Lagrangian which in its most general form is

$$L = T - V, \qquad (42)$$

where $T$ and $V$ are the kinetic and potential energies, respectively.

Even though the physical content of the system must remain constant, the action may take different forms. It is the purpose of our research to suggest a new action and to test it for several tasks.

The evolution, $\vec{q}(t)$, of every physical system is determined by the Hamilton's principle which states that $\vec{q}(t)$ between two specified states $\vec{q_1} = \vec{q}(t_1)$ and $\vec{q_1} = \vec{q}(t_1)$ at two specified times $t_1$ and $t_2$ is a stationary point (a point where the variation is zero) of the action functional (41).

That is

$$\frac{\delta S}{\delta \vec{q}(t)} = 0. \qquad (43)$$

The resulting equations of motion must be the same for whatever type of the action we take. Therefore, it is the criterion for accepting the action as valid. The initial action which any other action will be compared with is derived based on general requirements.

The key point is that the number of lagrangian variables can differ. Therefore, the number of the equations of motion which is equal to the number of lagrangian variables also differ in different formulations.

In this chapter we look through a textbook actions of a point particle and of a string, look at the two-measure string action and introduce the one that is totally new.

## 2.1 A point particle

A point particle being the most simple object we can imagine is a good start for all further research. The guiding principles and methods used here are rather similar when considering strings. Therefore, in this subsection we try them out on the elementary example.

We set requirements for the action based on fundamental conventions and then find a viable solution to fulfill these demands.

The first requirement is that the action, $S_{point}$, is Lorentz invariant. Therefore, $S_{point}$ has the same value for all Lorentz observers. This demand comes from the first postulate of special relativity on the invariance of the laws of physics in all initial frames of reference. Mathematically it means that $S_{point}$ transforms under a given representation of the Lorentz group, and in particular, it is a scalar.

All Lorentz observers agree on the value of the elapsed proper time, $d\tau$, i.e. the time on a clock carried by the moving particle. Then

$$S_{point} = \alpha \int d\tau, \qquad (44)$$

where $\alpha$ is some constant of proportionality.

The proper time is closely related to the interval, $ds^2$, which is an invariant too.

$$d\tau^2 = -ds^2, \qquad (45)$$

where, for example, in 4-dimensional flat background geometry with the Minkowski signature we have

$$-ds^2 = -c^2 dt^2 + (dx^1)^2 + (dx^2)^2 + (dx^3)^2. \qquad (46)$$

It is more common to write the action in terms of the interval. Then

$$S_{point} = -\alpha \int ds. \qquad (47)$$

We choose $ds^2$ to be positive for time-like trajectories. And the minus sign indicates that for time-like trajectories $ds$ is real.

The second requirement concerns the dimension of $\alpha$. Followed from (41) and (42), $S_{point}$ is a dimensionless quantity in units $c = \hbar = 1$, otherwise it has dimension of [energy] · [time].

By the definition, $ds$ is the invariant length of the particle's trajectory, $[ds] = L$, then $[\alpha]$ has the dimension of inverse length which is equivalent to $m$. That is

$$\alpha \sim m. \qquad (48)$$

The third requirement is the existence of the correct nonrelativistic limit, i.e. when the velocity of the particle, $\vec{v}$, is much lower than $c$ we are back to the nonrelativistic physics.

The lagrangian and the action of the nonrelativistic particle are

$$L_{NRpoint} = \frac{1}{2}m\vec{v}^2 \quad \rightarrow \quad S_{NRpoint} = \int dt \frac{1}{2}m\vec{v}^2. \tag{49}$$

The action (47) together with (46) leads to

$$S_{point} = -\alpha \int \sqrt{dt^2 - d\vec{x}^2} = -\alpha \int dt\sqrt{1-\vec{v}^2} \approx -\alpha \int dt(1 - \frac{1}{2}\vec{v}^2 + \dots). \tag{50}$$

The first term is a constant, so it doesn't affect the equations of motion. The second term coincides with $S_{NRpoint}$ if we set the following:

$$\alpha = m. \tag{51}$$

The fourth requirement is the reparameterization invariance, i.e. the possibility to parametrize the particle's world line in any possible way while the value of the action remains the same.

Let $\tau$ be a parameter. The only requirement is that it is real and increasing as the worldline evolves. Then $X^\mu(\tau)$ is the particle's trajectory.

Generally, we do not have to assume that the background spacetime is flat. Therefore, the interval (46) is

$$ds^2 = -\eta_{\mu\nu}dX^\mu dX^\nu \quad \rightarrow \quad ds^2 = -g_{\mu\nu}dX^\mu dX^\nu, \tag{52}$$

where $\eta_{\mu\nu}$ is the metric of the Minkowski spacetime which is flat and $g_{\mu\nu}$ is an arbitrary metric. From now on we assume that the background geometry is some curved spacetime.

Furthermore,

$$ds^2 = -g_{\mu\nu}\frac{dX^\mu}{d\tau}\frac{dX^\nu}{d\tau}d\tau^2. \tag{53}$$

Then the action (47) is

$$S_{point} = -m \int \sqrt{-g_{\mu\nu}\dot{X}^\mu \dot{X}^\nu} d\tau, \tag{54}$$

where the dot represents the derivative with respect to $\tau$.

The lagrangian variable is $X^\mu$.

The action (54) is independent of the choice of parametrization because of the following:

If

$$\tau \to \tau' \quad \rightarrow \quad \frac{dX^\mu}{d\tau} = \frac{dX^\mu}{d\tau'}\frac{d\tau'}{d\tau}, \tag{55}$$

then

$$S_{point} = -m \int \sqrt{-g_{\mu\nu} \frac{dX^\mu}{d\tau'} \frac{dX^\nu}{d\tau'} \frac{d\tau'}{d\tau}} d\tau = -m \int \sqrt{-g_{\mu\nu} \frac{dX^\mu}{d\tau'} \frac{dX^\nu}{d\tau'}} d\tau'. \tag{56}$$

The action (54) meets our requirements and is considered a main action of the free relativistic point particle. However, it is not unique. The major challenge of this whole research is to go beyond the conventional approach and reach new insights.

Despite being a leading action, (54) has two problems. The first problem is that it is useless for the massless particles and the second problem is the square root integrand which makes the quantization much more difficult. Both problems are solved by the introducing of the so-called auxiliary field, $e(\tau)$.

The new action is

$$S_{aux} = \frac{1}{2} \int d\tau (\frac{1}{e(\tau)} \dot{X}^2 - m^2 e(\tau)), \tag{57}$$

where $\dot{X}^2 = g_{\mu\nu} \dot{X}^\mu \dot{X}^\nu$.

The lagrangian variables are $X^\mu$ and $e(\tau)$.

The necessary step is to check the consistency of the actions (54) and (57). For this purpose we derive the equation of motion for $e(\tau)$, i.e. we vary the action (57) with respect to $e(\tau)$. Then we obtain

$$\dot{X}^2 + m^2 e^2(\tau) = 0. \tag{58}$$

Therefore,

$$e(\tau) = \sqrt{-\frac{\dot{X}^2}{m^2}}. \tag{59}$$

When substituting (59) back to (57) we get the action (54) which proves the equivalence of these two actions.

## 2.2 A standard string

There is a reason that we pay so much attention to the point particle. The central idea of the string theory is to consider extended objects instead of zero-dimensional ones. However, at an early stage we rely on the old known theory and embrace its methods and techniques.

We still require the action to be Lorentz and reparameterization invariant and have a correct non-relativistic limit. But now we consider a one-dimensional object that sweeps out a two-dimensional surface while evolving. A worldsheet replaces a worldline.

There are two possible topologies of a one-dimensional object, so that a string can be an open string or a closed string.

There are two kinds of the boundary conditions for the open strings. They are

Neumann boundary conditions

$$\partial_\sigma X^\mu(\tau, \sigma = 0) = \partial_\sigma X^\mu(\tau, \sigma = l) = 0 \tag{60}$$

and Dirichlet boundary conditions

$$\partial_\tau X^i(\tau, \sigma = 0) = \partial_\tau X^i(\tau, \sigma = l) = 0, \tag{61}$$

where $X^i$ are the spatial components.

Closed strings have periodic boundary conditions, namely

$$X^i(\tau, \sigma = 0) = X^i(\tau, \sigma = l). \tag{62}$$

### 2.2.1 The Nambu-Goto action

A proper area replaces a proper time.

$$S = -m \int ds \quad \rightarrow \quad S = \beta \int dA, \tag{63}$$

where $dA$ is the differential element of area on the string's worldsheet, $\beta$ is some constant of proportionality.

As the motion of a particle acts to extremize $ds$, the motion of a string acts to extremize $dA$.

The first step is to determine $dA$.

An area element of a parameterized spatial surface which has two parameters, $\xi^0$ and $\xi^1$, is

$$dA = d\xi^0 d\xi^1 \sqrt{(\frac{\partial \vec{x}}{\partial \xi^0} \frac{\partial \vec{x}}{\partial \xi^0})(\frac{\partial \vec{x}}{\partial \xi^1} \frac{\partial \vec{x}}{\partial \xi^1}) - (\frac{\partial \vec{x}}{\partial \xi^0} \frac{\partial \vec{x}}{\partial \xi^1})^2}, \tag{64}$$

where $\vec{x} = \vec{x}(\xi^0, \xi^1)$.

A proper area element which has two parameters, $\tau$ and $\sigma$, is

$$dA = \int d\tau d\sigma \sqrt{(\frac{\partial X^\mu}{\partial \tau} \frac{\partial X_\mu}{\partial \sigma})^2 - (\frac{\partial X^\mu}{\partial \tau} \frac{\partial X_\mu}{\partial \tau})^2 (\frac{\partial X^\nu}{\partial \sigma} \frac{\partial X_\nu}{\partial \sigma})^2}, \tag{65}$$

where $X^\mu = X^\mu(\tau, \sigma)$ and so on.

The fundamental difference between (64) and (65) is that the terms under the square root are reordered since the radicand in (65) must be positive.

Therefore, the string action, $S_{string}$, is

$$S_{string} = \beta \int d\tau d\sigma \sqrt{(\dot{X} \cdot X')^2 - (\dot{X})^2 (X')^2}, \tag{66}$$

where the dot represents the derivative with respect to $\tau$ and the prime represents the derivative with respect to $\sigma$.

The second step is to ensure that the action is a reparameterization invariant.

The action (66) is independent of the choice of parametrization because of the following:

If

$$\tau \to \tau' \qquad \to \qquad \frac{dX^\mu}{d\tau} = \frac{dX^\mu}{d\tau'} \frac{d\tau'}{d\tau}, \tag{67}$$

$$\sigma \to \sigma' \qquad \to \qquad \frac{dX^\mu}{d\sigma} = \frac{dX^\mu}{d\sigma'} \frac{d\sigma'}{d\sigma}, \tag{68}$$

then

$$S_{string} = \beta \int d\tau d\sigma \sqrt{\left(\frac{\partial X^\mu}{\partial \tau'} \frac{d\tau'}{d\tau} \frac{\partial X_\mu}{\partial \sigma'} \frac{d\sigma'}{d\sigma}\right)^2 - \left(\frac{\partial X^\mu}{\partial \tau'} \frac{d\tau'}{d\tau} \frac{\partial X_\mu}{\partial \tau'} \frac{d\tau'}{d\tau}\right)^2 \left(\frac{\partial X^\nu}{\partial \sigma'} \frac{d\sigma'}{d\sigma} \frac{\partial X_\nu}{\partial \sigma'} \frac{d\sigma'}{d\sigma}\right)^2} =$$

$$= \beta \int d\tau' d\sigma' \sqrt{\left(\frac{\partial X^\mu}{\partial \tau'} \frac{\partial X_\mu}{\partial \sigma'}\right)^2 - \left(\frac{\partial X^\mu}{\partial \tau'} \frac{\partial X_\mu}{\partial \tau'}\right)^2 \left(\frac{\partial X^\nu}{\partial \sigma'} \frac{\partial X_\nu}{\partial \sigma'}\right)^2}. \tag{69}$$

Even if $\tau \to \tau'(\tau, \sigma)$ and $\sigma \to \sigma'(\tau, \sigma)$ the reparameterization invariance is still confirmed. However, the calculations is not that straightforward. Now the idea to modify the action comes into play.

This time the new metric, $\gamma_{ab}$, on the worldsheet is introduced.

Similarly to (53),

$$ds^2 = -\gamma_{ab} d\xi^a d\xi^b, \tag{70}$$

where $\xi^a = (\xi^0, \xi^1) = (\tau, \sigma)$, $\xi^b = (\xi^0, \xi^1) = (\tau, \sigma)$ and

$$\gamma_{ab} = g_{\mu\nu} \frac{\partial X^\mu}{\partial \xi^a} \frac{\partial X^\nu}{\partial \xi^b}. \tag{71}$$

The dependence of $\gamma_{ab}$ on $g_{\mu\nu}$ makes it the so-called induced metric.

Then the Nambu-Goto action (yet with $\beta$ being some constant) is

$$S_{Nambu-Goto} = \beta \int d\tau d\sigma \sqrt{-\gamma}, \tag{72}$$

where $\gamma = \det(\gamma_{ab})$.

The reparameterization invariance becomes explicit for all possible replacements of parameters.

The third step is to determine $\beta$.

Let's do it on the example of the static string.

Note that with Neumann boundary conditions a classical string at rest will collapse to zero size. It can have a finite size only with rotation or when its ends are "nailed" with Dirichlet boundary conditions. It has a finite length, say $a$, that is dictated by the boundary conditions.

When the string is at rest, its energy (see (42)) is entirely potential. Then, its action, $S_{example}$, is

$$S_{example} = \int dt(-V). \tag{73}$$

On the other hand, when dealing with the static string we obviously use a static gauge, $X^0 = c\tau$, and notice that $X^1 = f(\sigma)$, $\sigma \in [\sigma_1, \sigma_2]$, while all the other components of $X$ vanish. Then the action (66) becomes

$$S_{example} = \beta \int dt(f(\sigma_2) - f(\sigma_1)) = \int dt(\beta a), \tag{74}$$

where $a$ is the length of the string.

Comparing (73) and (74) we obtain

$$\beta = -\frac{V}{a} = -T, \tag{75}$$

where the constant of proportionality is therefore, the tension, $T$, with the minus sign.

Finally, the Nambu - Goto action is

$$S_{Nambu-Goto} = -T \int d\tau d\sigma \sqrt{-\gamma}. \tag{76}$$

The lagrangian variable is $X^\mu$.

### 2.2.2 The sigma-model action

The reason to revise the action (76) is the technical inconvenience in quantization that appears from the square root of the induced metric. To solve the problem the following action, $S_{w/o\sqrt{...}}$, was presented.

$$S_{w/o\sqrt{\cdots}} = \frac{T}{2}\int d\tau d\sigma(\partial_\tau X^\mu \partial_\tau X_\mu - \partial_\sigma X^\mu \partial_\sigma X_\mu). \tag{77}$$

Indeed, this action contains no square root. However, it does not contain Virasoro constrains[1] either. One has to put it by hand. Not to trade one problem for another, we will not consider $S_{w/o\sqrt{\cdots}}$ for now. However, we will see that this action is the yet not derived sigma-model action in the conformal gauge, and we will accept it as a bosonic part of the superstring in Chapter 4.

And now we introduce the so-called intrinsic metric, $h_{ab}(\tau, \sigma)$, which is the kind of the auxiliary field as in the point particle action (57). $h_{ab}(\tau, \sigma)$ replaces $e(\tau)$.

Then the action is

$$S_{sigma-model} = -\frac{T}{2}\int d\tau d\sigma \sqrt{-h}h^{ab}\partial_a X^\mu \partial_b X^\nu g_{\mu\nu}, \tag{78}$$

where $h = \det(h_{ab})$.

This is the sigma-model action. This action is also called the Polyakov action because Polyakov was the first to use it for the path integral quantization of the string. But it was Deser and Zumino and independently Brink, Di Vecchia and Howe who actually proposed it.

The lagrangian variables are $X^\mu$, $h^{ab}$.

The actions (54) and (57) are equivalent. So the actions (76) and (78) must be equivalent too. Again, to see it explicitly we derive the equation of motion for $h^{ab}$, i.e. we vary the action (78) with respect to $h^{ab}$.

We remark that the worldsheet energy-momentum tensor is defined as

$$T_{ab} = -\frac{2}{\sqrt{-h}}\frac{\delta S}{\delta h^{ab}}, \tag{79}$$

which obviously vanishes, $T_{ab} = 0$.

In other words, the equation of motion for $h^{ab}$ is

$$T_{ab} = 0. \tag{80}$$

See Appendix 1 for the intermediate calculations.

Therefore,

$$T_{ab} = \partial_a X \cdot \partial_b X - \frac{1}{2}h_{ab}(h^{cd}\partial_c X \cdot \partial_d X) = 0. \tag{81}$$

Taking the square root of minus the determinant, we obtain

---

[1] These constraints come from fixing the conformal gauge and take the form of $T_{ab} = 0$

$$\gamma_{ab} - \frac{1}{2}h_{ab}(h^{cd}\gamma_{cd}) = 0. \tag{82}$$

We see that the intrinsic metric is proportional to the induced metric, $h_{ab} \sim \gamma_{ab}$. However, this time the constant of proportionality is undetermined because the presence of the product of the metric and its inverse makes both solutions, $h_{ab}$ and $\Omega^2 h_{ab}$ where $\Omega^2$ is an arbitrary function, possible. Then

$$h_{ab} = f^2(\xi)\gamma_{ab}, \tag{83}$$

where $f^2(\xi)$ is some constant of proportionality which is generally position dependent. As $f^2(\xi)$ is chosen to be positive, these two metrics are said to be conformal to each other.

Moving back to the problem of the equivalence, we see that $f^2(\xi)$ cancels out, namely

$$\sqrt{-h}h^{ab} = f^2(\xi)\sqrt{-\gamma}\frac{1}{f^2(\xi)}\gamma^{ab} = \sqrt{-\gamma}\gamma^{ab}. \tag{84}$$

Then, the action (78) transforms into the action (76).

$$S_{sigma-model} = -\frac{T}{2}\int d\tau d\sigma \sqrt{-h}h^{ab}\partial_a X^\mu \partial_b X^\nu g_{\mu\nu} =$$

$$= -\frac{T}{2}\int d\tau d\sigma \sqrt{-\gamma}\gamma^{ab}\gamma_{ab} = -T\int d\tau d\sigma \sqrt{-\gamma} = S_{Nambu-Goto}. \tag{85}$$

That is, the sigma-model action is equivalent to the Nambu - Goto action.

To double-check, let us consider another dynamical variable $X^\mu$ of the sigma-model action.

The equation of motion of $X^\mu$ is

$$\frac{1}{\sqrt{-h}}\partial_a(\sqrt{-h}h^{ab}\partial_b X^\mu) + h^{ab}\partial_a X^\nu \partial_b X^\lambda \Gamma^\mu_{\nu\lambda} = 0, \tag{86}$$

where $\Gamma^\mu_{\nu\lambda}$ is the affine connection for the external, $g_{\mu\nu}$, metric,

$$\Gamma^\mu_{\nu\lambda} = \frac{1}{2}g^{\mu\alpha}\left(\frac{\partial g_{\alpha\nu}}{\partial x^\lambda} + \frac{\partial g_{\alpha\lambda}}{\partial x^\nu} - \frac{\partial g_{\nu\lambda}}{\partial x^\alpha}\right). \tag{87}$$

If we vary the action (76) with respect to $X^\mu$, we obtain the same equation of motion (86).

## 2.3 A modified measure string

The Nambu-Goto action (76) and the sigma-model action (78) contain the string tension, $T$, explicitly. It brings a scale in the otherwise scale invariant theory. This is the reason to reformulate the action again.

So far our activity concerns the inclusion of some auxiliary function. The general form (41) needs a refinement. Let us introduce the lagrangian density, $\mathcal{L}$, in the way that

$$L = \int \mathcal{L} d^{N-1}x \quad \rightarrow \quad S = \int \mathcal{L} d^N x, \tag{88}$$

where $N$ is the number of spacetime dimensions.

The key thing to notice is that the volume element, $d^N x$, is not invariant. However, this is not an insurmountable problem. It is resolved by multiplying some scalar density to balance the Jacobian factors.

The square root of a minus the determinant of the metric, $\sqrt{-g}$, is ordinarily used for the task. The argument is the following:

If under the coordinate transformation

$$x^\mu \rightarrow x^{\mu'} \tag{89}$$

the transformation laws are

$$d^N x \rightarrow \det(\frac{\partial x^{\mu'}}{\partial x^\mu}) d^N x, \tag{90}$$

$$\sqrt{-g} \rightarrow (\det(\frac{\partial x^{\mu'}}{\partial x^\mu}))^{-1} \sqrt{-g}, \tag{91}$$

then

$$\sqrt{-g} d^N x \quad \rightarrow \quad \sqrt{-g} d^N x. \tag{92}$$

Therefore, the correct action is

$$S = \int \mathcal{L} \sqrt{-g} d^N x. \tag{93}$$

The Nambu-Goto action (76) is exactly the integral over the invariant area element.

Here $\sqrt{-g}$ is at its core the measure of integration. Let us denote the arbitrary measure of integration by $\Phi$.

We see that the only requirement for $\Phi$ is to be a density under diffeomorphic transformations. And $\sqrt{-g}$ is not the only physical quantity with such property. We are about to present two alternatives: the so-called two-measure which is already known and the so-called Galileon measure which is completely new. See chapter 4 for further details on the background.

### 2.3.1 The two-measure string action

The two-measure $\Phi(\varphi)$ is constructed out of two (the number of worldsheet dimensions) scalar fields, $\varphi^i, \varphi^j$, and it does not depend on the determinant of the metric

$$\Phi(\varphi) = \epsilon^{ab}\epsilon_{ij}\partial_a\varphi^i\partial_b\varphi^j, \qquad (94)$$

where $\epsilon$ is the Levi-Civita symbol.

Therefore, the two-measure string action, $S_{TM}$, is

$$S_{TM} = -\int d\tau d\sigma \Phi(\varphi)\mathcal{L}, \qquad (95)$$

where $\mathcal{L}$ is arbitrary.

The lagrangian variables are $\varphi^i$ and those that are contained in $\mathcal{L}$.

The variation with respect to $\varphi^i$ is

$$\delta S_{TM} = -\int d\tau d\sigma \delta(\Phi(\varphi)\mathcal{L}) = -\int d\tau d\sigma \delta(\epsilon^{ab}\epsilon_{ij}\partial_a\varphi^i\partial_b\varphi^j \mathcal{L}) = \qquad (96)$$

After the integration by parts we obtain

$$= \int d\tau d\sigma \delta(\epsilon^{ab}\epsilon_{ij}\varphi^i\partial_b\varphi^j\partial_a\mathcal{L}) = \int d\tau d\sigma (\epsilon^{ab}\epsilon_{ij}\partial_b\varphi^j\partial_a\mathcal{L})\delta\varphi^i \qquad (97)$$

By the variational principle ($\delta S_{TM} = 0$)

$$\epsilon^{ab}\epsilon_{ij}\partial_b\varphi^j\partial_a\mathcal{L} = 0 \qquad (98)$$

As $\Phi(\varphi)$ is not degenerate, then

$$\partial_a\mathcal{L} = 0. \qquad (99)$$

Therefore,

$$\mathcal{L} = const. \qquad (100)$$

### 2.3.2 The Galileon string action

The Galileon measure density, $\Phi(\chi)$, is constructed out of one scalar field, $\chi$, and depends on the determinant of the metric

$$\Phi(\chi) = \partial_a(\sqrt{-h}h^{ab}\partial_b\chi). \qquad (101)$$

Therefore, the Galileon string action, $S_{GM}$, is

$$S_{GM} = -\int d\tau d\sigma \Phi(\chi)\mathcal{L}. \qquad (102)$$

The scalar field $\chi$ is a Galileon scalar field since $\Phi(\chi)$ is invariant under a Galileon shift symmetry:

$$\partial_a \chi \to \partial_a \chi + b_a, \tag{103}$$

$$\chi \to \chi + b_a \sigma^a, \tag{104}$$

where $b_a$ is a constant vector and $\sigma^a = (\tau, \sigma)$.

This symmetry is available only in two dimensions, for only in two dimensions there exist a conformally flat frame for $h^{ab}$.

Indeed, in the conformally flat metric gauge

$$h^{ab} = \exp(-\phi)\eta^{ab}, \qquad \sqrt{-h} = \sqrt{\exp(2\phi)}, \tag{105}$$

where $\phi$ is some scalar function,

we obtain

$$h^{ab}\sqrt{-h} = \eta^{ab}. \tag{106}$$

Therefore,

$$\Phi(\chi) = \partial_a(\eta^{ab}\partial_b \chi). \tag{107}$$

The Galileon symmetry is thus an exact symmetry.

We should point out that a shift of the scalar by a linear function of coordinates resembles a "gauge symmetry" (although the gauge function is restricted to be a linear function of the coordinates, not a general function). These kind of "scalar gauge fields" were considered in [101, 102].

Next we move on to the equations of motion of the action (102).

The lagrangian variables are $h^{ab}$ and those that are contained in $\mathcal{L}$.

As with the sigma-model action, the variation of $S_{GM}$ with respect to $h_{ab}$ is the energy-momentum tensor $T^{ab}$.

$$T^{ab} = -\frac{2}{\sqrt{-h}}\frac{\delta(\Phi(\chi)\mathcal{L})}{\delta h_{ab}}. \tag{108}$$

Then

$$T^{ab} = -\frac{2}{\sqrt{-h}}\left(\frac{\partial(\Phi(\chi)\mathcal{L})}{\partial h_{ab}} - \partial_c\left(\frac{\partial(\Phi(\chi)\mathcal{L})}{\partial h_{ab,c}}\right)\right) =$$

$$= -\frac{2}{\sqrt{-h}}\left(\frac{\partial \mathcal{L}}{\partial h_{ab}}\Phi(\chi) + \mathcal{L}\frac{\partial \Phi(\chi)}{\partial h_{ab}} - \partial_c\left(\mathcal{L}\frac{\partial \Phi(\chi)}{\partial h_{ab,c}}\right) - \partial_c\left(\Phi(\chi)\frac{\partial \mathcal{L}}{\partial h_{ab,c}}\right)\right), \tag{109}$$

where $h_{ab,c}$ is a derivative of the metric with respect to the coordinates.

The only thing that is assumed from now on about $\mathcal{L}$ is that it is independent of $h_{ab,c}$, i.e. $\mathcal{L}$ is homogeneous of degree 1 in $h^{ab}$. Thereby the last term in (109) vanishes.

It is more convenient to rewrite the Galileon measure in the following form:

$$\Phi(\chi) = \partial_a(\sqrt{-h}h^{ab}\partial_b\chi) = \sqrt{-h}\nabla_a(h^{ab}\partial_b\chi) =$$

$$= h^{ab}\sqrt{-h}\nabla_a(\partial_b\chi) = h^{ab}\sqrt{-h}(\partial_a\partial_b\chi - \Gamma^c_{ab}\partial_c\chi), \tag{110}$$

Then, back to (109):

$$T^{ab} = -\frac{2}{\sqrt{-h}}[\frac{\partial \mathcal{L}}{\partial h_{ab}}\Phi(\chi) + \mathcal{L}h^{ef}\nabla_e(\partial_f\chi)\frac{\partial\sqrt{-h}}{\partial h_{ab}} + \mathcal{L}\sqrt{-h}\nabla_e(\partial_f\chi)\frac{\partial h^{ef}}{\partial h_{ab}} -$$

$$-\mathcal{L}\sqrt{-h}h^{ef}\partial_c\chi\frac{\partial \Gamma^c_{ef}}{\partial h_{ab}} + \partial_c(\mathcal{L}h^{ef}\sqrt{-h}\partial_g\chi\frac{\partial \Gamma^g_{ef}}{\partial h_{ab,c}})] =$$

$$= -\frac{2}{\sqrt{-h}}[\frac{\partial \mathcal{L}}{\partial h_{ab}}\Phi(\chi) + \frac{1}{2}\sqrt{-h}h^{ab}\mathcal{L}h^{cd}\nabla_c\partial_d\chi -$$

$$-\frac{1}{2}\mathcal{L}\sqrt{-h}(\nabla^a\nabla^b + \nabla^b\nabla^a)\chi + \frac{1}{2}(h^{af}\Gamma^b_{cd} + h^{bf}\Gamma^a_{cd})\mathcal{L}\sqrt{-h}h^{cd}\partial_f\chi -$$

$$+\partial_g(\frac{1}{4}\mathcal{L}\sqrt{-h}h^{cd}\partial_e\chi(h^{ae}(\delta^b_c\delta^g_d + \delta^b_d\delta^g_c) + \gamma^{eb}(\delta^a_c\delta^g_d + \delta^a_d\delta^g_c) - \gamma^{eg}(\delta^a_d\delta^b_c + \delta^a_c\delta^b_d)))]. \tag{111}$$

Just for ease of calculations and without the loss of generality we go to a local Lorentz frame of reference, i.e. an inertial reference frame.

A reference frame is said to be inertial in a certain region of space and time when, throughout that region of spacetime, and within some specified accuracy, every test particle that is initially at rest remains at rest, and every test particle that is initially in motion continues that motion without change in speed or in direction. In terms of this definition, inertial frames are necessarily always local ones, that is inertial in a limited region of spacetime [103].

Then it follows that in such a frame the Christoffel symbols are zero, and therefore, the covariant derivatives are partial derivatives.

$$T^{ab} = -\frac{2}{\sqrt{h}}((\frac{\partial \mathcal{L}}{\partial h_{ab}}\Phi(\chi) + \frac{1}{2}\mathcal{L}\sqrt{-h}h^{ab}h^{cd}\partial_c\partial_d\chi) - \frac{1}{2}\mathcal{L}\sqrt{-h}(\partial^a\partial^b + \partial^b\partial^a)\chi +$$

$$+\partial_g(\frac{1}{2}\mathcal{L}\sqrt{-h}\partial_e\chi(h^{bg}h^{ae} + h^{eb}h^{ag} - h^{eg}h^{ab}))) =$$

$$= -\frac{2}{\sqrt{-h}}((\frac{\partial \mathcal{L}}{\partial h_{ab}}\Phi(\chi) + \frac{1}{2}\mathcal{L}\sqrt{-h}h^{ab}h^{cd}\partial_c\partial_d\chi) + \frac{1}{2}\sqrt{-h}\partial^a\chi\partial^b\mathcal{L} + \frac{1}{2}\sqrt{-h}\partial^b\chi\partial^a\mathcal{L}-$$

$$-\frac{1}{2}\mathcal{L}\sqrt{-h}h^{ab}h^{cd}\partial_c\partial_d\chi - \frac{1}{2}\sqrt{-h}h^{ab}\partial_e\chi\partial^e\mathcal{L}) =$$

$$= -\frac{2}{\sqrt{-h}}\frac{\partial \mathcal{L}}{\partial h_{ab}}\Phi(\chi) - \partial^a\chi\partial^b\mathcal{L} - \partial^b\chi\partial^a\mathcal{L} + h^{ab}\partial_e\chi\partial^e\mathcal{L}. \tag{112}$$

In sum,

$$T^{ab} = -\frac{2}{\sqrt{-h}}\frac{\partial \mathcal{L}}{\partial h_{ab}}\Phi(\chi) - \partial^a\chi\partial^b\mathcal{L} - \partial^b\chi\partial^a\mathcal{L} + h^{ab}\partial_g\chi\partial^g\mathcal{L} = 0. \tag{113}$$

The trace equation is

$$h_{ab}T^{ab} = -h_{ab}\frac{2}{\sqrt{-h}}\frac{\delta(\Phi(\chi)\mathcal{L})}{\delta h_{ab}} = 0. \tag{114}$$

When taking a trace, the last three terms in (113) vanish.

$$h_{ab}(-\partial^a\chi\partial^b\mathcal{L} - \partial^b\chi\partial^a\mathcal{L} + h^{ab}\partial_c\chi\partial^c\mathcal{L}) = -\partial_b\chi\partial^b\mathcal{L} - \partial_a\chi\partial^a\mathcal{L} + 2\partial_c\chi\partial^c\mathcal{L} = 0. \tag{115}$$

Then

$$-h_{ab}\frac{2}{\sqrt{-h}}\frac{\partial \mathcal{L}}{\partial h_{ab}}\Phi(\chi) = 0. \tag{116}$$

Using the fact that $\frac{\partial \mathcal{L}}{\partial h_{ab}}h_{ab} = -\mathcal{L}$, we obtain

$$h_{ab}T^{ab} = \frac{2}{\sqrt{-h}}\mathcal{L}\Phi(\chi) = 0. \tag{117}$$

If the trivial case $\Phi(\chi) = 0$ is excluded, then

$$\mathcal{L} = 0. \tag{118}$$

The proof of conformal invariance requires the equations of motion, meaning that we prove the conformal invariance on the mass shell.

Naively one may conclude that if the lagrangian density is zero this means one has nothing – no equations of motion. This is not the case. In the next subsection the equations of motion are presented.

The lagrangian density being 0 is not a unique case. For example, the lagrangian density for the Dirac equation is also 0 once we use the equations of motion. So there is nothing wrong with the $\mathcal{L} = 0$ after the use of the equations of motion.

We then obtain the following:

First, it is the proof that the higher-derivative terms in (113) are canceled. And we are left with the known modified measure theory.

$$\frac{\delta(\Phi(\chi)\mathcal{L})}{\delta h_{ab}} = \frac{\partial \mathcal{L}}{\partial h_{ab}}\Phi(\chi). \tag{119}$$

Second, the relation (118) determines the connection point between the two-measure and the Galileon measure theories.

The last thing to consider is the behavior of the equations under conformal transformations parametrized by $\Omega$.

If

$$h_{ab} \to \Omega^2 h_{ab}, \qquad \sqrt{-h} \to \Omega^2\sqrt{-h} \tag{120}$$

and

$$\mathcal{L} \to \Omega^{-2}\mathcal{L}, \tag{121}$$

then

$$\delta S = \int d\tau d\sigma \Omega^2 \Phi(\chi)\mathcal{L} = 0. \tag{122}$$

Again we obtain the same result.

$$\mathcal{L} = 0. \tag{123}$$

This means that the conformal symmetry is confirmed.

### 2.3.3 The lagrangian density and the equations of motion.

We have defined $\Phi(\varphi)$ and $\Phi(\chi)$ and derived the respective restrictions on $\mathcal{L}$, (100) and (118). The next step is to construct $\mathcal{L}$ itself.

The simplest proposition is

$$\mathcal{L}_{simple} = h^{ab}\partial_a X^\mu \partial_b X^\nu g_{\mu\nu}. \tag{124}$$

It is actually the lagrangian density of the previously encountered action (77). Again it is not appropriate. It fails in the two-measure theory because the variation with respect to $h^{ab}$ leads to

$$\Phi \partial_a X^\mu \partial_b X^\nu g_{\mu\nu} = 0. \tag{125}$$

Then $\Phi(\varphi)$ or the induced metric must vanish. It fails in the Galileon measure theory because of the constraint (118).

However, we are not going to abandon (124) at all. Something must be added to it. That is entirely possible because any term that is a total derivative (when multiplied by a measure) could be added to $\mathcal{L}$ without any consequences for the equations of motion.

There is a defining difference between the measures $\sqrt{-g}$ and $\Phi$ (meaning both $\Phi(\varphi)$ and $\Phi(\chi)$).

$$S = \int d\tau d\sigma \sqrt{-g} \mathcal{L} \quad \rightarrow \quad S = \int d\tau d\sigma \Phi \mathcal{L}. \tag{126}$$

Some terms in $\mathcal{L}$ when multiplied by $\sqrt{-g}$ may form a total derivative. However, those terms may not form a total derivative when multiplied by $\Phi$.

So we define an auxiliary Abelian gauge field, $A_a$, in the worldsheet. Then the contribution is

$$\mathcal{L}_{additional} = \frac{\epsilon^{ab}}{\sqrt{-h}} F_{ab}, \tag{127}$$

where $F_{ab} = \partial_a A_b - \partial_b A_a$ is the field-strength of $A_a$.

Then the new action is

$$S = -\int d\tau d\sigma \Phi (\mathcal{L}_{simple} + \mathcal{L}_{additional}), \tag{128}$$

where $\Phi$ could be either $\Phi(\chi)$ or $\Phi(\varphi)$.

We see that one has to add two scalars and one world sheet vector and complicate the action. However, the advantage is that we get the string tension as a constant of integration and later on as a dynamical variable when we introduce charges at the string worldsheet.

We have to understand it classically first. We do not consider the quantum case.

The two-measure action is

$$S_{TM} = -\int d\tau d\sigma (h^{ab} \partial_a X^\mu \partial_b X^\nu g_{\mu\nu} - \frac{\epsilon^{cd}}{\sqrt{-h}} F_{cd}) \epsilon^{ab} \epsilon^{ij} \partial_a \varphi^i \partial_b \varphi^j. \tag{129}$$

The Galileon measure action is

$$S_{GM} = -\int d\tau d\sigma (h^{ab} \partial_a X^\mu \partial_b X^\nu g_{\mu\nu} - \frac{\epsilon^{cd}}{\sqrt{-h}} F_{cd}) \partial_e (h^{ef} \sqrt{-h} \partial_f \chi). \tag{130}$$

The restriction in the two-measure theory is

$$\mathcal{L}_{simple} + \mathcal{L}_{additional} = const. \tag{131}$$

The restriction in the Galileon measure theory is

$$\mathcal{L}_{simple} + \mathcal{L}_{additional} = 0. \tag{132}$$

To prove the validity of the modified measure theories we derive the equations of motion of $S_{TM}$ and $S_{GM}$ and compare them with those of $S_{sigma-model}$.

If two actions yield the same equations of motion in implies only that the two are classically equivalent and not quantum equivalent. For the latter one has to show that all the correlation function (of gauge invariant operators) are the same which means that the generating functionals of the two theories are the same. So indeed, it is not obvious that the modified actions are equivalent to the original one (the Nambu-Goto action and the Polyakov action) in quantum case. However, we have to understand it classically first.

There are processes described by modified measure strings that do not have an analogue in the the Nambu-Goto action and the Polyakov cases like the discontinuity of string tension induced by charges.

Let's start with the two-measure theory.

The constraint (100) implies

$$h^{ab}\partial_a X^\mu \partial_b X^\nu g_{\mu\nu} - \frac{\epsilon^{cd}}{\sqrt{-h}}F_{cd} = const = M. \tag{133}$$

The variation with respect to $h^{ab}$ is

$$\partial_a X^\mu \partial_b X^\nu g_{\mu\nu} - \frac{1}{2}h_{ab}\frac{\epsilon^{cd}}{\sqrt{-h}}F_{cd} = 0. \tag{134}$$

When we take the trace of (134) and compare the result with (133), we get that $M=0$.

Then

$$h^{ab}\partial_a X^\mu \partial_b X^\nu g_{\mu\nu} = \frac{\epsilon^{cd}}{\sqrt{-h}}F_{cd}. \tag{135}$$

When we introduce it in (134), we obtain the equation of motion which is exactly the same as (81).

The variation with respect to $X^\mu$ is

$$\partial_a(\Phi h^{ab}\partial_b X^\mu) + \Phi h^{ab}\partial_a X^\nu \partial_b X^\lambda \Gamma^\mu_{\nu\lambda} = 0. \tag{136}$$

The variation with respect to $A_a$ is

$$\epsilon^{ab}\partial_b\left(\frac{\Phi}{\sqrt{-h}}\right) = 0. \tag{137}$$

If $\Phi \neq 0$, then

$$\frac{\Phi}{\sqrt{-h}} = Const. \tag{138}$$

When we introduce it in (136), we obtain the equation of motion which is exactly the same as (86) provided that this constant is the string tension, $T$.

$$\frac{\Phi}{\sqrt{-h}} = T. \tag{139}$$

Therefore, the tension is spontaneously induced. It happens because of the extra lagrangian variable, $A_a$. It is the variation with respect to the gauge field that is responsible for the appearance of $T$ as an integration constant.

Now let's turn to the Galileon measure theory.

The constraint (118) implies

$$h^{ab}\partial_a X^\mu \partial_b X^\nu g_{\mu\nu} - \frac{\epsilon^{cd}}{\sqrt{-h}} F_{cd} = 0. \tag{140}$$

The variations with respect to $h^{ab}$, $X^\mu$ and $A_a$ and the later conclusion on $T$ are exactly the same as in the two-measure theory. See Appendix 1 for details.

# 3 The Standard and Modified Measure Superstring Theory.

Every physical system (or at least its simplified version) has some symmetries. Symmetries are the transformations that leave the action, $S$, invariant.

If $A \to f(A) = A'$, so that $S(A) = S(A')$, then $f$ is said to be a symmetry.

In addition to the technical simplifications symmetries are the direct path to conserved quantities and conserved currents.

The previous chapter discusses only the bosonic strings. The inclusion of fermions requires the supersymmetry.

In this chapter we look through a textbook symmetries of a string, look at the supersymmetry of a standard and modified measure string and introduce the new supersymmetric string action.

## 3.1 String symmetries

Once again, the sigma-model action is

$$S_{sigma-model} = -\frac{T}{2} \int d\tau d\sigma \sqrt{-h} h^{ab} \partial_a X^\mu \partial_b X^\nu g_{\mu\nu}. \tag{141}$$

When considering a string we always have two distinct locations. One of them is the spacetime we live in[2] with the metric $g_{\mu\nu}$ and the coordinates $X^\mu$. And the other one is the worldsheet with the metric $h^{ab}$ and the scalar fields $X^\mu$.

Hence, there are two types of symmetries.

The spacetime symmetry is the Poincare transformations.

$$X^\mu \to X'^\mu = \Lambda^\mu_\nu X^\nu + V^\mu, \tag{142}$$

where $\Lambda^\mu_\nu$ are indeed the Lorentz transformations and $V^\mu$ are spacetime translations.

It is the global symmetry.

The worldsheet symmetries are the following:

The first symmetry is the reparameterization invariance or diffeomorphisms.

If

$$\sigma^a \to \tilde{\sigma}^a(\sigma), \tag{143}$$

then

---

[2]Actually no, the dimension of that spacetime is 26 when fermions do not exist and 10 or 11 for the full picture.

$$X^\mu(\sigma) \to \tilde{X}^\mu(\tilde{\sigma}) = X^\mu(\sigma), \tag{144}$$

$$h_{ab}(\sigma) \to \tilde{h}_{ab}(\tilde{\sigma}) = \frac{\partial \sigma^c}{\partial \tilde{\sigma}^a} \frac{\partial \sigma^d}{\partial \tilde{\sigma}^b} h_{cd}(\sigma). \tag{145}$$

The second symmetry is the Weyl invariance.

$$X^\mu(\sigma) \to X^\mu(\sigma), \tag{146}$$

$$h_{ab}(\sigma) \to h'_{ab}(\sigma) = e^\phi h_{ab}(\sigma). \tag{147}$$

Both symmetries are local.

The Nambu-Goto action has two symmetries: Poincare invariance and reparameterization invariance.

The Galileon measure action has four symmetries: the sigma-model action symmetries plus the Galileon symmetry.

## 3.2 A standard superstring

There is no experimental evidence that supersymmetry exists. However, we realize that every theory is only half-full if only the bosons are presented. Fermions must be introduced. Commonly, the supersymmetry is involved, that is, the symmetry that relates bosonic and fermionic degrees of freedom.

There are two standard approaches, the Ramond - Neveu - Schwarz approach and the Green - Schwarz approach. The main difference between them is where we assume the supersymmetry to be, namely, on the spacetime or on the worldsheet.

### 3.2.1 The Ramond - Neveu - Schwarz approach

The distinctive feature of the Ramond - Neveu - Schwarz approach is that the fermionic degrees of freedom are defined on the worldsheet.

There is no need to change the bosonic sector. So we take the usual sigma-model action, however, in the conformal gauge. Despite that the Virasoro constraints are put by hand, this is the simplest way to see what the superstring theory is about.

The sigma-model action (78) in the conformal gauge is

$$S = -\frac{1}{2\pi} \int d^2\sigma \, \partial_a X_\mu \partial^a X^\mu. \tag{148}$$

Fermions have half-integer spin and obey the Pauli statistics. Therefore, they must be described by anticommuting variables, i.e. the fermionic worldsheet field, $\psi^\mu$, is constructed out of Grassmann numbers, so that

$$\{\psi^\mu, \psi^\nu\} = 0. \tag{149}$$

The field $\psi^\mu$ is a two-component spinor on the worldsheet and at the same time is a vector on the background spacetime. As a spinor, $\psi^\mu$ has two components, $\psi^\mu_A$, where $A_\pm$.

In sum,

$$X^\mu(\tau, \sigma) \quad \to \quad (X^\mu(\tau, \sigma), \psi^\mu(\tau, \sigma)). \tag{150}$$

The action is

$$S_{SUSY\,string} = -\frac{1}{2\pi} \int d^2\sigma (\partial_a X_\mu \partial^a X^\mu + \bar{\psi}^\mu \rho^a \partial_a \psi_\mu), \tag{151}$$

where $\bar{\psi}^\mu$ is the Dirac conjugate to $\psi_\mu$

$$\bar{\psi} = \psi^\dagger b, \qquad b = i\rho^0 \tag{152}$$

and $\rho^\alpha$ are the Dirac matrices. See Appendix 2 for details.

The supersymmetry infinitesimal transformations are

$$\delta X^\mu = \bar{\epsilon}\psi^\mu, \tag{153}$$

$$\delta \psi^\mu = \rho^a \partial_a X^\mu \epsilon, \tag{154}$$

where $\epsilon$ is a constant infinitesimal Majorana spinor. As $\epsilon \neq \epsilon(\tau, \sigma)$, then this symmetry is global.

The action (151) is invariant under such transformations. However, the symmetry holds up only to a total derivative.

The problem is that this symmetry is not manifest. In the previous chapter we encountered that the reparameterization invariance was not manifest, and then we have rewritten the action to make it manifest.

To solve this problem, we introduce the superspace.

In order to do this, we extend the ordinary spacetime to include additional coordinates, which must be anticommuting ones to describe the fermions.

The superworldsheet coordinates are $(\sigma^a, \theta_A)$, where $\theta_A$ being Grassmann coordinates, $\{\theta_A, \theta_B\} = 0$, form a Majorana spinor.

In sum,

$$(\tau, \sigma) \quad \to \quad (\tau, \sigma, \theta^1, \theta^2). \tag{155}$$

The supersymmetry transformations are

$$\delta \theta^A = [\bar{\epsilon} Q, \theta^A] = \epsilon^A, \tag{156}$$

$$\delta \sigma^a = [\bar{\epsilon} Q, \sigma^a] = -\bar{\epsilon} \rho^a \theta, \tag{157}$$

where $Q$ are the supercharges.

$$Q_A = \frac{\partial}{\partial \bar{\theta}^A} - (\rho^a \theta)_A \partial_a. \tag{158}$$

Fields that are defined on the superspace are called superfields. The superfield in its most general form is

$$Y^\mu(\sigma^\alpha, \theta) = X^\mu(\sigma^\alpha) + \bar{\theta} \psi^\mu(\sigma^\alpha) + \frac{1}{2} \bar{\theta} \theta B^\mu(\sigma^\alpha), \tag{159}$$

where $B^\mu(\sigma^\alpha)$ is an auxiliary field.

The supercharge acts on the superfield as

$$\delta Y^\mu = [\bar{\epsilon} Q, Y^\mu] = \bar{\epsilon} Q Y^\mu. \tag{160}$$

Therefore, the supersymmetry transformations are

$$\delta X^\mu = \bar{\epsilon} \psi^\mu, \tag{161}$$

$$\delta \psi^\mu = \rho^a \partial_a X^\mu \epsilon + B^\mu \epsilon, \tag{162}$$

$$\delta B^\mu = \bar{\epsilon} \rho^a \partial_a \psi^\mu. \tag{163}$$

The supersymmetric action is

$$S_{RNS\_SUSY} = -\frac{1}{2\pi} \int d\tau d\sigma (\partial_a X_\mu \partial^a X^\mu + \bar{\psi}^\mu \rho^a \partial_a \psi_\mu - B_\mu B^\mu). \tag{164}$$

The transformations (161), (162) and (163) reduce to (153) and (154), provided that the equation of motion is $B^\mu = 0$.

So again we get things done with the help of the auxiliary field.

### 3.2.2 The Green - Schwarz approach

The distinctive feature of the Green - Schwarz approach is that the fermionic degrees of freedom are defined on the spacetime.

In addition to the bosonic field $X^\mu(\tau, \sigma)$ there is the fermionic field $\Theta^A(\tau, \sigma)$ with $A$-Grassmann-valued coordinates. They both map the worldsheet into superspace.

$$X^\mu(\tau, \sigma) \quad \to \quad (X^\mu(\tau, \sigma), \Theta^A(\tau, \sigma)). \tag{165}$$

The supersymmetry transformations are

$$\delta X^\mu = -i(\epsilon^A \Gamma^\mu \Theta^A), \tag{166}$$

$$\delta \Theta^A = \epsilon^A, \tag{167}$$

where $\Gamma^\mu \equiv \Gamma^\mu_{\alpha\beta}$ are the Dirac matrices.

Our goal is to construct the spacetime supersymmetric action.

First, we define the supersymmetric combination, $\Pi^\mu_a$,

$$\Pi^\mu_a = \partial_a X^\mu - i(\Theta^A \Gamma^\mu \partial_a \Theta^A). \tag{168}$$

in a way that it is invariant under the transformations (166) and (167).

Second, we suggest the replacement.

$$\partial_a X^\mu \to \Pi^\mu_a \tag{169}$$

Third, if we proceed on the basis of the Nambu-Goto action (76), then the spacetime supersymmetric action, $S_{NGsimpleSUSY}$, is

$$S_{NGsimpleSUSY} = -T \int d\tau d\sigma \mathcal{L}_{NGsimpleSUSY}, \tag{170}$$

where

$$\mathcal{L}_{NGsimpleSUSY} = \sqrt{-\det(\Pi^\mu_a \Pi_{b\mu})}. \tag{171}$$

As we planned, this action is invariant under super-Poincare transformations and under diffeomorphisms. However, there is a new symmetry which must be preserved to balance the number of independent equations of motion with the number of $\Theta$ components. This symmetry is called a local fermionic symmetry or the kappa symmetry. To take it into account, the Wess-Zumino term is introduced.

$$\mathcal{L}_{additionalSUSY} = i\epsilon^{ab} \partial_a X^\mu (\bar{\Theta}^1 \Gamma_\mu \partial_b \Theta^1 + \bar{\Theta}^2 \Gamma_\mu \partial_b \Theta^2) - \epsilon^{ab} \bar{\Theta}^1 \Gamma^\mu \partial_a \Theta^1 \bar{\Theta}^2 \Gamma_\mu \partial_b \Theta^2. \tag{172}$$

Finally, the Green-Schwarz superstring action is

$$S_{Green-SchwarzNG} = -T \int d\tau d\sigma (\mathcal{L}_{NGsimpleSUSY} + \mathcal{L}_{additionalSUSY}). \tag{173}$$

The problem is that $\mathcal{L}_{additionalSUSY}$ is invariant under global transformations only up to total derivatives.

If we proceed on the basis of the sigma-model action (78), then we obtain the similar result with the same problem. Namely, if we replace

$$\mathcal{L}_{NGsimpleSUSY} \to \mathcal{L}_{SMsimpleSUSY}, \tag{174}$$

where

$$\mathcal{L}_{SMsimpleSUSY} = \frac{1}{2}\sqrt{-h}h^{ab}\Pi_a^\mu \Pi_{b\mu}, \tag{175}$$

then the spacetime supersymmetric action, $S_{SMsimpleSUSY}$, is

$$S_{Green-SchwarzSM} = -T \int d\tau d\sigma (\mathcal{L}_{SMsimpleSUSY} + \mathcal{L}_{additionalSUSY}). \tag{176}$$

### 3.2.3 The Siegel action

As before, the problem is solved by rewriting the action. In the present case the reformulation is based on the idea of the supersymmetric currents. The desired action is

$$S_{Siegel} = -T \int d\tau d\sigma \sqrt{-h}(\mathcal{L}_{simpleSiegel} + \mathcal{L}_{additionalSiegel}), \tag{177}$$

where

$$\mathcal{L}_{simpleSiegel} = \frac{1}{2}h^{ab}\Pi_a^\mu \Pi_{b\mu}, \tag{178}$$

$$\mathcal{L}_{additionalSiegel} = i\frac{\epsilon^{cd}}{\sqrt{-h}}J_c^\alpha J_{\alpha d}, \tag{179}$$

where

$$J_a^\alpha = \partial_a \Theta^\alpha, \tag{180}$$

$$J_{\alpha b} = \partial_b \phi_\alpha - 2i(\partial_b X^\mu)\Gamma_{\mu\alpha\beta}\Theta^\beta - \frac{2}{3}(\partial_b \Theta^\beta)\Gamma^\mu_{\beta\delta}\Theta^\delta \Gamma_{\mu\alpha\epsilon}\Theta^\epsilon. \tag{181}$$

It was proposed by Siegel in [54]. The global symmetry is exact now, not just up to a total divergence. It is $\phi_\alpha$ that allows the Wess-Zumino term to be expressed in a manifestly supersymmetric way.

The additional supersymmetric transformation is

$$\delta\phi_\alpha = 2i\epsilon^\beta \Gamma_{\mu\alpha\beta} X^\mu + \frac{2}{3}(\epsilon^\beta \Gamma^\mu_{\beta\epsilon}\Theta^\epsilon)\Gamma_{\mu\alpha\kappa}\Theta^\kappa. \tag{182}$$

However, the new problem is that the $\phi_\alpha$ fields are not dynamical. That is the reason to involve the modified measure theories which are extremely useful in giving meaning to the yet unknown constants.

## 3.3 A modified measure superstring

In the previous chapter we have introduced two different measures, $\Phi(\varphi)$ and $\Phi(\chi)$. In this chapter we implement them for superstrings. The two-measure superstring was already studied, the Galileon superstring is completely new.

We have already addressed the gauge field, $A_\delta$, to modify the string action. By doing it again, we define

$$-i\epsilon^{cd}\partial_c \Theta^\alpha \partial_d \phi_\alpha \equiv \epsilon^{cd}\partial_c A_d. \tag{183}$$

It is solved by

$$A_d \equiv -i\Theta^\alpha \partial_d \phi_\alpha. \tag{184}$$

Therefore, the additional fields $\phi_\alpha$ induce the abelian gauge field, $A_d$.

### 3.3.1 The two-measure superstring

The two-measure superstring action is

$$S_{TMSUSY} = \int d\tau d\sigma \Phi(\varphi)(\mathcal{L}_{simpleTM} + \mathcal{L}_{additionalTM}), \tag{185}$$

where

$$\mathcal{L}_{additionalTM} = \frac{\epsilon^{ab}}{\sqrt{-h}}(\Pi^\mu_a(\Theta\Gamma_\mu\partial_b\Theta) + \frac{1}{2}F_{ab}(A)). \tag{186}$$

The part $\mathcal{L}_{simpleTM}$ coincides with $\mathcal{L}_{simpleSiegel}$. The metric $h^{ab}$ does not participate in supersymmetric transformations.

The spacetime supersymmetric transformations are

$$\delta X^\mu = -i(\epsilon\Gamma^\mu\Theta), \tag{187}$$

$$\delta\Theta^A = \epsilon^A, \tag{188}$$

$$\delta\phi_\alpha = 2i\epsilon^\beta \Gamma_{\mu\alpha\beta} X^\mu + \frac{2}{3}(\epsilon^\beta \Gamma^\mu_{\beta\epsilon}\Theta^\epsilon)\Gamma_{\mu\alpha\kappa}\Theta^\kappa, \tag{189}$$

$$\delta A_a = i(\epsilon \Gamma_\mu \Theta)(\partial_a X^\mu + \frac{i}{3}\Theta \Gamma^\mu \partial_a \Theta). \tag{190}$$

As in Siegel case, the lagrangian density is invariant under such transformations. The measure is invariant in our case and in Siegel case. The string metric is invariant.

The fields $\phi_\alpha$ and $A_a$ are not the physical fields and are not associated with particles but the auxiliary fields. The field strength that is built out of $A_a$ from the equation $\mathcal{L} = 0$ is solved in terms of the other fields.

The action (185) is invariant under those transformations.

### 3.3.2 A Galileon superstring

The Galileon measure superstring action is

$$S_{GMTSUSY} = -\int d\tau d\sigma (\mathcal{L}_{simpleGM} + \mathcal{L}_{additionalGM})\partial_a(h^{ab}\sqrt{-h}\partial_b\chi), \tag{191}$$

where

$$\mathcal{L}_{additionalGM} = i\frac{\epsilon^{cd}}{\sqrt{-h}} J_c^\alpha J_{\alpha d}. \tag{192}$$

The part $\mathcal{L}_{simpleGM}$ coincides with $\mathcal{L}_{simpleTM}$.

The constraint is already inside the action. Namely,

$$\frac{1}{2}h^{ab}\Pi_a^\mu \Pi_b^\nu + i\frac{\epsilon^{cd}}{\sqrt{-h}} J_c^\alpha J_{\alpha d} = 0. \tag{193}$$

The equations of motion are the same as obtained for (177), the consistency is proved.

However, what matters is that we have the additional equation of motion that determines $\phi_\alpha$. This same result was derived in the two-measure theory but as in the string case the steps were different.

The variation with respect to $\phi_\alpha$ is

$$\epsilon^{ab}\partial_a \Theta^\alpha \partial_b(\frac{\Phi(\chi)}{\sqrt{-h}}) = 0. \tag{194}$$

Therefore, in the nondegenerate case ($\partial_a \Theta^\alpha \neq 0$), we have

$$\frac{\Phi(\chi)}{\sqrt{-h}} = const. \tag{195}$$

The result is very similar to the one obtained for the bosonic string in the previous chapter. Then again the constant is a tension of the string.

Another formulation of supersymmetric strings is also possible. When applied to the Galileon measure to (185), we obtain

$$S_{GMSUSY} = \int d\tau d\sigma \Phi(\chi)(\mathcal{L}_{simpleGM} + \mathcal{L}_{additionalGM}), \tag{196}$$

The part $\mathcal{L}_{simpleGM}$ coincides with $\mathcal{L}_{simpleTM}$. The part $\mathcal{L}_{additionalGM}$ coincides with $\mathcal{L}_{additionalGM}$.

The key point is that the equations of motion are identical, the legitimacy is confirmed.

# 4 A Modified Measure Theory and a Scale Invariance

Every physical quantity is defined by its transformation properties. A scalar field, $\phi$, being the most simple field we can imagine, is a single real function of spacetime that behaves as a scalar under Lorentz transformations, that is, $\phi(x) \to \phi'(x')$. Its dynamics is determined by the kinetic and potential energy densities.

In this chapter we momentarily pause the discussion of strings and consider the modified measure theory by itself. We construct two different systems and study their properties, including the scale invariance.

The scale invariance is the requirement that physics must be the same at all scales, i.e. the system must be invariant under the global scale transformations ($\omega$ is a constant):

$$g_{\mu\nu} \to \omega g_{\mu\nu}. \tag{197}$$

However, the physical universe definitely does not have such property, and different scales show different behavior. Therefore, to approach reality the scale invariance must be broken.

According to Noether's theorem, the symmetries and conservation laws are tightly connected. Despite being a symmetry, the scale invariance does not lead to the conservation of the scale charge.

The consideration of a two scalar system in two dimensions in the framework of two-measure theory is a new thing. The two scalar system with potentials in four dimensions is already known. The latter system is tightly connected with gravity which is beyond the scope of our research, and is given here rather to complete the picture and to show where the notion "two-measure" comes from.

## 4.1 The modified measure theory in two dimensions

We consider a two-dimensional spacetime to avoid any unnecessary complications.

The general form of the action is

$$S = \int \Phi(\varphi) \mathcal{L} d^2 x, \tag{198}$$

where the modified measure is

$$\Phi(\varphi) = \epsilon^{ab} \epsilon_{ij} \partial_a \varphi^i \partial_b \varphi^j, \tag{199}$$

where $\varphi^i, \varphi^j$ are additional scalar fields that have nothing to do with the original $\phi$.

The modified measure leads to the dynamical violation of the scale invariance. To see that, we consider the variation of (198) with respect to $\varphi^j$:

$$A_i^a \partial_a \mathcal{L} = 0, \tag{200}$$

where $A_i^a = \epsilon^{ab}\epsilon_{ij}\partial_b\varphi^j$. The only assumption on $\mathcal{L}$ is its independence from $\varphi^j$'s.

If $\det(A_i^a) \sim \Phi(\varphi)$ is non-trivial, then the solution is

$$\mathcal{L} = M = constant. \tag{201}$$

The appearance of the constant in (201) breaks the scale invariance.

### 4.1.1 The two scalar field system

To break the scale invariance later, the action must be scale invariant initially. Without loss of generality, we assume that the scalar field possesses only the kinetic energy.

$$S = \frac{1}{2}\int \Phi(\varphi)\partial_a\phi\partial_b\phi g^{ab}d^2x. \tag{202}$$

The theory has the scale invariance with the following choice of the rescaling of the fields.

$$\phi \to \lambda^{-\frac{1}{2}}\phi, \quad \varphi^i \to \lambda^{\frac{1}{2}}\varphi^i, \tag{203}$$

where $\lambda$ is the rescaling parameter that applies to the scalar fields and the measure only (and the metric remains invariant).

However, it turns out that this model is not able to bring the enviable results. Then we add one more scalar to the lagrangian.

The final lagrangian is

$$\mathcal{L} = \frac{1}{2}(\partial_a\phi_1\partial_b\phi_1 g^{ab} + \partial_a\phi_2\partial_b\phi_2 g^{ab}), \tag{204}$$

where $\phi_1$ is the former scalar field $\phi$ and $\phi_2$ is the supplemented one.

The final action is

$$S = \frac{1}{2}\int \Phi(\varphi)(\partial_a\phi_1\partial_b\phi_1 g^{ab} + \partial_a\phi_2\partial_b\phi_2 g^{ab})d^2x, \tag{205}$$

where for $S$ to be scale invariant we choose the rescaling of the additional field as

$$\phi_2 \to \lambda^{-\frac{1}{2}}\phi_2. \tag{206}$$

So at that level the dynamics of the initial scalar field $\phi$ is defined by the action (205). We add to that action three more scalar fields. Namely, one scalar field $\phi_2$ is physically equivalent to $\phi$ and enter the lagrangian in the same footing. Two more scalar fields $\varphi^j$ are the base for the newly constructed modified measure in two dimensions. In the following we show where such complexity leads.

A step further is to obtain the equations of motion, i.e. the variations of $S$ with respect to the lagrangian variables.

For simplicity we consider flat two dimensional Minkowski spacetime so that

$$g_{ab} = \eta_{ab}, \tag{207}$$

with the signature $(+-)$.

### 4.1.2 A Born-Infeld dynamics sector

The essence of the Born-Infeld theory is the requirement of finitness of a physical quantities. So we are moving towards a naturally arising restraints on our scalar field.

The lagrangian variables of (205) are $\phi_1$, $\phi_2$, $\varphi^j$.

The variation with respect to $\varphi^j$ leads to

$$\mathcal{L} = \frac{1}{2}(\partial_a\phi_1\partial_b\phi_1 g^{ab} + \partial_a\phi_2\partial_b\phi_2 g^{ab}) = M. \tag{208}$$

Then we can see (even before studying the Born-Infeld scalar sector) that for static case the gradients of the two scalar fields are bounded by $\sqrt{2M}$.

The equation (208) is rewritten to give

$$(\partial_a\phi_1)^2 + (\partial_a\phi_2)^2 = 2M. \tag{209}$$

It is interesting to notice that this is a kind of "nonlinear gradient $\sigma$ model".

The variation with respect to $\phi_2$ is

$$\partial_a(\Phi\partial^a\phi_2) = 0. \tag{210}$$

We assume that $\phi_2$ depends only on the spatial coordinate $x$, $\phi_2 = \phi_2(x)$. Then (210) becomes

$$\partial_1(\Phi\partial_1\phi_2) = 0, \tag{211}$$

which can be integrated to give

$$\Phi\partial_1\phi_2 = J = constant. \tag{212}$$

We observe that the action has the additional shift symmetry:

$$\phi_2 \to \phi_2 + c_2, \tag{213}$$

where $c_2$ is a constant.

This symmetry leads to the conservation law. Then $J$ has the interpretation of a constant current flowing in the $x$-direction. Then

$$\partial_1\phi_2 = \frac{J}{\Phi}. \tag{214}$$

Inserting this into (208), we get

$$\partial_a\phi_1\partial^a\phi_1 + \frac{J^2}{\Phi^2} = 2M, \qquad (215)$$

which can be used to solve for the measure $\Phi$, giving

$$\Phi = \frac{J}{\sqrt{2M - \partial_a\phi_1\partial^a\phi_1}}. \qquad (216)$$

The variation with respect to $\phi_1$ is

$$\partial_a(\Phi\partial^a\phi_1) = 0. \qquad (217)$$

Making the same assumptions as for the $\phi_2$, namely $\phi_1 = \phi_1(x)$ and

$$\phi_1 \to \phi_1 + c_1, \qquad (218)$$

where $c_1$ is a constant, we obtain

$$\partial_a(\Phi\partial^a\phi_1) = 0. \qquad (219)$$

Inserting (216) into (217), we get the Born-Infeld scalar equation (for $M > 0$)

$$J\partial_a\left(\frac{\partial^a\phi_1}{\sqrt{2M - \partial_a\phi_1\partial^a\phi_1}}\right) = 0, \qquad (220)$$

which is also obtained from the effective Born-Infeld action for this kind of solutions.

$$S_{eff} = \int \sqrt{2M - \partial_a\phi_1\partial^a\phi_1}\, d^2x. \qquad (221)$$

The dynamics of $\phi_1$ defined by the equation (220) is the same as the dynamics of $\phi_1$ derived from the variation of (221). This means that $\partial_a\phi_1\partial^a\phi_1$ is bounded in this sector of the theory (the Born-Infeld scalar sector) Notice, however, that we are now considering only a sector of the theory.

### 4.1.3 A dynamical violation of the scale invariance

The action (205) is invariant under the scale transformations (203) and (206). By the Noether's theorem a conserved quantity must appear, namely, the scale charge, $Q$. Then the continuity equation must be satisfied.

$$\frac{\partial\rho}{\partial t} + \nabla \cdot j = 0, \qquad (222)$$

where $\rho$ is a density of $Q$ and $j$ is the flux of $Q$.

By the the integration

$$\int_{x_1}^{x_2} \frac{\partial\rho}{\partial t} d^2x + \int_{x_1}^{x_2} \frac{\partial j^1}{\partial x} d^2x = 0 \qquad (223)$$

we obtain

$$\frac{dQ}{dt} + j^1(x_2) - j^1(x_1) = 0. \tag{224}$$

Therefore, the conservation of the total charge requires

$$j^1(x_1 \to -\infty) - j^1(x_2 \to +\infty) = 0. \tag{225}$$

However, it does not always happen. Our case is one of the exceptions.

The conserved current is given by

$$j^a = \frac{\partial \Phi(\varphi)\mathcal{L}}{\partial(\partial_a \varphi^j)}\delta\varphi^j + \frac{\partial \Phi(\varphi)\mathcal{L}}{\partial(\partial_a \phi_1)}\delta\phi_1 + \frac{\partial \Phi(\varphi)\mathcal{L}}{\partial(\partial_a \phi_2)}\delta\phi_2. \tag{226}$$

We consider a scale transformations infinitesimally closed to the identity: $\lambda = (1+\theta)$, so that (203) and (206) turn into

$$\varphi^j \to (1+\theta)^{\frac{1}{2}}\varphi^j \simeq \varphi^j + \frac{\theta}{2}\varphi^j, \tag{227}$$

$$\phi_1 \to (1+\theta)^{-\frac{1}{2}}\phi_1 \simeq \phi_1 - \frac{\theta}{2}\phi_1, \tag{228}$$

$$\phi_2 \to (1+\theta)^{-\frac{1}{2}}\phi_2 \simeq \phi_2 - \frac{\theta}{2}\phi_2. \tag{229}$$

Therefore,

$$\delta\varphi^j = \frac{\theta}{2}\varphi^j, \quad \delta\phi_1 = -\frac{\theta}{2}\phi_1, \quad \delta\phi_1 = -\frac{\theta}{2}\phi_2. \tag{230}$$

Then (226) becomes

$$j^a = M\frac{\theta}{2}\epsilon^{ab}\epsilon_{ij}\varphi^i\partial_b\varphi^j - \frac{\theta}{2}\Phi\partial^a\phi_1 - \frac{\theta}{2}\Phi\partial^a\phi_2. \tag{231}$$

Let's go back to (220) and find the static solutions ($\partial_0\phi_1 = 0$):

$$\partial_1\left(\frac{\partial_1\phi_1}{\sqrt{2M - (\partial_1\phi_1)^2}}\right) = 0. \tag{232}$$

By integration we get

$$\frac{\partial_1\phi_1}{\sqrt{2M - (\partial_1\phi_1)^2}} = c_3, \tag{233}$$

where $c_3$ is a constant.

Then

$$\phi_1 = \frac{\sqrt{2M}|c_3|}{\sqrt{1+|c_3|^2}}(x_2 - x_1). \tag{234}$$

We have done all the calculations for $\phi_1$. However, the same is relevant for $\phi_2$. So that

$$\phi_2 = \frac{\sqrt{2M}|c_4|}{\sqrt{1+|c_4|^2}}(x_2 - x_1), \tag{235}$$

where $c_4$ is a constant.

Inserting this solution to (216), we obtain

$$\Phi = J\sqrt{\frac{1+c_3^2}{(1-2M)c_3^2 + 1}}. \tag{236}$$

Then we see

$$\Phi = \Phi_0 = constant. \tag{237}$$

It is satisfied for

$$\varphi^1 = c_5 t, \quad \varphi^2 = c_6 x, \tag{238}$$

where $c_5$ and $c_6$ are constants.

Then indeed

$$\Phi = c_5 c_6. \tag{239}$$

Now by inserting the solutions for $\varphi^j$, $\phi_1$ and $\phi_2$ into (278), we obtain for $j^1$

$$j^1 = \frac{\theta}{2}Mc_5c_6 x - \frac{\theta}{2}c_5c_6\frac{\sqrt{2M}|c_3|}{\sqrt{1+|c_3|^2}} - \frac{\theta}{2}c_5c_6\frac{\sqrt{2M}|c_4|}{\sqrt{1+|c_4|^2}}. \tag{240}$$

We see that $j^1$ is a constant plus a term proportional to $x$ and therefore, $j^1(\infty) - j^1(-\infty) \neq 0$ and in fact diverges. Therefore, $Q$ is not conserved.

Let's calculate $j^0$ explicitly.

$$j^0 = -\frac{\theta}{2}Mc_5c_6 t. \tag{241}$$

Then we see that

$$Q = \int_{x_1}^{x_2} j^0 dx = -(x_2 - x_1)\frac{\theta}{2}Mc_5c_6 t. \tag{242}$$

We checked that $Q$ is not conserved.

## 4.2 The modified measure theory in four dimensions

We consider a four-dimensional spacetime to apply it later to the cosmological problem.

The general form of the action is

$$S = \int \Phi(\varphi)\mathcal{L}d^4x, \tag{243}$$

where the modified measure is

$$\Phi(\varphi) = \epsilon^{\mu\nu\alpha\beta}\epsilon_{\gamma\delta\sigma\kappa}\partial_\mu\varphi^\gamma\partial_\nu\varphi^\delta\partial_\alpha\varphi^\sigma\partial_\beta\varphi^\kappa. \tag{244}$$

If we rewrite the measure $\Phi(\varphi)$ in the following way:

$$\Phi(\varphi) = \partial_\mu(\epsilon^{\mu\nu\alpha\beta}\epsilon_{\gamma\delta\sigma\kappa}\varphi^\gamma\partial_\nu\varphi^\delta\partial_\alpha\varphi^\sigma\partial_\beta\varphi^\kappa), \tag{245}$$

then we see that $\Phi(\varphi)$ is the total derivative. Therefore, a shift of the form

$$\mathcal{L} \to \mathcal{L} + const \tag{246}$$

does not affect the equations of motion since it adds just the integral of a total divergence. The theory with such property is called the Non Gravitating Vacuum Energy Theory.

Indeed, if we take the standard measure of integration, i.e. $\Phi(\varphi) = \sqrt{-g}$, then the shift (246) leads to the cosmological constant.

The next step is to consider two different measures at the same time. Then the action is

$$S = \int \mathcal{L}_1\Phi(\varphi)d^4x + \int \mathcal{L}_2\sqrt{-g}d^4x. \tag{247}$$

That is where the title "two measure" originated from. That is why $\Phi(\varphi)$ is called the modified measure from the two-measure theory.

### 4.2.1 The two scalar field system with potentials

The scalar field $\phi$ interacts with gravity. Then the lagrangians are

$$\mathcal{L}_1 = -\frac{1}{\kappa}R(\Gamma, g) + \frac{1}{2}g^{\mu\nu}\partial_\mu\phi\partial_\nu\phi - V(\phi), \tag{248}$$

where $\kappa = 8\pi G$, $G$ is the gravitational constant, $R$ is the Ricci scalar and $V(\phi)$ is a potential,

$$\mathcal{L}_2 = U(\phi), \tag{249}$$

where $U(\phi)$ is a potential. The potentials $V(\phi)$ and $U(\phi)$ are arbitrary for now.

The lagrangian variables are $\varphi^\gamma$, $\phi$, $g_{\mu\nu}$ and $\Gamma^\lambda_{\mu\nu}$ which are the connections.

The variation with respect to $\varphi^\gamma$ leads to

$$\mathcal{L}_1 = -\frac{1}{\kappa}R(\Gamma, g) + \frac{1}{2}g^{\mu\nu}\partial_\mu\phi\partial_\nu\phi - V = M, \tag{250}$$

where $M$ is a constant.

The variation with respect to $\Gamma^\lambda_{\mu\nu}$ leads to

$$-\Gamma^\lambda_{\mu\nu} - \Gamma^\alpha_{\beta\mu}g^{\beta\lambda}g_{\alpha\nu} + \delta^\lambda_\nu \Gamma^\alpha_{\mu\alpha} + \delta^\lambda_\mu g^{\alpha\beta}\Gamma^\gamma_{\alpha\beta}g_{\gamma\nu}-$$

$$-g_{\alpha\nu}\partial_\mu g^{\alpha\lambda} + \delta^\lambda_\mu g_{\alpha\nu}\partial_\beta g^{\alpha\beta} - \delta^\lambda_\nu \frac{\Phi(\varphi)_{,\mu}}{\Phi(\varphi)} + \delta^\lambda_\mu \frac{\Phi(\varphi)_{,\nu}}{\Phi(\varphi)} = 0. \tag{251}$$

The variation with respect to $g^{\mu\nu}$ leads to

$$\Phi(\varphi) = (-\frac{1}{\kappa}R_{\mu\nu} + \frac{1}{2}\phi_{,\mu}\phi_{,\nu}) - \frac{1}{2}\sqrt{-g}U(\phi)g_{\mu\nu} = 0 \tag{252}$$

If we combine (251) and (252), then we obtain

$$M + V(\phi) - 2U(\phi)\frac{\sqrt{-g}}{\Phi(\varphi)} = 0. \tag{253}$$

Therefore,

$$\frac{\Phi}{\sqrt{-g}} = \frac{2U\phi}{M + V(\phi)}. \tag{254}$$

In the Einstein conformal frame, in which

$$\bar{g}_{\mu\nu} = \frac{\Phi}{\sqrt{-g}}g_{\mu\nu}. \tag{255}$$

the equations are

$$R_{\mu\nu}(\bar{g}_{\mu\nu}) - \frac{1}{2}\bar{g}_{\mu\nu}R(\bar{g}_{\mu\nu}) = \frac{\kappa}{2}T^{eff}_{\mu\nu}(\phi), \tag{256}$$

where

$$T^{eff}_{\mu\nu}(\phi) = \phi_{,\mu}\phi_{,\nu} - \frac{1}{2}\bar{g}_{\mu\nu}\phi_{,\alpha}\phi_{,\beta}\bar{g}_{\alpha\beta} + \bar{g}_{\mu\nu}V_{eff}(\phi) \tag{257}$$

and

$$V_{eff}(\phi) = \frac{1}{4U(\phi)}(V(\phi) + M)^2. \tag{258}$$

The equation of motion of the scalar field $\phi$ is

$$\frac{1}{\sqrt{-\bar{g}}}\partial_\mu(\bar{g}^{\mu\nu}\sqrt{-\bar{g}}\partial_\nu\phi) + V'_{eff}(\phi) = 0. \tag{259}$$

If $V + M = 0$, then $V_{eff} = 0$ and $V'_{eff} = 0$ and, therefore, $V'$ is finite and $U \neq 0$.

The main benefit of the non gravitating vacuum energy theory is that the zero cosmological constant state is achieved without any fine tuning.

### 4.2.2 A cosmological application

The global scale invariance, $g_{\mu\nu} \to e^\theta g_{\mu\nu}$, requires the potentials to be of the following form:

$$V(\phi) = f_1 e^{\alpha\phi}, \qquad U(\phi) = f_2 e^{2\alpha\phi}. \tag{260}$$

Then

$$V_{eff} = \frac{1}{4f_2}(f_1 + Me^{-\alpha\phi})^2. \tag{261}$$

If $\phi \to \infty$, then $V_{eff} \to \frac{f_1^2}{4f_2} = const$ and therefore, there exists an infinite flat region.

The minimum is achieved at zero cosmological constant at the following point:

$$\phi_{min} = -\frac{1}{\alpha} \ln\left|\frac{f_1}{M}\right|, \tag{262}$$

if $\frac{f_1}{M} < 0$.

We obtain at the minimum

$$V''_{eff} = \frac{\alpha^2}{2f_2}|f_1|^2 > 0, \tag{263}$$

if $f_2 > 0$.

Then we see that the scalar field potential energy, $V_{eff}(\phi)$, has two critical features. First, a standard scalar field potential is achieved. Second, there is an infinite region of flat potential for $\phi \to \infty$ and, therefore, a slowly rolling scenario is viable, provided the universe is started at a sufficiently large value if the scalar field $\phi$.

# 5 The String Model of Hadrons.

By the definition, hadrons are subatomic particles that are made up of quarks and, therefore, interact by a strong force. Mesons being two-quarks particles may be approximated by a single string, and baryons being three-quarks particles have more freedom in that regard.

In this chapter we present our own models of stringy hadrons and discuss their benefits.

## 5.1 The String Meson Model

A meson being a quark-antiquark bound system is reproduced by an open string with the opposite charged endpoints, see Fig.1.

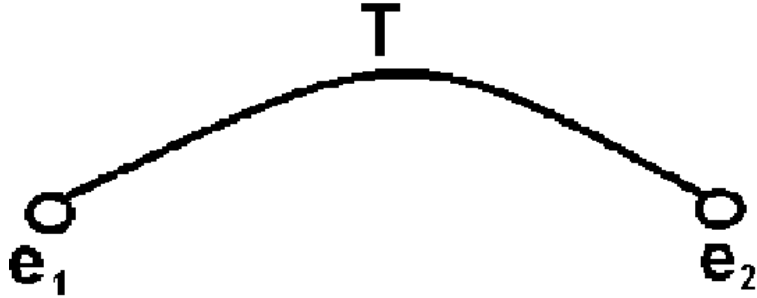

Figure 1: The string meson. Circles are the charged endpoints. $T$ is the tension of the string. $e_1$ and $e_2$ are charges with the condition $e_1 = -e_2$.

Once again, the two-measure action of a single string is

$$S_{single-string} = -\int d\tau d\sigma \Phi(\varphi)(h^{ab}\partial_a X^\mu \partial_b X^\nu g_{\mu\nu} - \frac{\epsilon^{cd}}{\sqrt{-h}}F_{cd}), \tag{264}$$

where $\tau, \sigma$ are the worldsheet parameters, $h^{ab}$ is the intrinsic metric on the worldsheet, $h$ is its determinant, $g_{\mu\nu}$ is the spacetime metric, $X^\mu$ are coordinate functions, $X^\mu = X^\mu(\tau, \sigma)$, $\epsilon^{cd}$ is the Levi-Civita symbol, $F_{cd} = \partial_c A_d - \partial_d A_c$ is the field-strength of the auxiliary Abelian gauge field $A_d$ and the two-measure is

$$\Phi(\varphi) = \epsilon^{ab}\epsilon_{ij}\partial_a \varphi^i \partial_b \varphi^j, \tag{265}$$

where $\varphi^i$ are two (by the number of dimensions) additional worldsheet scalar fields.

This string is not infinite. A string tension, $T$, is a constant along the string and vanishes at the endpoints. The string terminates when its tension discontinues as can be seen from the following:

The modified measure string with tension that can dynamically end at the endpoints has an action defined by

$$S_{single-endpoints} = S_{single-string} + \int d\sigma d\tau A_a j^a, \tag{266}$$

where $j_a$ is the current of point-like charges setting at the endpoints.

The additional term in the action (266) contains the gauge field, $A_a$, interacting with point charges. Then the equations of motion with respect to the gauge field are modified compared to (137). They are

$$\epsilon^{ab}\partial_b\left(\frac{\Phi}{\sqrt{-h}}\right) = j^a. \tag{267}$$

As the meson is in a static configuration then the current becomes

$$j^0 = \sum_i e_i \delta(\sigma - \sigma_i), \tag{268}$$

where $e_i$ are charges that are associated with the gauge field $A^a$ and $\sigma_i$, $i = 1, 2$ are their locations, i.e. the endpoints.

Note, that the string with Neumann boundary conditions has zero size if it is static. Classically one can discuss rotating strings and quantum mechanically excited states (not the ground state) can have non-trivial length even without angular momentum.

Therefore,

$$\int d\sigma d\tau A_a j^a = \int d\sigma d\tau A_0(\tau,\sigma) j^0 = \sum_i e_i \int d\tau A_0(\tau,\sigma_i). \tag{269}$$

Also (267) turns to

$$\epsilon^{01}\partial_1\left(\frac{\Phi}{\sqrt{-h}}\right) = j^0, \tag{270}$$

where $(0, 1)$ means $(\tau, \sigma)$.

Then, instead of (139) we obtain

$$\frac{\Phi}{\sqrt{-h}} = \sum_i e_i \theta(\sigma - \sigma_i). \tag{271}$$

For proof we first consider the endpoint on the left with $e_1$ which is located at $\sigma_1$

$$\partial_\sigma\left(\frac{\Phi}{\sqrt{-h}}\right) = e_1 \delta(\sigma - \sigma_1) \tag{272}$$

and integrate the right-hand-side (rhs)

$$\int_{\sigma_1-\epsilon}^{\sigma_1+\epsilon} e_1 \delta(\sigma - \sigma_1) d\sigma = e_1, \tag{273}$$

where $\epsilon$ is some positive constant.

The integration of the left-hand-side (lhs) gives

$$\int_{\sigma_1-\epsilon}^{\sigma_1+\epsilon} \partial_\sigma(\frac{\Phi}{\sqrt{-h}})d\sigma = (\frac{\Phi}{\sqrt{-h}})|_{\sigma_1+\epsilon} - (\frac{\Phi}{\sqrt{-h}})|_{\sigma_1-\epsilon}. \tag{274}$$

The first rhs term of (274) is equal to $T$ because of (139) and the second one vanishes because the string starts at $\sigma_1$ and does not exist to $\sigma_1$'s left. Therefore, $e_1 = T$.

Now we consider the endpoint on the right with $e_2$ which is located at $\sigma_2$

$$\partial_\sigma(\frac{\Phi}{\sqrt{-h}}) = e_2\delta(\sigma - \sigma_2). \tag{275}$$

In analogue with the endpoint on the left, we obtain

$$(\frac{\Phi}{\sqrt{-h}})|_{\sigma_2+\epsilon} = e_2 + (\frac{\Phi}{\sqrt{-h}})|_{\sigma_2-\epsilon} = 0. \tag{276}$$

Then

$$T = -e_2. \tag{277}$$

The tension is changed discontinuously from zero at one end of the string to some constant and back to zero at the other end of the string. This is the condition for the string to terminate. The opposite charges at the ends of the string guarantee it.

In the standard string theory the endpoints are free until the boundary conditions are applied. In the modified measure theory the boundary conditions are derived as is seen from the following consideration:

The variation with respect to $X^\mu$ gives the same equations of motion as those obtained from the action (129). We need only the first term from the lhs of (136) because only this term could be singular. The external space is well defined and therefore, so is $\Gamma^\mu_{\nu\alpha}$, and $\frac{\Phi}{\sqrt{-h}}$ can jump but still remains finite, however, $\partial_a(\frac{\Phi}{\sqrt{-h}})$ will be singular

$$\partial_a(\frac{\Phi}{\sqrt{-h}})\sqrt{-h}h^{ab}\partial_b X^\mu = 0. \tag{278}$$

Inserting (271) in (278) we obtain

$$e_i\delta(\sigma - \sigma_i)\delta^\sigma_a\sqrt{-h}h^{ab}\partial_b X^\mu = 0. \tag{279}$$

The worldsheet metric $h^{ab}$ can always be taken in a certain gauge to be conformally flat. Then in the conformal gauge in which $h^{ab}\sqrt{-h} = \eta^{ab}$, we obtain

$$\partial_\sigma X^\mu(\tau, \sigma_i) = 0. \tag{280}$$

The equation (280) is the Neumann boundary conditions, which are in fact the constraints on momentum components. They are obtained at the points where charges are located. Being originated from the discontinuity of the dynamical tension these conditions arise naturally in the framework of the modified measure theory. It is even impossible to violate them when having in hand only one string.

There is no reason to believe that the modified string model is non-tachyonic. The equations of motion are the same in the classical case. The quantum theory should be similar. However, we do not consider the quantum case.

In the hadronic string models the glueballs are closed strings. The action, $S_{glueball}$, is

$$S_{glueball} = -\int d\tau d\sigma \Phi(\varphi)(h^{ab}\partial_a X^\mu \partial_b X^\nu g_{\mu\nu} - \frac{\epsilon^{cd}}{\sqrt{-h}}F_{cd}). \tag{281}$$

No discontinuity of a tension is required. Therefore, no charges are required. The tension is constant and appears as a constant of integration, as the ratio between the two measures:

$$\frac{\Phi}{\sqrt{-h}} = const. \tag{282}$$

## 5.2 The String Baryon Model

A baryon being a three-quark bound system is reproduced by two open strings with charged endpoints each and an additional charge within one of the strings, see Fig.2.

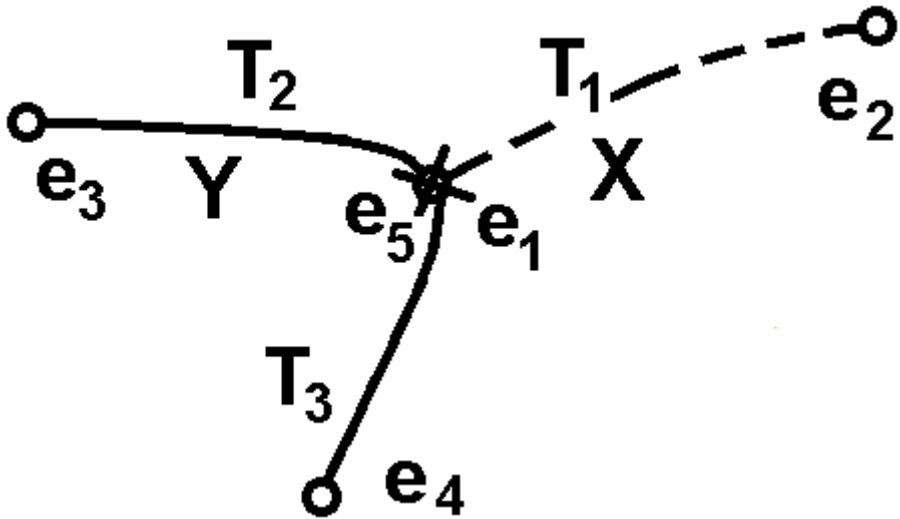

Figure 2: The string baryon. Dotted and curved lines denote two strings, $X$ and $Y$, respectively. A cross is the intersection point. $T_1$, $T_2$, $T_3$ are the tensions. The charge's number corresponds to the number of its location, for example, the charge $e_3$ is located at the point $\sigma_3$, for example, the charge $e_3$ is located at the point $\sigma_3$.

Note that the X-string stretches from $e_2$ to $e_1$ and the Y-string stretches from $e_3$ via $e_1$ to $e_4$. If one considers the line between $e_4$ (or $e_3$) via $e_1$ to $e_2$ as a Y-string and the line between $e_3$ (or $e_4$) and $e_1$ as an X-string, it will lead through the same calculations to the same results. Those choices are equivalent. However, once the choice is made, the picture is not symmetric anymore. We are working further with the one particular case because the charges are related among themselves and to the tensions in a particular way each time a choice is made.

The X-string has two endpoints with charges $e_1$ and $e_2$. In order for this string to terminate, the charges must be opposite: $e_1 = -e_2$. The Y-string too has two endpoints with charges $e_3$ and $e_4$. However, it has the additional charge $e_5$, which appears within the Y-string. Therefore, these strings do not enter equally. The charges $e_1$ and $e_5$ are located at the point $\sigma_1$ which is then the intersection point of two strings. At the point $\sigma_1$ the Y-string tension is changed from $T_2$ to $T_3$. Therefore, although $e_3 = T_2$ (see (271)) but $T_3 = e_3 + e_5 = -e_4$. The mathematical formulation is coming.

Each string comes with its own internal metric $h_X^{ab}$ or $h_Y^{ab}$, its own measure $\Phi_X$ or $\Phi_Y$ and its own gauge field $A_a$ or $B_a$.

The additional terms in the X-string and Y-string actions are $\sum_i \int d\sigma d\tau A_a j_A^{ia}$ and $\sum_j \int d\sigma d\tau B_a j_B^{ja}$, respectively. Note that $i = 1, 2$ while $j = 3, 4, 5$. The interaction takes place at $\sigma = \sigma_1$ as viewed from the X-string or $\sigma = \sigma_5$ as viewed from the Y-string. Then at this point ($\sigma = \sigma_1 = \sigma_5$) we obtain $j_A^0 = e_1 \delta(\sigma - \sigma_1)$ and $j_B^0 = e_5 \delta(\sigma - \sigma_1)$.

The X-string charges are similar to the string meson charges (280) that were considered in the previous subsection. So are the Y-string endpoint charges. But the charge $e_5$ has its own Neumann boundary conditions at the point $\sigma_1$

$$\partial_\sigma \left(\frac{\Phi}{\sqrt{-h}}\right) = e_5 \delta(\sigma - \sigma_1). \tag{283}$$

Integrating both parts we obtain

$$\int_{\sigma_1 - \epsilon}^{\sigma_1 + \epsilon} e_5 \delta(\sigma - \sigma_1) d\sigma = e_5, \tag{284}$$

$$\int_{\sigma_1 - \epsilon}^{\sigma_1 + \epsilon} \partial_\sigma \left(\frac{\Phi}{\sqrt{-h}}\right) d\sigma = \left(\frac{\Phi}{\sqrt{-h}}\right)\big|_{\sigma_1 + \epsilon} - \left(\frac{\Phi}{\sqrt{-h}}\right)\big|_{\sigma_1 - \epsilon} = T_3 - T_2. \tag{285}$$

Therefore

$$e_5 = T_3 - T_2. \tag{286}$$

The equations of motion with respect to $Y^\mu$ gives us as previously the Neumann boundary conditions

$$\partial_\sigma Y^\mu(\tau, \sigma_1) = 0. \tag{287}$$

All together the Neumann boundary conditions for the Y-string are

$$\partial_\sigma Y^\mu(\tau, \sigma_j) = 0, \tag{288}$$

where $j = 3, 4, 5$.

Then we see that the Neumann boundary conditions are applied not only at the endpoints but at the intersection point too. They signalize that the tension undergoes alterations: it becomes zero at the endpoints thereby terminating the string while at the intersection point it changes its value thereby dividing the string into two strings with different tensions. This is how we construct a baryon out of two strings as opposed to the more standard three string construction.

Just like the Neumann boundary conditions exist already in the modified measure theory, the Dirichlet boundary conditions are also contained within the theory and are derived in this subsection.

In order to obtain that $X$-string $= Y$-string at $\sigma = \sigma_1$ the conditions for the intersection point are needed.

Our guiding principle is that the part of the action that is responsible for the interaction must be conformal invariant, generalizing the case of a single string equations.

The interaction term for two strings that leads to such conditions is

$$S_{interaction} = \int d\sigma d\tau (\lambda_1 \sqrt{-h_X} h_X^{ab} + \lambda_2 \sqrt{-h_Y} h_Y^{ab}) \partial_a(\frac{\Phi_X}{\sqrt{-h_X}}) \partial_b(\frac{\Phi_Y}{\sqrt{-h_Y}}) V(X,Y), \tag{289}$$

where $\lambda_1, \lambda_2$ are positive coefficients and $V(X,Y)$ is some potential that is defined later. The range of integration over $\sigma$ is taken to be $[-\infty, \infty]$, because, nonetheless, the physical range is only the parts where the tensions are not zero. The actual limits of integration are set dynamically.

Note that the $(\tau, \sigma)$-space being the common space of two strings is not the worldsheet of any string, and therefore, $\sigma$ is not the worldsheet coordinate. The actual space where the strings live is determined by measures, $\Phi_X$ and $\Phi_Y$.

The equations of motion provide us with the constraints on $V(X,Y)$.

Notice that in order to be effective $S_{interaction}$ requires both tensions, $\partial_a(\frac{\Phi_X}{\sqrt{-h_X}})$ and $\partial_b(\frac{\Phi_Y}{\sqrt{-h_Y}})$, to have a jump at the same point, otherwise $S_{interaction}$ vanishes. So that is why there is the need for the gauge fields charges at the point $\sigma_1$.

The equation (289) is indeed conformal invariant because

$$h_X^{ab} \to \Omega^{-2} h_X^{ab}, \quad \sqrt{-h_X} \to \Omega^2 \sqrt{-h_X}. \tag{290}$$

Then

$$h_X^{ab}\sqrt{-h_X} \to h_X^{ab}\sqrt{-h_X}. \tag{291}$$

Using $\Phi_X \to \Omega^2 \Phi_X$, we obtain

$$h_X^{ab}\Phi_X \to h_X^{ab}\Phi_X \tag{292}$$

and

$$\left(\frac{\Phi_X}{\sqrt{-h_X}}\right) \to \left(\frac{\Phi_X}{\sqrt{-h_X}}\right). \tag{293}$$

Also

$$h_Y^{ab}\sqrt{-h_Y} \to h_Y^{ab}\sqrt{-h_Y}. \tag{294}$$

Note that even though the full system has separate conformal invariance for each string, $h_X^{ab} \to \Omega_X^{-2}h_X^{ab}$, $h_Y^{ab} \to \Omega_Y^{-2}h_Y^{ab}$, the reparameterization invariance is still common. As we work in a worldsheet that is common to both strings, a separate reparameterization invariance is not possible in principle .

The equations of motion with respect to $X^\mu$ acquire an extra term comparing to (136). It is

$$(\lambda_1\sqrt{-h_X}h_X^{ab} + \lambda_2\sqrt{-h_Y}h_Y^{ab})\partial_a\left(\frac{\Phi_X}{\sqrt{-h_X}}\right)\partial_b\left(\frac{\Phi_Y}{\sqrt{-h_Y}}\right)\frac{\partial V(X,Y)}{\partial X^\mu}. \tag{295}$$

From the analogues to (271) we see that

$$\partial_\sigma\left(\frac{\Phi_X}{\sqrt{-h_X}}\right) = e_1\delta(\sigma - \sigma_1), \tag{296}$$

$$\partial_\sigma\left(\frac{\Phi_Y}{\sqrt{-h_Y}}\right) = e_5\delta(\sigma - \sigma_1). \tag{297}$$

These terms produce two delta-functions, $\delta^2(\sigma - \sigma_1)$, which should be eliminated. Then

$$\frac{\partial V(X,Y)}{\partial X^\mu}\bigg|_{\sigma=\sigma_1} = 0. \tag{298}$$

If not the X-string but Y-string is considered, then variation with respect to $Y^\mu$ gives us the similar condition

$$\frac{\partial V(X,Y)}{\partial Y^\mu}\bigg|_{\sigma=\sigma_1} = 0. \tag{299}$$

The equations of motion with respect to $h_X^{cd}$ are altered too comparing with (134). The additional term is

$$\lambda_1 h_X^{ab}\frac{1}{2}\sqrt{-h_X}h_{Xcd}\partial_a\left(\frac{\Phi_X}{\sqrt{-h_X}}\right)\partial_b\left(\frac{\Phi_Y}{\sqrt{-h_Y}}\right)V(X,Y)+$$

$$+\lambda_1\frac{1}{2}(\delta_a^c\delta_b^d + \delta_b^c\delta_a^d)\sqrt{-h_X}\partial_a\left(\frac{\Phi_X}{\sqrt{-h_X}}\right)\partial_b\left(\frac{\Phi_Y}{\sqrt{-h_Y}}\right)V(X,Y)+$$

$$+\frac{1}{2}\frac{\Phi_X}{\sqrt{-h_X}}h_{Xcd}\partial_a((\lambda_1\sqrt{-h_X}h_X^{ab}+\lambda_2\sqrt{-h_Y}h_Y^{ab})\partial_b(\frac{\Phi_Y}{\sqrt{-h_Y}})V(X,Y))=0. \tag{300}$$

Since (298) is established, then the additional constraint on $V(X,Y)$ is

$$V(X,Y)|_{\sigma=\sigma_1}=0. \tag{301}$$

The intersection point is fixed now. The next task is to define the potential $V(X,Y)$ itself.

It is a function that is defined at the point where we demand the intersection of two strings to occur. The intersection condition is

$$X^\mu|_{\sigma=\sigma_1}=Y^\mu|_{\sigma=\sigma_1}. \tag{302}$$

The most simple form it can take in the case of a flat spacetime background is

$$V=(X-Y)^2=\eta_{\mu\nu}(X^\mu-Y^\mu)(X^\nu-Y^\nu). \tag{303}$$

The constraint (301) leads exactly to

$$X^\mu|_{\sigma=\sigma_1}=Y^\mu|_{\sigma=\sigma_1}. \tag{304}$$

We obtain these conditions dynamically by adding the term to the action and specifying the potential $V(X,Y)$. These are exactly the Dirichlet boundary conditions.

All in all, the action governing the string baryon configuration is

$$S_{baryon}=-\int d\tau d\sigma \Phi(\varphi)_X[\frac{1}{2}h_X^{ab}\partial_a X^\mu\partial_b X^\nu g_{\mu\nu}-\frac{\epsilon^{cd}}{2\sqrt{-h_X}}F_{cd}]+\sum_{i=1,2}\int d\tau d\sigma A_i j_A^i+$$

$$-\int d\tau d\sigma \Phi(\varphi)_Y[\frac{1}{2}h_Y^{ab}\partial_a Y^\mu\partial_b Y^\nu g_{\mu\nu}-\frac{\epsilon^{cd}}{2\sqrt{-h_Y}}F_{cd}]+\sum_{j=3,4,5}\int d\tau d\sigma B_j j_B^j+$$

$$+\int d\tau d\sigma(\lambda_1\sqrt{-h_X}h_X^{ab}+\lambda_2\sqrt{-h_Y}h_Y^{ab})\partial_a(\frac{\Phi(\varphi)_X}{\sqrt{-h_X}})\partial_b(\frac{\Phi(\varphi)_Y}{\sqrt{-h_Y}})V(X,Y). \tag{305}$$

This equation is invariant under local scale transformations.

The single string action is invariant. The terms with the charges are not transformed. The interaction term is invariant, see 290.

## 5.3 The String Baryon Model by G. 't Hooft

The following string baryon model was proposed by G. 't Hooft in [91].

Three strings $(X^{\mu,1}, X^{\mu,2}, X^{\mu,3})$ intersect at the point $\sigma = 0$. The Lagrange multipliers $(l_1^\mu(\tau), l_2^\mu(\tau))$ are introduced in the interaction term.

$$S_{interaction-tHooft} = \int d\tau (l_1^\mu(\tau)(X^{\mu,1}(0,\tau) - X^{\mu,3}(0,\tau)) + l_2^\mu(\tau)(X^{\mu,2}(0,\tau) - X^{\mu,3}(0,\tau))). \tag{306}$$

The boundary conditions for the intersection point $\sigma = 0$ are

$$\partial_\sigma(X^{\mu,1} + X^{\mu,2} + X^{\mu,3}) = 0, \quad X^{\mu,1} - X^{\mu,3} = X^{\mu,2} - X^{\mu,3} = 0, \tag{307}$$

and for each endpoint ($\sigma = L^k(\tau)$) are:

$$\partial_\sigma X^{\mu,k} = 0, \tag{308}$$

where $k = 1,2,3$ and $L^k(\tau)$ are the lengths.

So the Neumann boundary conditions hold only for the sum $\sum_{k=1}^{3} X^{\mu,k}$ at the intersection point.

By choosing conformal gauge, where $h_{ab} = \Omega^2 \eta_{ab}$, the wave equation $\Box X^\mu = 0$ holds outside the intersection point or endpoints. Therefore

$$X^\mu = X_L^\mu(\tau + \sigma) + X_R^\mu(\tau - \sigma). \tag{309}$$

Then the boundary conditions at the endpoints (308) are

$$X_L^{\mu,k}(\tau, L^k(\tau)) = X_R^{\mu,k}(\tau, L^k(\tau)). \tag{310}$$

But in G. 't Hooft's treatment things are more complicated at the intersection point. The signal is propagated to the endpoints and reflects back to the intersection point. So that the boundary conditions (307) are nonlocal in time and take the form

$$X_L^{\mu,k}(\tau, 0) = X_R^{\mu,k}(\tau_k, 0). \tag{311}$$

As is seen, $X_L$ and $X_R$ are evaluated at different times, and $\tau_k(\tau)$ are the solutions of

$$\tau - \tau_k = 2L^k(\tau_k'), \quad \tau_k' \equiv \frac{\tau + \tau_k}{2}. \tag{312}$$

In the next subsection we show that this nonlocality is absent in our approach.

G. t' Hooft analysis shows that the $Y$ shape collapses to a string between a quark and a diquark due to classical fluctuations. The string advocated by this thesis could be a $Y$ shape and that does not contradict the analysis of G. t' Hooft. In our research we consider two separate strings that are close to each other. However, we do not indicate how close they are to each other.

It could be also the collapsed case. However, the general construction is still valid. The analysis is not affected by the shape.

## 5.4 The Resolution of the Nonlocality

Starting from here $\sigma_0$ denotes the intersection point as previously interchangeably $\sigma_1$ and $\sigma_5$. We put $\sigma_0$ to $0$ for the comparison with G. 't Hooft's results.

The key feature of our model is that the Neumann boundary conditions (280), (288) hold not only at the endpoints but at the intersection point too.

Again by choosing the conformal gauge the wave equations $\Box X^\mu = 0$, $\Box Y^\mu = 0$ hold outside the intersection point or endpoints. Therefore

$$X^\mu = X_L^\mu(\tau + \sigma) + X_R^\mu(\tau - \sigma), \quad Y^\mu = Y_L^\mu(\tau + \sigma) + Y_R^\mu(\tau - \sigma). \tag{313}$$

Then we directly obtain for the $X$-string:

$$X_L^{\mu,i}(\tau, 0) = X_R^{\mu,i}(\tau, 0) \tag{314}$$

and for the $Y$-string:

$$Y_L^{\mu,j}(\tau, 0) = Y_R^{\mu,j}(\tau, 0). \tag{315}$$

It is true up to a constant term that can be ignored while considered as either a function of $(\tau + \sigma)$ or $(\tau - \sigma)$ irrespectively.

As is seen, $X_L$ and $X_R$ are evaluated at the same time $\tau$. Therefore, we have locality at the intersection point as opposed to G. 't Hooft condition (311).

At the endpoints we get the same conditions (310).

## 5.5 The Solutions for the Equations of Motion in a Minkowski Background Spacetime

We continue to assume that the endpoints are massless as opposed to [92, 93], where the massive endpoints cases are investigated. Our analysis can be generalized for massive endpoints.

Here the rotation of the strings comes into play. We consider the motion on the plane, and two points are enough to define it. Any other motion demands higher dimensions and will unnecessary complicate our rotating configuration that is fully described in three dimensions.

The solution in this chapter is of a baryon with orbital angular momentum. The baryons, like the proton or the neutron with zero orbital angular momentum should be treated quantum mechanically, otherwise they will collapse.

As we are dealing with stringy particles, let's take the embedding spacetime to be the Minkowski spacetime. The signature of $\eta_{\mu\nu}$ is $(+1, -1, -1)$.

The equations of motion are

$$\Box X^\mu = \frac{1}{\sqrt{-h}} \partial_a(\sqrt{-h} h^{ab} \partial_b X^\mu) = 0, \tag{316}$$

where $\mu = 0, 1, 2$ denotes the components of $X$. As previously, all the calculations are correct for both branches of the $Y$-string too.

Note that since $T_1, T_2, T_3$ are not the same, then the wave vectors, $k_1, k_2, k_3$ ($k = \frac{2\pi}{\lambda}$), that will appear later, are not the same.

Variation of the sigma-model action with respect to $h_{ab}$, the equation (81), can be rewritten as

$$h_{ab} = \frac{2\eta_{\mu\nu} \partial_a X^\mu \partial_b X^\nu}{h^{cd} \partial_c X^\mu \partial_d X^\nu \eta_{\mu\nu}} = \Omega^2 h_{ab}, \tag{317}$$

where $h_{ab} = \eta_{\mu\nu} \partial_a X^\mu \partial_b X^\nu$ is the induced metric. As $\sqrt{-h} h^{ab}$ is invariant under conformal transformations (see equation (291)), then (316) reduces to

$$\Box X^\mu = \frac{1}{\sqrt{-h}} \partial_a(\sqrt{-h} h^{ab} \partial_b X^\mu) = 0. \tag{318}$$

As we will see there are solutions of the form

$$X^0 = c_1 \tau + c_2 \sigma; \tag{319}$$

$$X^1 = R(\sigma) \cos(\omega \tau); \tag{320}$$

$$X^2 = R(\sigma) \sin(\omega \tau), \tag{321}$$

where $c_1, c_2$ are some constants.

The Neumann boundary conditions are imposed at the intersection point ($\sigma = 0$). Therefore, $c_2 = 0$, and $X^0$ is a monotonic function of $\tau$:

$$X^0 = \tau. \tag{322}$$

The Neumann boundary conditions are also imposed at the endpoints and provide that $\sigma = 0$. Then again

$$X^0 = \tau. \tag{323}$$

Before the boundary conditions are imposed to $X^1$ and $X^2$, let's check that $R(\sigma)$ is an arbitrary function of $\sigma$.

The matrix elements are

$$h_{\tau\tau} = \eta_{00}\partial_\tau X^0 \partial_\tau X^0 + \eta_{11}\partial_\tau X^1 \partial_\tau X^1 + \eta_{22}\partial_\tau X^2 \partial_\tau X^2 =$$

$$= 1 - (-R(\sigma)\sin(\omega\tau)\omega)^2 - (R(\sigma)\cos(\omega\tau)\omega)^2 = 1 - R^2(\sigma)\omega^2; \tag{324}$$

$$h_{\sigma\sigma} = \eta_{00}\partial_\sigma X^0 \partial_\sigma X^0 + \eta_{11}\partial_\sigma X^1 \partial_\sigma X^1 + \eta_{22}\partial_\sigma X^2 \partial_\sigma X^2 =$$

$$= -(-(\frac{\partial R}{\partial \sigma})\sin(\omega\tau))^2 - ((\frac{\partial R}{\partial \sigma})\cos(\omega\tau))^2 = -(\frac{\partial R}{\partial \sigma})^2; \tag{325}$$

$$h_{\tau\sigma} = h_{\sigma\tau} = 0. \tag{326}$$

The inverse matrix elements are

$$h^{\tau\tau} = \frac{1}{1 - R^2(\sigma)\omega^2}; \tag{327}$$

$$h^{\sigma\sigma} = \frac{-1}{(\frac{\partial R}{\partial \sigma})^2}; \tag{328}$$

$$h^{\tau\sigma} = h^{\sigma\tau} = 0. \tag{329}$$

Then

$$\det h_{ab} = -(1 - R^2(\sigma)\omega^2)(\frac{\partial R}{\partial \sigma})^2. \tag{330}$$

$$\sqrt{-\det h_{ab}} \equiv \sqrt{-h} = \sqrt{(1 - R^2(\sigma)\omega^2)}(\frac{\partial R}{\partial \sigma}). \tag{331}$$

Therefore, the equations of motion

$$\partial_\tau(\sqrt{-h}h^{\tau\tau}\partial_\tau X^\mu) + \partial_\sigma(\sqrt{-h}h^{\sigma\sigma}\partial_\sigma X^\mu) + \partial_\tau(\sqrt{-h}h^{\tau\sigma}\partial_\sigma X^\mu) + \partial_\sigma(\sqrt{-h}h^{\sigma\tau}\partial_\tau X^\mu) = 0 \tag{332}$$

reduce to

$$\partial_\tau(\sqrt{-h}h^{\tau\tau}\partial_\tau X^\mu) + \partial_\sigma(\sqrt{-h}h^{\sigma\sigma}\partial_\sigma X^\mu) = 0. \tag{333}$$

Then for $\mu = 0$:

$$\partial_\tau(\sqrt{1 - R^2(\sigma)\omega^2}(\frac{\partial R}{\partial \sigma})\frac{1}{1 - R^2(\sigma)\omega^2}) = 0 \tag{334}$$

for $\mu = 1$:

$$\partial_\tau(\sqrt{1 - R^2(\sigma)\omega^2}(\frac{\partial R}{\partial \sigma})\frac{1}{1 - R^2(\sigma)\omega^2}(-R(\sigma)\omega\sin(\omega\tau))) +$$

$$+ \partial_\sigma(-\sqrt{1 - R^2(\sigma)\omega^2}(\frac{\partial R}{\partial \sigma})\frac{1}{(\frac{\partial R}{\partial \sigma})^2}(\frac{\partial R}{\partial \sigma})\cos(\omega\tau)) = 0. \tag{335}$$

for $\mu = 2$:

$$\partial_\tau(\frac{\partial R}{\partial \sigma}\frac{1}{\sqrt{1-R^2(\sigma)\omega^2}}R(\sigma)\omega\cos(\omega\tau))+$$
$$+\partial_\sigma(\sqrt{1-R^2(\sigma)\omega^2}(\frac{\partial R}{\partial \sigma})\frac{-1}{(\frac{\partial R}{\partial \sigma})^2}(\frac{\partial R}{\partial \sigma})\sin(\omega\tau)) = 0. \tag{336}$$

We do not specify $R(\sigma)$. However, the equations of motion (333) are satisfied.

As $R(\sigma)$ is arbitrary, we take it to be

$$R(\sigma) = \sigma - \frac{1}{k}\sin(k\sigma). \tag{337}$$

Next, the Neumann boundary conditions, $\frac{\partial X^\mu}{\partial \sigma} = 0$, are going to be set at the intersection point and then at the endpoints.

Again, at the intersection point $\sigma$ is assumed to be equal to $0$. Then

for $\mu = 1$ and $\mu = 2$ we obtain

$$X^1 = (\sigma - \frac{1}{k}\sin(k\sigma))\cos(\omega\tau); \tag{338}$$

$$X^2 = (\sigma - \frac{1}{k}\sin(k\sigma))\sin(\omega\tau). \tag{339}$$

At the endpoints $\sigma$ can not be assumed to be equal to $0$. Then

for $\mu = 1$ and $\mu = 2$ we obtain

$$X^1 = (\sigma - \frac{1}{k}\sin(k\sigma))\cos(\omega\tau); \tag{340}$$

$$X^2 = (\sigma - \frac{1}{k}\sin(k\sigma))\sin(\omega\tau). \tag{341}$$

Therefore, the Neumann boundary conditions at the endpoints give us the constraints on $k\sigma$ at the endpoints:

$$1 - \cos(k\sigma) = 0. \tag{342}$$

Then the conditions for the wave vectors are

$$k\sigma = 2\pi n, \tag{343}$$

where $n$ is an integer. The intersection point is excluded, then $n \neq 0$.

The energy of a single string is

$$E = -p_0 = -\int P_0^\tau d\sigma, \tag{344}$$

where $P_0^\tau \equiv \frac{\partial L}{\partial(\partial_\tau X^0)}$ and $L = -T\sqrt{(\partial_\tau X \partial_\sigma X)^2 - (\partial_\tau X)^2(\partial_\sigma X)^2}$. Then in our case

$$P_0^\tau = -T\frac{-(\partial_\sigma X)^2}{\sqrt{(\partial_\tau X \partial_\sigma X)^2 - (\partial_\tau X)^2(\partial_\sigma X)^2}} =$$
$$= -T(\frac{\partial R}{\partial \sigma})^2 \frac{1}{\sqrt{1 - R^2(\sigma)\omega^2(\frac{\partial R}{\partial \sigma})}} = -T(\frac{\partial R}{\partial \sigma})\frac{1}{\sqrt{1 - R^2(\sigma)\omega^2}}. \tag{345}$$

Therefore,

$$E_{single} = T\int (\frac{\partial R}{\partial \sigma})\frac{1}{\sqrt{1 - R^2(\sigma)\omega^2}}d\sigma = T\int_{R(\sigma_0=0)}^{R(\sigma_{endpoint})} \frac{dR}{\sqrt{1 - R^2(\sigma)\omega^2}} =$$
$$= -T\frac{1}{\omega}\arcsin(-\omega R(\sigma))|_{R(\sigma_0=0)}^{R(\sigma_{endpoint})} = T\frac{1}{\omega}\arcsin(\omega R(\sigma_{endpoint})). \tag{346}$$

The energy of our string baryon configuration is

$$E_{system} = T_1 \int_{R(\sigma_0=0)}^{R(\sigma_2)} \frac{dR}{\sqrt{1 - R^2(\sigma)\omega^2}} +$$
$$+T_2 \int_{R(\sigma_0=0)}^{R(\sigma_3)} \frac{dR}{\sqrt{1 - R^2(\sigma)\omega^2}} + T_3 \int_{R(\sigma_0=0)}^{R(\sigma_4)} \frac{dR}{\sqrt{1 - R^2(\sigma)\omega^2}} =$$
$$= T_1\frac{1}{\omega}\arcsin(\omega R(\sigma_2)) + T_2\frac{1}{\omega}\arcsin(\omega R(\sigma_3)) + T_3\frac{1}{\omega}\arcsin(\omega R(\sigma_4)). \tag{347}$$

The angular momentum of a single string is

$$J_{single}^{\mu\nu} = -\int (X^\mu P_\nu^\tau - X^\nu P_\mu^\tau)d\sigma. \tag{348}$$

In our case

$$J_{single}^{12} = -\int (X^1 P_2^\tau - X^2 P_1^\tau)d\sigma, \tag{349}$$

where

$$P_{1,2}^\tau = -T\frac{(\partial_\tau X \partial_\sigma X)\partial_\sigma X_{1,2} - (\partial_\sigma X)^2 \partial_\tau X_{1,2}}{\sqrt{(\partial_\tau X \partial_\sigma X)^2 - (\partial_\tau X)^2(\partial_\sigma X)^2}}. \tag{350}$$

Then

$$P_1^\tau = T(\frac{\partial R}{\partial \sigma})\frac{R(\sigma)\omega \sin(\omega\tau)}{\sqrt{1 - R^2(\sigma)\omega^2}}; \tag{351}$$

$$P_2^\tau = -T(\frac{\partial R}{\partial \sigma})\frac{R(\sigma)\omega \cos(\omega\tau)}{\sqrt{1 - R^2(\sigma)\omega^2}}. \tag{352}$$

Therefore,

$$J^{12}_{single} = -\int (-R(\sigma)\cos(\omega\tau)T(\frac{\partial R}{\partial \sigma})\frac{R(\sigma)\omega\cos(\omega\tau)}{\sqrt{1-R^2(\sigma)\omega^2}}-$$

$$-R(\sigma)\sin(\omega\tau)T(\frac{\partial R}{\partial \sigma})\frac{R(\sigma)\omega\sin(\omega\tau)}{\sqrt{1-R^2(\sigma)\omega^2}})d\sigma = T\omega\int_{R(\sigma_0=0)}^{R(\sigma_{endpoint})}\frac{R^2(\sigma)dR}{\sqrt{1-R^2(\sigma)\omega^2}} =$$

$$= T\omega(\frac{R(\sigma_{endpoint})}{-2\omega^2}\sqrt{1-R^2(\sigma_{endpoint})\omega^2} + \frac{1}{2\omega^2}(-\frac{1}{\omega}\arcsin(-\omega R(\sigma_{endpoint})))) =$$

$$= T\frac{1}{2\omega^2}(\arcsin(\omega R(\sigma_{endpoint})) - R(\sigma_{endpoint})\sqrt{1-R^2(\sigma_{endpoint})\omega^2}). \quad (353)$$

The angular momentum of our string baryon configuration is

$$J^{12}_{system} = T_1\frac{1}{2\omega^2}(\arcsin(\omega R(\sigma_2)) - R(\sigma_2)\sqrt{1-R^2(\sigma_2)\omega^2})+$$

$$+T_2\frac{1}{2\omega^2}(\arcsin(\omega R(\sigma_3)) - R(\sigma_3)\sqrt{1-R^2(\sigma_3)\omega^2})+$$

$$+T_3\frac{1}{2\omega^2}(\arcsin(\omega R(\sigma_4)) - R(\sigma_4)\sqrt{1-R^2(\sigma_4)\omega^2}). \quad (354)$$

The Regge trajectory is a relation between the angular momentum and the square of the energy. Let's see it here.

Up to this point we did not assume any specific dynamics of the point charges at $\sigma_2, \sigma_3, \sigma_4$. Here we assign a massless dynamics to these endpoints. Therefore, $R(\sigma_2)\omega = R(\sigma_3)\omega = R(\sigma_4)\omega = 1$, that is, each endpoint moves with the speed of light.

From our choice of $R(\sigma)$, (337), and the following condition on $k\sigma$, (342), we get

$$R(\sigma)|_{endpoint} = \sigma_{endpoint} = \frac{2\pi n}{k}. \quad (355)$$

Then for the lowest mode ($n = 1$), we obtain

$$R(\sigma)|_{endpoint} = \frac{1}{\omega}. \quad (356)$$

Then

$$E_{system} = (T_1 + T_2 + T_3)\frac{1}{\omega}(\frac{\pi}{2}); \quad (357)$$

$$J^{12}_{system} = (T_1 + T_2 + T_3)\frac{1}{2\omega^2}(\frac{\pi}{2}). \quad (358)$$

Then the Regge trajectory of our system is

$$J = 2\alpha' E^2, \quad (359)$$

where $\alpha' = \frac{1}{2\pi(T_1+T_2+T_3)}$ is the slope parameter.

## 5.6 Quantum Discussions

In our research we have ignored the structure of the quantum version of the theory. In this respect we propose this model as an effective model that may not be considered beyond the tree level, so the question of quantization may not be relevant. In any case any string model applied to hadron phenomenology has to be understood as an effective theory, since the fundamental theory is QCD.

It is interesting to notice that the structure of the action we have considered is linear in each measure. For example, the dependence on the measure $\Phi_X(\varphi)$ is (after integration by parts)

$$S = \int d\sigma d\tau \Phi_X(\varphi) \mathcal{L}_X + \ldots, \tag{360}$$

With $\mathcal{L}_X$ being independent of $\varphi_X^i$ ($\Phi_X(\varphi) = \epsilon^{ab}\epsilon_{ij}\partial_a\varphi_X^i\partial_b\varphi_X^j$), we get that the following infinite dimensional symmetry exists, up to a total divergence

$$\varphi_X^i \to \varphi_X^i + f^i(\mathcal{L}_X), \tag{361}$$

which preserves the linear structure of the action with respect to $\Phi_X(\varphi)$. This symmetry is infinite dimensional because it holds for any function $f^i(\mathcal{L}_X)$, and the set of all functions is an infinite dimensional set.

Similar arguments can be made to justify the linearity of $\Phi_Y(\varphi)$.

Without any assumptions on $\mathcal{L}$, we show now that $\varphi_i \to \varphi_i + f_i(\mathcal{L})$ is a symmetry. We do not specify whether $X$-string or $Y$-string is under consideration because the proof is similar for both strings.

$$\Phi(\varphi)\mathcal{L} = \epsilon^{ab}\epsilon_{ij}\partial_a\varphi_i\partial_b\varphi_j\mathcal{L} \to \epsilon^{ab}\epsilon_{ij}\partial_a(\varphi_i + f_i(\mathcal{L}))\partial_b(\varphi_j + f_j(\mathcal{L}))\mathcal{L} =$$

$$= \epsilon^{ab}\epsilon_{ij}\partial_a\varphi_i\partial_b\varphi_j\mathcal{L} + \epsilon^{ab}\epsilon_{ij}\partial_a f_i(\mathcal{L})\partial_b\varphi_j\mathcal{L} +$$

$$+\epsilon^{ab}\epsilon_{ij}\partial_a\varphi_i\partial_b f_j(\mathcal{L})\mathcal{L} + \epsilon^{ab}\epsilon_{ij}\partial_a f_i(\mathcal{L})\partial_b f_j(\mathcal{L})\mathcal{L}. \tag{362}$$

Let's consider the last three terms separately:

The fourth term is

$$\epsilon^{ab}\epsilon_{ij}\partial_a f_i(\mathcal{L})\partial_b f_j(\mathcal{L})\mathcal{L} = \epsilon^{ab}\epsilon_{ij} f'_i \partial_a\mathcal{L} f'_j \partial_b\mathcal{L} = 0, \tag{363}$$

where $f'_i(\mathcal{L}) = \frac{\partial f_i}{\partial \mathcal{L}}$. It is equal to zero because $\partial_a\mathcal{L}\partial_b\mathcal{L}$ is symmetric, while $\epsilon^{ab}$ is antisymmetric.

The second and the third terms are

Let's make the following definitions:

$$\epsilon^{ab}\epsilon_{ij}\partial_b\varphi_j = A_i^a, \qquad (364)$$

$$\epsilon^{ba}\epsilon_{ji}\partial_a\varphi_i = A_j^b. \qquad (365)$$

Then

$$A_i^a f_i'(\mathcal{L})\partial_a\mathcal{L}\mathcal{L} + A_j^b f_j'(\mathcal{L})\partial_b\mathcal{L}\mathcal{L} = 2A_i^a f_i'(\mathcal{L})\partial_a\mathcal{L}\mathcal{L} =$$

$$= \partial_a(A_i^a g_i(\mathcal{L})) = (\partial_a A_i^a)g_i(\mathcal{L}) + A_i^a g_i'(\mathcal{L})\partial_a\mathcal{L}, \qquad (366)$$

so since $\partial_a A_i^a = 0$, this is satisfied for

$$g_i'(\mathcal{L}) = 2f_i'(\mathcal{L})\mathcal{L}, \qquad (367)$$

$$g_i(\mathcal{L}) = 2\int d\mathcal{L} f_i'(\mathcal{L})\mathcal{L}. \qquad (368)$$

For every $f_i$ there exists such $g_i$. And the additional part of $\mathcal{L}$ after the transformation is a total derivative.

This infinite dimensional symmetry (since the function $f_i(\mathcal{L})$ is arbitrary) is defined for all configurations, including those that do not satisfy equations of motion. It must be so because in quantum theory we integrate over configurations that are off shell.

The existence of symmetries is usually used as a way to protect the theory, so that it keeps its basic structure even after quantum effects. Here the linearity on the measure, $\Phi$, in the action (so that it remains a measure) is protected under quantum corrections, in the case this symmetry (or a subgroup of this symmetry) is not plagued with anomalies.

Finally, other terms that do not contribute in the case of static configurations could be considered, like

$$\int d\sigma d\tau \epsilon^{ab}\partial_a(\frac{\Phi_X(\varphi)}{\sqrt{-h_X}})\partial_b(\frac{\Phi_Y(\varphi)}{\sqrt{-h_Y}})V(X,Y). \qquad (369)$$

This term could be relevant for quantum effects, like quantum creation processes, etc.

# 6 Higher Dimensional Extended Objects

Just like a string replaces a point particle, a p-brane replaces a string. A p-brane is a general notion of a p-dimensional extended object. In particular, a point particle is a zero-brane, a string is a one-brane, a membrane is a two-brane.

Just like a worldsheet replaces a worldline, a worldvolume replaces a worldsheet. A p-brane sweeps out a $(p+1)$-dimensional worldvolume.

In this chapter we look through a textbook p-brane, look at the two-measure p-brane and discuss the generalization of the Galileon measure string.

## 6.1 A standard p-brane

Just like a proper area replaces a proper time, a proper volume replaces a proper area .

$$S = -m \int ds \quad \to \quad S = -T \int dA \quad \to \quad S = -T_p \int d\mu_p, \tag{370}$$

where $d\mu_p$ is the $(p+1)$-dimensional worldvolume, $T_p$ is the p-brane tension.

The Nambu-Goto action for the p-brane replaces the Nambu-Goto action for the string.

$$S_{stringNG} = -T \int d\tau d\sigma \sqrt{-\det \gamma} \quad \to \quad S_{p-braneNG} = -T_p \int d^{p+1}\sigma \sqrt{-\det G_{xy}}, \tag{371}$$

where the induced metric $G_{xy}$ replaces the induced metric $\gamma_{ab}$

$$\gamma_{ab} = g_{\mu\nu} \partial_a X^\mu \partial_b X^\nu \quad \to \quad G_{xy} = g_{\mu\nu} \partial_x X^\mu \partial_y X^\nu. \tag{372}$$

The worldvolume is parameterized by $\sigma^{p+1}$ parameters, where $\sigma^0$ is time-like and all $\sigma^p$ are space-like.

The string Nambu-Goto action is invariant under reparameterization. The p-brane Nambu-Goto action is also invariant under reparameterization.

However, the string sigma-model action doesn't have such direct analogue. In order to obtain consistent classical equations of motion the cosmological constant term must be included.

$$S_{stringSM} = -\frac{T}{2} \int d\tau d\sigma \sqrt{-h} h^{ab} \partial_a X^\mu \partial_b X^\nu g_{\mu\nu} \tag{373}$$

$$\downarrow$$

$$S_{p-braneSM} = -\frac{T_p}{2} \int d^{p+1}\sigma \sqrt{-h} h^{xy} \partial_x X^\mu \partial_y X^\nu g_{\mu\nu} + \Lambda_p \int d^{p+1}\sigma \sqrt{-h}. \tag{374}$$

However, the addition of the cosmological constant term to the string action (the case $p = 1$) is not required. If we do so, then we obtain the equations of motion that are not the same as before, (81), (86).

$$S_{stringSM} = -\frac{T_1}{2} \int d^2\sigma \sqrt{-h} h^{xy} \partial_x X^\mu \partial_y X^\nu g_{\mu\nu} + \Lambda_1 \int d^2\sigma \sqrt{-h}. \tag{375}$$

Let us consider it in details.

The variation of the action (375) with respect to $h^{xy}$ is

$$T_1(\partial_x X^\mu \partial_y X_\mu - \frac{1}{2} h_{xy}(h^{uz} \partial_u X^\mu \partial_z X^\mu)) + \Lambda_1 h_{xy} = 0. \tag{376}$$

As $h_{xy} h^{xy} = 2$, the contraction with $h^{xy}$ leads to

$$h^{xy} h^{xy} \Lambda = T(\frac{1}{2} h_{xy} h^{xy}) h^{uz} \partial_u X^\mu \partial_z X^\mu = T(\frac{2}{2} - 1) h^{uz} \partial_u X^\mu \partial_z X^\mu = 0. \tag{377}$$

Therefore, as $h \neq 0$, $\Lambda_1 = 0$.

It proves that the addition of the cosmological constant term to the string action is not required. Otherwise, it brings inconsistency.

However, the addition of the cosmological constant term to the p-brane action (all the cases except $p = 1$) is required. If we do so, then we obtain the equations of motion that are the same as before, (81), (86).

$$S_{p-braneSM} = -\frac{T_p}{2} \int d^{p+1}\sigma \sqrt{-h} h^{xy} \partial_x X^\mu \partial_y X^\nu g_{\mu\nu} + \Lambda_p \int d^{p+1}\sigma \sqrt{-h}. \tag{378}$$

Let us consider it in details.

The variation of the action (378) with respect to $h^{xy}$ is

$$T_p(\partial_x X^\mu \partial_y X_\mu - \frac{1}{2} h_{xy}(h^{uz} \partial_u X^\mu \partial_z X^\mu)) + \Lambda_p h_{xy} = 0. \tag{379}$$

As $h_{uz} = \partial_u X^\mu \partial_z X^\mu g_{\mu\nu}$ (see the discussion on the induced and spacetime metrics in Section 2.2 for the details), we obtain

$$T_p(h_{xy} - \frac{1}{2} h_{xy}(h^{uz} h_{uz})) + \Lambda_p h_{xy} = 0. \tag{380}$$

The metric $h_{xy}$ being a common factor is dropped. Then

$$T_p(1 - \frac{1}{2} h^{uz} h_{uz}) + \Lambda_p = 0. \tag{381}$$

As $h^{uz} h_{uz} = p + 1$, then

$$T_p(1 - \frac{1}{2}(p+1)) + \Lambda_p = 0. \tag{382}$$

Therefore,

$$\Lambda_p = \frac{1}{2}(p-1)T_p. \qquad (383)$$

Indeed, for the string case $\Lambda_1 = 0$.

If we choose the cosmological constant term that way, then we obtain the Nambu-Goto action for the p-brane, $S_{p-braneNG}$.

$$S_{p-braneSM} = -\frac{T_p}{2}\int d^{p+1}\sigma\sqrt{-h}h^{xy}h_{xy} + \frac{1}{2}(p-1)T_p\int d^{p+1}\sigma\sqrt{-h} =$$

$$= -\frac{T_p}{2}\int d^{p+1}\sigma\sqrt{-h}(p+1) + \frac{1}{2}(p-1)T_p\int d^{p+1}\sigma\sqrt{-h} =$$

$$= (-\frac{p+1}{2} + \frac{p-1}{2})T_p\int d^{p+1}\sigma\sqrt{-h} = -T_p\int d^{p+1}\sigma\sqrt{-h}. \qquad (384)$$

## 6.2 The two-measure p-branes

The two-measure $\Phi(\varphi)$ is constructed out of $p+1$ (the number of worldvolume dimensions) scalar fields

$$\Phi_{p+1}(\varphi) = \epsilon^{x_1 x_2 \ldots x_{p+1}} \epsilon_{y_1 y_2 \ldots y_{p+1}} \partial_{x_1}\varphi^{y_1} \partial_{x_2}\varphi^{y_2} \ldots \partial_{x_{p+1}}\varphi^{y_{p+1}}. \qquad (385)$$

Therefore, the two-measure p-brane action is

$$S_{TMp-brane} = -\int d^{p+1}\sigma \Phi_{p+1}(\varphi)(h^{xy}\partial_x X^\mu \partial_y X^\nu g_{\mu\nu} + \frac{\epsilon^{x_1 x_2 \ldots x_{p+1}}}{\sqrt{-h}}\partial_{[x_1} A_{x_2 \ldots x_{p+1}]}). \qquad (386)$$

As in the string case, the $S_{TMp-brane}$ action consists of two terms.

As opposed to the string case, $S_{TMp-brane}$ action is not invariant under a set of diffeomorphisms in the space of measure fields coupled with a conformal transformation of the metric. That is, if

$$\varphi^y \to \varphi^{y'}, \qquad (387)$$

then

$$\Phi \to \Phi' = J\Phi \qquad (388)$$

and

$$h_{xy} \to h'_{xy} = Jh_{xy}, \qquad (389)$$

where $J$ is the Jacobian of the transformation (387).

However, the p-brane action is invariant under a global scaling symmetry

$$h_{xy} \longrightarrow e^{\theta} h_{xy}, \tag{390}$$

where $\theta$ is a constant, provided that

$$\varphi^y \longrightarrow \lambda^y \varphi^y, \tag{391}$$

which means

$$\Phi_{p+1} \longrightarrow \lambda \Phi_{p+1}. \tag{392}$$

The additional requirements for this symmetry to exist are

$$\lambda = e^{\theta} \tag{393}$$

and

$$A_{x_2...x_{p+1}} \longrightarrow \lambda^{\frac{p-1}{2}} A_{x_2...x_{p+1}}. \tag{394}$$

The variation with respect to $\varphi^y$ is

$$-h^{xy} \partial_x X^{\mu} \partial_y X^{\nu} g_{\mu\nu} + \frac{\epsilon^{x_1 x_2...x_{d+1}}}{\sqrt{-h}} \partial_{[x_1} A_{x_2...x_{d+1}]} = M. \tag{395}$$

As opposed to the string case, $M$ is a constant that is $\neq 0$.

The variation with respect to $h^{xy}$ is

$$-\Phi(\varphi)(\partial_x X^{\mu} \partial_y X^{\nu} g_{\mu\nu} - \frac{1}{2} h_{xy} \frac{\epsilon^{x_1 x_2...x_{d+1}}}{\sqrt{-h}} \partial_{[x_1} A_{x_2...x_{d+1}]}) = 0. \tag{396}$$

If we combine (395) and (396), then

$$\partial_x X^{\mu} \partial_y X^{\nu} g_{\mu\nu} = \frac{1}{2} h_{xy} (h^{uz} \partial_u X^{\mu} \partial_z X^{\nu} g_{\mu\nu} + M). \tag{397}$$

The sigma-model action with the cosmological constant term leads to the same equation.

The exact form of $M$ is

$$M = \frac{h^{uz} \partial_u X^{\mu} \partial_z X^{\nu} g_{\mu\nu} (1-p)}{1+p}. \tag{398}$$

If we combine (397) and (398), then

$$h_{xy} = \frac{1-p}{M} \partial_x X^{\mu} \partial_y X^{\nu} g_{\mu\nu}. \tag{399}$$

Then $h_{xy}$ is equal to the induced metric up to the constant $\frac{1-p}{M}$. However, the global scaling symmetry allows us to set $h_{xy}$ equal to the induced metric.

Therefore, the constant $M$ is

$$M = 1 - p. \tag{400}$$

Therefore, the main advantage of the two-measure p-brane is that the $S_{TMp-brane}$ does not contain a cosmological constant term as opposed to the $S_{p-braneSM}$. All p-branes (including strings) are treated on the same footing, i.e. the $S_{TMp-brane}$ is the same for all p-branes as opposed to the sigma-model action which is differ by the cosmological constant term for strings and other p-branes.

The variation with respect to $A_{x_2...x_{d+1}}$ is

$$\epsilon^{x_1...x_{p+1}}\partial_{x_1}\frac{\Phi(\varphi)}{\sqrt{-h}} = 0, \tag{401}$$

then

$$\frac{\Phi(\varphi)}{\sqrt{-h}} = Const. \tag{402}$$

It follows from the requirement of the consistency with $S_{p-braneSM}$ that $Const = T_p$. Then as in the string case, the brane tension is not put by hand but is generated spontaneously.

## 6.3 The Galileon p-branes

Once again, the Galileon measure is

$$\Phi(\chi) = \partial_x(\sqrt{-h}h^{xy}\partial_y\chi), \tag{403}$$

where $\chi$ is a Galileon scalar field since $\Phi(\chi)$ is invariant under a Galileon shift symmetry, (103), (104).

The conformally flat frame for the metric $h^{ab}$ exists only in two-dimensions. Therefore, the Galileon symmetry too exists only in two-dimensions. And there is no way to generalize a Galileon symmetry to the higher dimensions.

The key point is that the modified measure $\Phi(\chi)$ is conformal invariant and has the Galileon symmetry at the same time only in two dimensions, $p = 1$. So in fact the Galileon p-brane (except $p = 1$) does not exist.

However, if we modify $\Phi(\chi)$ in a special way, then the conformal invariance is restored in higher dimensions.

The modified ex-Galileon measure, $\Phi_{branes}(\chi)$, in D-dimensions is

$$\Phi_{branes}(\chi) = \partial_x(h^{xy}\sqrt{-h}\partial_y\chi(-2h^{uz}\partial_u\chi\partial_z\chi)^{\frac{D-2}{2}}). \tag{404}$$

If we repeat the sequence of actions from Chapter 2 (a Galileon string), then we obtain the same result

$$\mathcal{L} = 0, \tag{405}$$

provided that $\mathcal{L}$ is homogeneous.

The measure (404) is not the unique generalization.

The modified ex-Galileon measure, $\Phi_{branes}(\chi)$, in D-dimensions is

$$\Phi_{branes}(\chi) = \partial_x(h^{xy}\sqrt{-h}\partial_y\chi(F_{uz}F^{uz})^{\frac{D-2}{4}}). \tag{406}$$

The constraint (405) is obtained again, and the internal structure of measure does not reveal itself in the equations of motion.

The conformal invariant modified measure, $\Phi_{branes4}$, in four-dimensions is

$$\Phi_{branes4} = \epsilon^{uxyz} F_{ux} F_{yz}. \tag{407}$$

The conformal invariant modified measure, $\Phi_{branes6}$, in six-dimensions is

$$\Phi_{branes6} = \epsilon^{x_1 x_2 x_3 x_4 x_5 x_6} \epsilon_{y_1 y_2 y_3 y_4} \partial_{x_1}\varphi_{y_1} \partial_{x_2}\varphi_{y_2} \partial_{x_3}\varphi_{y_3} \partial_{x_4}\varphi_{y_4} F_{x_5 x_6}. \tag{408}$$

On the other hand, if we give up the conformal invariance and take the same $\Phi(\chi)$ as in two-dimensions for the higher dimensions, then the constraint $\mathcal{L} = 0$ is not feasible anymore.

Instead we will have

$$h_{xy}(-\frac{2}{\sqrt{-h}}\frac{\partial \mathcal{L}}{\partial h_{xy}}\Phi(\chi) - \partial^x\chi\partial^y\mathcal{L} - \partial^y\chi\partial^x\mathcal{L} + h^{xy}\partial_z\chi\partial^z\mathcal{L}) \neq 0 \tag{409}$$

The scalar field $\chi$ is a free dynamical degree of freedom. In the context of cosmology, this kind of models can provide interacting Dark Energy/Dark Matter scenarios [45].

# 7 Conclusions

The main purpose of our research is to suggest a modified action and to test it for several tasks. The whole thesis deals only with the classical actions.

Our research is divided into five parts.

In the first part we have constructed a new measure of integration $\Phi(\chi)$ out of the galileon field $\chi$. The Galileon measure, $\Phi(\chi)$, is

$$\Phi(\chi) = \partial_a(\sqrt{-h}h^{ab}\partial_b\chi). \tag{410}$$

The measure $\Phi(\chi)$ may indeed be called a Galileon measure because it is invariant under a Galileon shift symmetry in the conformally flat frame. The scalar field $\chi$ is a Galileon since $\Phi(\chi)$ is invariant under a Galileon shift symmetry in the conformally flat metric gauge:

$$\partial_a\chi \to \partial_a\chi + b_a, \tag{411}$$

$$\chi \to \chi + b_a\sigma^a, \tag{412}$$

where $b_a$ is a constant vector and $\sigma^a = (\tau, \sigma)$.

Previously, in the framework of the two measure theory the string tension was derived as a constant of integration, that is, it comes as an additional dynamical degree of freedom. Here a new $\Phi(\chi)$ gives the same result. The Galileon measure action, $S_{GM}$, is

$$S_{GM} = -\int d\tau d\sigma (h^{ab}\partial_a X^\mu \partial_b X^\nu g_{\mu\nu} - \frac{\epsilon^{cd}}{\sqrt{-h}}F_{cd})\partial_e(h^{ef}\sqrt{-h}\partial_f\chi), \tag{413}$$

where $\epsilon^{cd}$ is the Levi-Civita symbol, $F_{cd} = \partial_c A_d - \partial_d A_c$ is the field-strength of $A_c$.

Moreover, the degrees of freedom that appear at the action level do not appear in the equations of motion. Namely, the Galileon higher derivative theory leads to the second order equations of motion.

In the second part we have derived the supersymmetric string action with the Galileon one-scalar measure. It is

$$S_{GM\_SUSY} = -\int d\tau d\sigma \partial_a(h^{ab}\sqrt{-h}\partial_b\chi)(\mathcal{L}_{simpleGM} + \mathcal{L}_{additionalGM}), \tag{414}$$

where

$$\mathcal{L}_{simpleGM} = \frac{1}{2}h^{ab}\Pi_a^\mu\Pi_{b\mu}, \tag{415}$$

$$\mathcal{L}_{additionalGM} = i\frac{\epsilon^{cd}}{\sqrt{-h}}J_c^\alpha J_{\alpha d}, \tag{416}$$

where

$$\Pi_a^\mu = \partial_a X^\mu - i(\Theta^A \Gamma^\mu \partial_a \Theta^A), \tag{417}$$

$$J_a^\alpha = \partial_a \Theta^\alpha, \tag{418}$$

$$J_{\alpha a} = \partial_a \phi_\alpha - 2i(\partial_a X^\mu)\Gamma_{\mu\alpha\beta}\Theta^\beta - \frac{2}{3}(\partial_a \Theta^\beta)\Gamma^\mu_{\beta\delta}\Theta^\delta \Gamma_{\mu\alpha\epsilon}\Theta^\epsilon. \tag{419}$$

The important thing to notice is that we were guided by symmetry principles in every step of derivation. The very idea of inclusion fermions to the bosonic string totally reposes on the supersymmetry.

The integration measure $\Phi(\chi)$ is a density under diffeomorphisms on the worldsheet. It is constructed in such a way as to possess the Galileon shift symmetry. Moreover, when considering an action, the measure does not break the conformal transformation symmetry.

Then, since $\Phi(\chi)$ is a total derivative, $\mathcal{L} \to \mathcal{L} + const$ is a symmetry. It makes it possible to add $\mathcal{L}_{additionalGM}$ which is crucial for the whole theory.

As opposed to the Green-Schwarz superstring the term $\mathcal{L}_{additionalGM}$ which is the Wess-Zumino term is manifestly supersymmetric. Moreover as opposed to the research done by W. Siegel in [hep-th/9403144] our Galileon superstring action is presented with all terms being derived from the equations of motion.

In the third part we step aside and turn to the background of the modified measure theories. We consider the modified two-measure theory applied to the two scalar field system. It is an extension of the research done by E. Guendelman in [gr-qc/9901017].

We start with the scalar field, surround it with three supplementary scalar fields and investigate the resulting action. The two scalar field system is

$$S = \frac{1}{2}\int \Phi(\varphi)\mathcal{L}d^2x, \qquad \mathcal{L} = \frac{1}{2}(\partial_\mu\phi_1\partial_\nu\phi_1 g^{\mu\nu} + \partial_\mu\phi_2\partial_\nu\phi_2 g^{\mu\nu}), \tag{420}$$

where $\phi_1$ is the former scalar field $\phi$ and $\phi_2$ is the supplemented one.

One scalar field is physically equivalent to the former scalar field. However, the new measure of integration is constructed from the other two scalars.

The two-measure, $\Phi(\varphi)$, in two dimensions is

$$\Phi(\varphi) = \epsilon^{ab}\epsilon_{ij}\partial_a\varphi^i\partial_b\varphi^j, \tag{421}$$

where $\varphi^i, \varphi^j$ are scalar fields.

The source of the following findings is this modified measure. First, we show that the gradient of this initial scalar field is finite and in particular there is a sector which can be presented in the form of the Born-Infeld scalar.

$$S_{eff} = \int \sqrt{const - \partial_\mu \phi_1 \partial^\mu \phi_1} d^2x; \qquad (422)$$

Second, the initial action is scale invariant, however, the invariance gets spontaneously broken. In addition to having spontaneous symmetry breaking, our physical system serves as an example of a system with the symmetry that does not lead to the conserved charge. In the cosmological cases when the scale symmetry is spontaneously broken, there is a conserved current and since no singular behavior of the conserved current is obtained, so there is a conserved scale charge and the Goldstone theorem holds.

Note that in our case of a scalar field there remains a massless field which is a Goldstone boson of the shift symmetry ($\phi_1 \to \phi_1 + constant$), not of the scale symmetry, because for the scale symmetry that theorem cannot be applied since the dilatation charge is not conserved.

In the fourth part we implement the modified two-measure theory into the string models of mesons and baryons.

We start our consideration with a single string. But instead of using $\sqrt{-h}$ as a measure as it is done in a sigma-model action, we take a measure $\Phi$, which is constructed out of two scalar fields. We are permitted to do it as long as it is a density under arbitrary diffeomorphisms on the world-sheet spacetime, which is indeed the way we have constructed our measure.

Subsequently, the string tension appears as a constant of integration. In this framework it is not a scale that is put ad hoc but an additional dynamical degree of freedom. In [P.K. Townsend, Phys.Lett. B277, 285-288, (1992)] besides the supersymmetric extension, the gauge field and the new density in the action are quadratic and inverse, correspondingly, as opposed to the linear ones in our case. While our initial settings are different, the string tension appears there as an integration constant too. In principle the mechanism studied could be formulated also in the framework of the [P.K. Townsend, Phys.Lett. B277, 285-288, (1992)]'s approach, the action would be then modified by adding sources, etc.

In the string meson model a single string is considered. The charges at the endpoints of the string lead via the tension discontinuity to the Neumann boundary conditions. The string meson action, $S_{meson}$, is

$$S_{meson} = -\int d\sigma d\tau \Phi(\varphi)[\frac{1}{2}h^{ab}\partial_a X^\mu \partial_b X^\nu g_{\mu\nu} - \frac{\epsilon^{cd}}{2\sqrt{-h}}F_{cd}] + \int d\sigma d\tau A_i j^i, \qquad (423)$$

where $j^i$ is the current of point-like charges.

In the string baryon model a two strings are considered. Then we consider two strings. The endpoint of one string is connected to the internal part of the other one. The charge in the internal part of the string lead via the tension alterations to the Neumann boundary conditions. By the addition of an interaction term to the modified action we obtain the conditions for the intersection that are the Dirichlet boundary conditions.

The action governing the string baryon configuration is

$$S_{baryon} = -\int d\tau d\sigma \Phi(\varphi)_X[\frac{1}{2}h_X^{ab}\partial_a X^\mu \partial_b X^\nu g_{\mu\nu} - \frac{\epsilon^{cd}}{2\sqrt{-h_X}}F_{cd}] + \sum_{i=1,2}\int d\tau d\sigma A_i j_A^i +$$

$$-\int d\tau d\sigma \Phi(\varphi)_Y[\frac{1}{2}h_Y^{ab}\partial_a Y^\mu \partial_b Y^\nu g_{\mu\nu} - \frac{\epsilon^{cd}}{2\sqrt{-h_Y}}F_{cd}] + \sum_{j=3,4,5}\int d\tau d\sigma B_j j_B^j +$$

$$+ \int d\tau d\sigma (\lambda_1 \sqrt{-h_X}h_X^{ab} + \lambda_2 \sqrt{-h_Y}h_Y^{ab})\partial_a(\frac{\Phi(\varphi)_X}{\sqrt{-h_X}})\partial_b(\frac{\Phi(\varphi)_Y}{\sqrt{-h_Y}})V(X,Y), \qquad (424)$$

where $\lambda_1$, $\lambda_2$ are positive coefficients and $V(X,Y)$ is a potential which in its most simple form is $V = (X-Y)^2$.

The $A$-gauge field couples directly to the measure of the $X$-string, and the $B$-gauge field couples directly to the measure of $Y$-string. The charges $e_1$ and $e_2$ belong to the $A$-field, the charges $e_3$, $e_4$ and $e_5$ belong to the $B$-field. The field strength $F_{ab}^A$ arises from the $A$ gauge field, and the field strength $F_{ab}^B$ arises from the $B$ gauge field.

Neumann boundary conditions are presented at all endpoints, $l = 1, 2, m = 3, 4$:

$$\partial_\sigma X^\mu(\tau, \sigma_l) = 0, \quad \partial_\sigma Y^\mu(\tau, \sigma_m) = 0. \qquad (425)$$

Both Dirichlet and Neumann boundary conditions are presented at the single intersection point.

$$X^\mu|_{\sigma=\sigma_5} = Y^\mu|_{\sigma=\sigma_5}, \quad \partial_\sigma Y^\mu(\tau, \sigma_5) = 0. \qquad (426)$$

To avoid any confusion: generally, $\sigma$ denotes the location in the string. Especially, $\sigma_1$ and $\sigma_2$ denote the $X$-string endpoints with the charges $e_1$ and $e_2$, $\sigma_3$ and $\sigma_4$ denote the $Y$-string endpoints with the charges $e_3$ and $e_4$, $\sigma_5$ denotes the point in the $Y$-string where the charge $e_5$ is located. Strings intersect and by the construction, $\sigma_1$ and $\sigma_5$ denote the same location, the point of intersection. The charge $e_1$ being the endpoint of the $X$-string terminates the tension of the $X$-string and, as any other endpoint charge, raises the Neumann boundary conditions. The charge $e_5$ being the internal charge of the $Y$-string changes the value of the tension of the $Y$-string and raises the Neumann boundary conditions. Anytime, the tension of any string changes (including reducing to zero) the Neumann boundary conditions arise. In order for strings to intersect, the Dirichlet boundary conditions at the point $\sigma_1$ (the same as to say $\sigma_5$) are obtained. Starting from Section 6, where we make a comparison and later, when we solve the equations of motion, we denote the point of intersection as $\sigma_0$, that is $\sigma_1 = \sigma_5 = \sigma_0$ and without loss of generality set it to 0.

The Neumann boundary conditions differ from the ones obtained by G. 't Hooft in [hep-th/0408148]. There the Neumann boundary conditions at the intersection point hold only for the sum $\sum_{k=1}^{3} X^{\mu,k}$.

It leads to a nonlocality.

A remarkable difference with G. 't Hooft in [hep-th/0408148] is that in our case the boundary conditions at the intersection point become local in time:

$$X_R^{\mu,i}(\tau,0) = X_L^{\mu,i}(\tau,0) \tag{427}$$

and

$$Y_R^{\mu,j}(\tau,0) = Y_L^{\mu,j}(\tau,0). \tag{428}$$

Here the same $\tau$ is involved. This is due to the specific physics introduced to induce the boundary conditions: the dynamical tension mechanism which determines how strings end, interact etc. As we have seen, the approach with introduced Lagrange multipliers to enforce the strings to meet at some point $\sigma = 0$ is not equivalent.

The latter, as it is pointed out by 't Hooft himself, may require additional boundary conditions in particular because the method for propagating the signal from $\sigma = 0$ to $\sigma = L^k(\tau)$ and back can become ill-defined in some limits. Such problem is absent in our approach due to the locality in time of the boundary conditions.

In QCD which is the underlying microscopic theory in our case the chromo-electric field is generated by static quarks and leads to tube-like structures. String-like behavior in the chromo-electric field is well known. In our model charges too are responsible for the very existence of the string. It is demonstrated through the value of the string tension. In the string baryon model we use two types of abelian gauge fields. It is an indication that a more sophisticated theory that includes non-abelian gauge fields (and therefore many gauge fields automatically) would be more suitable.

Note that when the tensions are taken to be in a way that two of them are much greater than the third one then the diquark model arises. The introduction of the diquark provides us with a possibility to construct a more effective scheme for highly excited states, with less degrees of freedom and less number of highly excited states.

We have studied some rotating string solutions of equations of motion with the new boundary conditions for the sake of completeness. We obtain the energy and the angular momentum of our system, then assuming that each endpoint is a dynamical massless particle, the Regge trajectory with the slope parameter that depends on three different tensions is obtained.

In the fifth part we extend the research to higher dimensional extended objects with the Galileon measure, $\Phi(\chi)$. It turns out that only strings, being two-dimensional objects, have the Galileon symmetry and are conformal invariant at the same time. Therefore, we construct either conformal invariant extended object with one of the modified measures

$$\Phi(\chi) = \partial_u(h^{ux}\sqrt{-h}\partial_x\chi(-2h^{yz}\partial_y\chi\partial_z\chi)^{\frac{D-2}{2}}), \qquad \Phi(\chi) = \partial_u(h^{ux}\sqrt{-h}\partial_x\chi(F_{yz}F^{yz})^{\frac{D-2}{4}}), \tag{429}$$

or non-conformal invariant but still with the Galileon symmetry. However, the original equations of motion are not obtained.

# References

[1] B. Zwiebach, A First Course in String Theory (2006)

[2] K. Becker, M. Becker, J. Schwarz, String Theory and M-Theory (2007)

[3] J. Polchinski, String theory (1998)

[4] M.B. Green, J.H. Schwarz, E. Witten, Superstring Theory (1987)

[5] T. Banks, W. Fischler, S.H. Shenker, L. Susskind, Phys.Rev. D55, 5112-5128, (1997)

[6] Y. Hyakutake, JHEP 1409, 075 (2014)

[7] D. O'Connor, V. G. Filev, PoS CORFU 2015, 111 (2016)

[8] M.Hanada, Int.J.Mod.Phys. A31, no.22, 1643006 (2016)

[9] A. Chaney, A. Stern, Phys. Rev. D 95, no. 4, 046001 (2017)

[10] G.'t Hooft, Nucl.Phys. B72, 461, (1974)

[11] J.M. Maldacena, Int.J.Theor.Phys. 38, 1113-1133 (1999), Adv.Theor.Math.Phys. 2, 231-252, (1998)

[12] S. S. Gubser, M. Heydeman, C. Jepsen, M. Marcolli, S. Parikh, I. Saberi, B. Stoica, B. Trundy, JHEP 1706, 157, (2017)

[13] V.E. Didenko, M.A. Vasiliev, Phys.Lett. B775 (2017) 352-360

[14] C. Couzens, C. Lawrie, D. Martelli, S. Schafer-Nameki, Jin-Mann Wong, JHEP 1708, 043, (2017)

[15] P. Caputa, N. Kundu, M. Miyaji, T. Takayanagi, K. Watanabe, JHEP 1711, 097, (2017)

[16] S. Giombi, R. Roiban, A. A. Tseytlin, Nucl.Phys. B922, 499-527, (2017)

[17] J. de Boer, E. Llabres, J. F. Pedraza, D. Vegh, arXiv:1709.01052, (2017)

[18] J. Erdmenger, N. Miekley, MPP-2017-236 (2017)

[19] R. Gopakumar, C. Vafa, Adv.Theor.Math.Phys. 3 (1999) 1415-1443

[20] A. Gorsky, B. Le Floch, A. Milekhin, N. Sopenko, Nucl.Phys. B920, 122-156, (2017)

[21] C. Krishnan, K.V.P. Kumar, JHEP 1710 (2017) 099, (2017)

[22] T.D. Brennan, F. Carta, C. Vafa, IFT-UAM-CSIC-17-105

[23] A. Strominger, Nucl.Phys. B274, 253, (1986)

[24] B. de Wit, D.J. Smit, N.D.H. Dass, Nucl.Phys. B283, 165, (1987)

[25] N. T. Macpherson, A. Tomasiello, JHEP 1709, 126, (2017)

[26] A. Otal, L. Ugarte, R. Villacampa, Nucl.Phys. B920, 442-474, (2017)

[27] N. Halmagyi, D. Israel, M. Sarkis, E.E. Svanes, JHEP 1708, 138, (2017)

[28] A. Strominger, C. Vafa, Phys.Lett. B379, 99-104, (1996)

[29] S. Ferrara, R. Kallosh, A. Strominger, Phys.Rev. D52, R5412-R5416, (1995)

[30] P. A. Cano, P. Meessen, T. Ortin, P. F. Ramirez, JHEP 1712 (2017) 092

[31] H. Afshar, D. Grumiller, M.M. Sheikh-Jabbari, H. Yavartanoo, JHEP 1708, 087, (2017)

[32] F. Benini, H. Khachatryan, P. Milan, Class.Quant.Grav. 35, no.3, 035004, (2018)

[33] M. Guo, S. Song, H. Yan, Phys. Rev. D 101, 024055 (2020)

[34] G. Veneziano, Nuovo Cim. A57, 190-197, (1968)

[35] Y. Nambu, In *Detroit 1969, Symmetries and quark models* 269-278. In *Eguchi, T. (ed.) et al.: Broken symmetry* 258-267 Conference: C69-06-18, p.269-278, (1969)

[36] H.B. Nielsen, Submitted to the 15th International Conference on High Energy Physics (Kiev) (1970)

[37] L. Susskind, Nuovo Cim. A69 (1970) 457-496

[38] Y. Nambu, Notes prepared for the Copenhagen High Energy Symposium, (1970)

[39] T. Goto, Prog.Theor.Phys. 46 (1971) 1560-1569, (1971)

[40] O. Hara, Prog.Theor.Phys. 46 (1971) 1549-1559, (1971)

[41] E.I. Guendelman, A.B. Kaganovich, Phys.Rev.D55:5970 (1996)

[42] E.I. Guendelman, Mod.Phys.Lett. A14, 1043-1052, (1999)

[43] E.I. Guendelman, A.B. Kaganovich, Annals Phys. 323, 866-882 (2008)

[44] S. del Campo, E.I. Guendelman, R. Herrera, P. Labrana, JCAP, 1006, 026 (2010)

[45] D. Benisty, E.I. Guendelman, Eur.Phys.J.C77,n.6, 396 (2017)

[46] D. Benisty, E.I. Guendelman, Z. Haba, Phys.Rev.D 99, 12, 123521(2019), Phys.Rev.D 101, 4, 049901 (2020) (erratum)

[47] E.I. Guendelman, Class.Quant.Grav. 17, 3673-3680 (2000)

[48] E.I. Guendelman, A.B. Kaganovich, E. Nissimov, S. Pacheva, Phys.Rev.D66:046003 (2002)

[49] E.I. Guendelman, Phys.Lett., B580, 87-92 (2004)

[50] G.R. Dvali, G. Gabadadze, M. Porrati, Phys.Lett., B485, 208-214 (2000)

[51] A. Nicolis, R. Rattazzi, E. Trincherini, Phys.Rev.D79:064036 (2009)

[52] C. Deffayet, G. Esposito-Farese, A. Vikman, Phys.Rev.D79:084003 (2009)

[53] M.B. Green, J.H. Schwarz, Nucl.Phys.B243, 285 (1984)

[54] W. Siegel, Phys.Rev.D50:2799 (1994)

[55] E.I. Guendelman, A.B. Kaganovich, Phys.Rev.D 53, 7020-7025, (1996)

[56] E.I. Guendelman, Mod.Phys.Lett.A 14, 1397, (1999)

[57] M. Born, L. Infeld, Proc.Roy.Soc.Lond. A144, no.852, 425-451, (1934)

[58] C.G. Callan, J.M. Maldacena, Nucl.Phys. B513, 198-212, (1998)

[59] G.W. Gibbons, Nucl.Phys. B514, 603-639, (1998)

[60] J.A. Feigenbaum, P.G.O. Freund, M. Pigli, Phys.Rev. D57, 4738-4744, (1998)

[61] S. Deser, G.W. Gibbons, Class.Quant.Grav. 15, L35-L39, (1998)

[62] G.N. Felder, L. Kofman, , A. Starobinsky, JHEP 0209, 026, (2002)

[63] D.N. Vollick, Gen.Rel.Grav. 35, 1511-1516, (2003)

[64] Jian-gang Hao, Xin-zhou Li, Phys.Rev. D68, 043501, (2003)

[65] Dan N. Vollick, Phys.Rev. D72, 084026, (2005)

[66] W. Fang, H.Q. Lu, Z.G. Huang, K.F. Zhang, Int.J.Mod.Phys. D15, 199-214, (2006)

[67] S. Jana, S. Kar, Phys.Rev. D94, no.6, 064016, (2016)

[68] V.I. Afonso, G.J. Olmo, D. Rubiera-Garcia, JCAP 1708, no.08, 031, (2017)

[69] S. Jana, S. Kar, Phys.Rev. D96, no.2, 024050, (2017)

[70] G. 't Hooft, Phys.Rept. 142, 357-387, (1986)

[71] E.I. Guendelman, Class.Quant.Grav. 17, 361-372, (2000)

[72] E.I. Guendelman, R. Herrera, P. Labrana, E. Nissimov, S. Pacheva, Gen.Rel.Grav. 47, no.2, 10, (2015)

[73] E.I. Guendelman, H. Nishino, S. Rajpoot, Phys.Lett. B732, 156-160, (2014)

[74] E.I. Guendelman, E. Nissimov, S. Pacheva, M. Vasihoun, Bulg.J.Phys. 40, 121-126, (2013)

[75] S. del Campo, E.I. Guendelman, A.B. Kaganovich, R. Herrera, P. Labrana, Phys.Lett. B699, 211-216, (2011)

[76] S. del Campo, E.I. Guendelman, R. Herrera, P. Labrana, JCAP 1006, 026, (2010)

[77] E.I. Guendelman, A.B. Kaganovich, Annals Phys. 323, 866-882, (2008)

[78] E.I. Guendelman, A.B. Kaganovich, Phys.Rev. D75, 083505, (2007)

[79] E.I. Guendelman, O. Katz, Class.Quant.Grav. 20, 1715-1728, (2003)

[80] J. Garcia-Bellido, J. Rubio, M. Shaposhnikov, D. Zenhausern, Phys.Rev. D84, 123504, (2011)

[81] F. Bezrukov, G.K. Karananas, J. Rubio, M. Shaposhnikov, Phys.Rev. D87 (2013) no.9, 096001, (2013)

[82] P.G. Ferreira, C.T. Hill, G.G. Ross, Phys.Rev. D98, no.11, 116012, (2018)

[83] P. G. Ferreira, C.T. Hill, J. Noller, G.G. Ross, Phys.Rev. D97, no.12, 123516, (2018)

[84] A. Chodos, C. B. Thorn, Nucl.Phys. B72, 509 (1974)

[85] X. Artru, Nucl.Phys. B85, 442 (1975)

[86] P.A. Collins, J.F.L. Hopkinson, R.W. Tucker, Nucl.Phys. B100, 157 (1975)

[87] K. Sundermeyer, A. de la Torre, Phys.Rev.D15, 1745 (1977)

[88] K. Kikkawa, Masa-aki Sato, Phys.Rev.Lett. 38, 1309 (1977)

[89] G. S. Sharov, Phys.Rev.D58, 114009(1998)

[90] G. S. Sharov, Phys.Rev.D62, 094015 (2000)

[91] G.'t Hooft, ITP-UU-04-17, SPIN-04-10 (2004)

[92] J. Sonnenschein, D. Weissman, JHEP 1408, 013, (2014)

[93] J. Sonnenschein, D. Weissman, JHEP 1502, 147, (2015)

[94] J. Sonnenschein, D. Weissman, Nucl.Phys. B920, 319-344, (2017)

[95] J. Sonnenschein, D. Weissman, Nucl.Phys. B927, 368-454, (2018)

[96] A.V. Anisovich, V.V. Anisovich, M.A. Matveev, V.A. Nikonov, A.V. Sarantsev, T.O. Vulfs, Int.J.Mod.Phys. A25, 2965-2995 (2010)

[97] E. I. Guendelman, A. B. Kaganovich, Phys.Rev.D56:3548-3554, (1997)

[98] E. I. Guendelman, Phys.Lett. B580, 87-92, (2004)

[99] E. I. Guendelman, A. Kaganovich, E. Nissimov, S. Pacheva, Phys.Rev.D72:086011, (2005)

[100] S. Ansoldi, E.I. Guendelman, E. Spallucci, Mod.Phys.Lett. A21, 2055-2065, (2006)

[101] A. Kato, D. Singleton, Int.J.Theor.Phys. 41, 1563-1572 (2002)

[102] D. Singleton, A. Kato, A. Yoshida, Phys.Lett. A330, 326-337 (2004)

[103] E. Taylor, J. Wheeler, Spacetime Physics (2006)

# 8 Appendix 1.

It is rather common when the metric in the action appears under some mathematical operations. When we vary the action with respect to the metric we may face difficulties. Here we present some variations that are frequently used in the research.

The metric is $h^{ab}$, and it is arbitrary, unless otherwise stated.

- 
$$\delta h_{cd} = -\delta h^{ab} h_{bd} h_{ac}. \tag{430}$$

Proof:

By the definition of the inverse metric:

$$h^{ea} h_{ab} = \delta^e_b, \tag{431}$$

$$(h^{ea} + \delta h^{ea})(h_{ab} + \delta h_{ab}) = \delta^e_b, \tag{432}$$

$$\delta h^{ea} h_{ab} + h^{ea} \delta h_{ab} = 0, \tag{433}$$

$$h^{ea} \delta h_{ab} = -\delta h^{ea} h_{ab} = -\delta h^{ec} h_{cb}, \tag{434}$$

$$\delta h_{ab} = -\delta h^{ec} h_{cb} h_{ea}. \tag{435}$$

- 
$$\delta h = -h h_{ab} \delta h^{ab}. \tag{436}$$

Proof:

From linear algebra:

$$\det(Matrix) = e^{Tr \log(Matrix)}. \tag{437}$$

$$\delta h = \delta \det(h_{ab}) = h \delta Tr(\log h_{ab}) = h h^{ab} \delta h_{ab}. \tag{438}$$

- 
$$\delta\sqrt{-h} = -\frac{1}{2}\sqrt{-h}h_{ab}\delta h^{ab}. \tag{439}$$

Proof:

$$\delta\sqrt{-h} = \delta(-h)^{\frac{1}{2}} = -\frac{1}{2}\frac{1}{\sqrt{-h}}\delta h = \frac{1}{2}\frac{1}{\sqrt{-h}}hh_{ab}\delta h^{ab}. \tag{440}$$

- Let see in details the variation of $S_{GM}$ with respect to the two-dimensional worldsheet metric $\delta h^{ab}$.

$$\delta[(h^{ab}\partial_a X^c \partial_b X^d g_{cd} - \frac{\epsilon^{eg}}{\sqrt{-h}}F_{eg})\partial_k(h^{kf}\sqrt{-h}\partial_f\chi)] = \tag{441}$$

$$= \partial_a X^c \partial_b X^d g_{cd}\partial_k(h^{kf}\sqrt{-h}\partial_f\chi)gh^{ab}+$$

$$+\partial_k(h^{ab}\partial_a X^c \partial_b X^d g_{cd})\sqrt{-h}\partial_f\chi gh^{kf}+$$

$$+\partial_k(h^{ab}\partial_a X^c \partial_b X^d g_{cd})h^{kf}\partial_f\chi(-\frac{1}{2})\sqrt{-h}h_{ab}gh^{ab}-$$

$$-\partial_k(h^{kf}\sqrt{-h}\partial_f\chi)\epsilon^{eg}F_{eg}(\sqrt{-h})^{-2}\frac{1}{2}\sqrt{-h}h_{ab}gh^{ab}-$$

$$-\epsilon^{eg}F_{eg}(\sqrt{-h})^{-1}(\sqrt{-h}\partial_f\chi gh^{kf} + h^{kf}\partial_f\chi(-\frac{1}{2}\sqrt{-h}h_{ab}gh^{ab})) = \tag{442}$$

$$= \partial_a X^c \partial_b X^d g_{cd}\partial_k(h^{kf}\sqrt{-h}\partial_f\chi)gh^{ab}-$$

$$-\partial_k(h^{kf}\sqrt{-h}\partial_f\chi)\epsilon^{eg}F_{eg}(\sqrt{-h})^{-1}\frac{1}{2}h_{ab}gh^{ab}. \tag{443}$$

Then

$$\partial_k(h^{kf}\sqrt{-h}\partial_f\chi)(\partial_a X^c \partial_b X^d g_{cd} - \frac{1}{2}\frac{\epsilon^{eg}}{\sqrt{-h}}F_{eg}h_{ab})gh^{ab} = 0. \tag{444}$$

Then we get exactly the variation of $S_{TM}$ with respect to $h^{ab}$.

# 9 Appendix 2.

- The Dirac matrices, $\rho^a$, obey the following anticommutation relation:

$$\{\rho^a, \rho^b\} = 2\eta^{ab}. \tag{445}$$

The Dirac matrices, $\rho^a$, are

$$\rho^0 = \begin{pmatrix} 0 & -1 \\ 1 & 0 \end{pmatrix}, \qquad \rho^1 = \begin{pmatrix} 0 & 1 \\ 1 & 0 \end{pmatrix} \tag{446}$$

This is a Majorana representation.

The Majorana spinors $\psi^c$ are real: $\psi^\star_\pm = \psi_\pm$.

- Due to the anticommuting nature, the most general function of a Grassman variable $\theta$ is

$$f(\theta) = a\theta + b. \tag{447}$$

The Grassman integral over a single Grassman variable is

$$\int d\theta (a + \theta b) = b \tag{448}$$

and by the convention

$$\int d\theta = 0, \qquad \int d\theta\, \theta = 1. \tag{449}$$

The Grassman integral over two Grassman variables is

$$\int d^2\theta\, \bar{\theta}\theta = -2i. \tag{450}$$

- The Fierz transformation is

$$\theta_A \bar{\theta}_B = -\frac{1}{2}\delta_{AB} \bar{\theta}_C \theta_C. \tag{451}$$